\documentclass[11pt,a4paper]{article}
\usepackage{epsfig}
\usepackage[T1]{fontenc}    % Accents cods dans la fonte.
\usepackage{graphicx}
\usepackage{pstricks,pst-coil,pst-fill,pst-plot}
\usepackage{cite,indentfirst}
\usepackage[fleqn]{amsmath}    % Les symboles les plus frquents.
\usepackage{amssymb,amsfonts}   % Des fontes, eg pour \mathbb.
\usepackage{verbatim}   % Pour les codes sources en informatique.
\usepackage{mathrsfs}   % Des lettres majuscules cursives (\mathscr).
\usepackage{dsfont}
\usepackage{enumerate}     % met en place l'environement enumerate pour les listes
\usepackage{txfonts}
\usepackage{vmargin}        % Regler la taille de la feuille.

\usepackage{color}

\setmarginsrb{1.8cm}{2cm}{1.8cm}{2cm}{1cm}{1cm}{1cm}{1.6cm}
 \makeatletter
 \@addtoreset{equation}{section}
 \makeatother

\providecommand{\bysame}{\leavevmode\hbox to3em{\hrulefill}\thinspace}
\providecommand{\MR}{\relax\ifhmode\unskip\space\fi MR }
\providecommand{\href}[2]{#2}

\newcommand{\bra}[1]{\big< \,#1\,|}
\newcommand{\ket}[1]{|\,#1\, \big>}
\newcommand{\braket}[2]{\big<\,#1 \, \big|\,  #2  \big>}

\let\ua=\uparrow
\let\da=\downarrow
\let\tend=\rightarrow

\long\def\symbolfootnote[#1]#2{\begingroup%
\def\thefootnote{\fnsymbol{footnote}}\footnote[#1]{#2}\endgroup}

\newtheorem{theorem}{Theorem}[section]
\newtheorem{prop}[theorem]{Proposition}
\newtheorem{cor}[theorem]{Corollary}

\newtheorem{conj}[theorem]{Conjecture}

\newtheorem{lemme}[theorem]{Lemma}

\def\Proof{\medskip\noindent {\it Proof --- \ }}

\def\qed{\hfill\rule{2mm}{2mm}}

\newcommand\beq{\begin{equation}}
\newcommand\enq{\end{equation}}
\newcommand\bem{\begin{multline}}
\newcommand\enm{\end{multline}}

\def\beqa{\begin{eqnarray}}
\def\eeqa{\end{eqnarray}}
\def\ba{\begin{array}}
\def\ea{\end{array}}
\def\det{\operatorname{det}}

\newcommand{\f}[2]{{\ensuremath{%
    \mathchoice%
    {\dfrac{#1}{#2}}
    {\dfrac{#1}{#2}}
    {\frac{#1}{#2}}
    {\frac{#1}{#2}}
}}}
\newcommand{\tf}[2]{\ensuremath{#1/#2}}
\def\a{\alpha}

\def\be{\beta}

\def\Ga{\Gamma}

\def\de{\delta}

\def\De{\Delta}
\def\eps{\epsilon}
\def\veps{\varepsilon}
\def\la{\lambda}
\def\La{\Lambda}

\def\Ups{\Upsilon}
\def\th{\theta}

\def\vth{\vartheta}

\def\Om{\Omega}
\def\om{\omega}

\newcommand{\mc}[1]{\ensuremath{\mathcal{#1}}}
\newcommand{\mf}[1]{\ensuremath{\mathfrak{#1}}}
\newcommand{\msc}[1]{\ensuremath{\mathscr{#1}}}

\newcommand{\bs}[1]{\ensuremath{\boldsymbol{#1}}}
\def \i{ \mathrm i}

\newcommand{\ov}[1]{\ensuremath{\overline{#1}}}
\newcommand{\wt}[1]{\ensuremath{\widetilde{#1}}}
\newcommand{\wh}[1]{\ensuremath{\widehat{#1}}}

\newcommand{\Int}[2]{\ensuremath{\int\limits_{#1}^{#2}}}

\newcommand{\sul}[2]{\ensuremath{\sum\limits_{#1}^{#2}}}
\newcommand{\pl}[2]{\ensuremath{\prod\limits_{#1}^{#2}}}

\newcommand{\R}{\ensuremath{\mathbb{R}}}
\newcommand{\Cx}{\ensuremath{\mathbb{C}}}

\newcommand{\Dp}[1]{\ensuremath{\partial_{#1}}}
\newcommand{\limit}[2]{\ensuremath{\underset{#1 \tend #2}{\longrightarrow} }}
\newcommand{\ex}[1]{\ensuremath{\e{e}^{#1}}}

\newcommand{\op}[1]{ \boldsymbol{ \texttt{#1} } }

\newcommand{\norm}[1]{\ensuremath{|| #1 ||}}

\newcommand{\sla}[1]{\ensuremath{ #1 \hspace{-2mm}\backslash }}

\newcommand{\dd}{\mathrm{d}}
\newcommand{\e}[1]{\ensuremath{\mathrm{#1}}}

\newcommand{\intff}[2]{\ensuremath{ [  #1 \,; #2 ] }}

\newcommand{\intoo}[2]{\ensuremath{ ]  #1 \,; #2 [ }}

\newcommand{\intn}[2]{\ensuremath{[\![ \, #1 \,;\, #2 \,]\!]}}

\def\ff{{f}}

\begin{document}

\begin{center}
\begin{LARGE}
{\bf Asymptotic behaviour of two-point functions\\[1ex]
 in multi-species models}
\end{LARGE}

\vspace{30pt}

\begin{large}

{\large Karol K. Kozlowski \footnote{e-mail: karol.kozlowski@ens-lyon.fr}}%
\\[1ex]
Laboratoire de Physique, ENS de Lyon, Universit\'e de Lyon, UMR 5672 du CNRS, Lyon, France. \\[2.5ex]

{\large Eric Ragoucy \footnote{e-mail: ragoucy@lapth.cnrs.fr}}%
\\[1ex]
LAPTH, UMR 5108 du CNRS,
Universit\'e Savoie Mont Blanc, France.

\par

\end{large}

\vspace{40pt}

\centerline{\bf Abstract} \vspace{1cm}
\parbox{12cm}{\small
We extract the long-distance asymptotic behaviour  of two-point correlation functions in massless quantum integrable models containing multi-species excitations. 
For such a purpose, 
we extend  to these models the method of a large-distance regime re-summation of the form factor expansion of correlation functions. 
The key feature of our analysis is a technical hypothesis on the large-volume behaviour of the 
form factors of local operators in such models. We check the validity of this hypothesis on the example of the $SU(3)$-invariant XXX magnet 
by means of the  determinant representations for the form factors of local operators in this model. 
Our approach confirms the structure of the critical exponents obtained previously for numerous models solvable by the nested Bethe Ansatz.}

\end{center}
\vspace{20pt}

\rightline{LAPTH-001/16\qquad\qquad\qquad\qquad\ }

\vspace{40pt}

\section*{Introduction}

Form factor expansions, and hence form factors, play an important role in the characterisation of  correlation functions. 
Over the last few decades, there has been a significant progress in describing the form factors and the associated expansions for 
so-called quantum integrable systems. 
First progress in characterising the form factors has been achieved for massive models directly in the infinite volume. 
Archetypes of such models are given by the massive integrable quantum field theories in $1+1$ dimensions. 
In such a setting, the form factors were characterised as solutions to a set of bootstrap equations \cite{KarowskiWeiszFormFactorsFromSymetryAndSMatrices,KirillovSmirnovFirstCompleteSetBootstrapAxiomsForQIFT,SmirnovMatrixElementsforTodaFromSemiClassics}.
The resolution of the bootstrap program allowed for an explicit description of the form factors of local operators in numerous models
and of some of their intrinsic properties \cite{BabujianFoersterKarowskiSomeReviewOfBootstrap,CardyMussardoSameDimensionOfSolSpaceFFProgramAndVirasoro,FringMussardoSimonettiFFSOmeLocalObsSinhGordon,
KarowskiWeiszFormFactorsFromSymetryAndSMatrices,KoubekMussardoFFForMoreOpInSinhGordon,MussardoReviewOfBoostrap,SmirnovMatrixElementsforTodaFromSemiClassics}.
One should also mention the significant progress in conforming the representation theory of quantum affine algebras to the description of the spectrum of the XXZ spin chain Hamiltonian
\cite{DavisFodaJimboMiwaNakayashikiDiagonalizationXXZinfiniteDelta>1}. This progress allowed to access to multiple integral representations 
for the form factors in various massive spin chain models \cite{BaseilhacKozumiFormFactorXXZChainWithTriangBdry,JimboMiwaFormFactorsInMassiveXXZ,KoyamaStaggeredPolarisationForUqSlNHat},
again directly in the infinite volume. 

More recently, the calculation of form factors of finite volume quantum integrable models associated to rank one Lie algebras was undertaken within the algebraic Bethe Ansatz \cite{FaddeevSklyaninTakhtajanSineGordonFieldModel}. 
The approach builds on two major ingredients: on the one hand the solution of the quantum inverse scattering problem  \cite{KMTFormfactorsperiodicXXZ,MailletTerrasGeneralsolutionInverseProblem}
and, on the other hand, determinant representations for the norms \cite{KorepinNormBetheStates6-Vertex} and scalar products \cite{SlavnovScalarProductsXXZ} of Bethe vectors. 
Within such a setting one obtains finite-size determinant representations for the form factors, see \textit{e.g.} \cite{KMTFormfactorsperiodicXXZ,KorepinSlavnovFormFactorsNLSEasDeterminants}. 
Typically, when the model's volume goes to infinity -the so-called thermodynamic limit-, so does the size of the matrix whose determinant is evaluated. 
The very structure of the limit depends strongly on whether the underlying model exhibits a massless or a massive spectrum. 
The massive case is easier to deal with in that, then, individual form factors decay as integer powers of the volume $L$. 
This integer power in $L$ decay allows one to replace discrete sums appearing in a finite volume form factor expansion 
by a series of multiple integrals, once that the thermodynamic limit is taken. 
The investigation of the large-volume behaviour of a specific form factor in the massive regime of the 
XXZ chain was carried out in  \cite{IzerginKitMailTerSpontaneousMagnetizationMassiveXXZ}. Later, the analysis was  extended to all form factors of the longitudinal 
spin operator in \cite{KozDugaveGohmannSuzukiLargeVolumeBehaviourFFMassiveXXZ}. 
The main complication associated with a massless spectrum is that the form factors are expected to vanish as some, generically, non-integer power of the volume \cite{CardyConformalDimensionsFromLowLSpectrum}. 
The presence of such vanishing strongly complicates the analysis. First results relative to extracting the leading in $L$ behaviour out of the determinant representations
were obtained in \cite{SlavnovFormFactorsNLSE} and concerned the form factors of the current operator in the non-linear Schr\"odinger model. The technique of analysis developed there was improved and extended in 
\cite{KozKitMailSlaTerEffectiveFormFactorsForXXZ,KozKitMailSlaTerThermoLimPartHoleFormFactorsForXXZ} where the large-volume behaviour of
so-called particle-hole form factors in the massless regime of the XXZ chain was obtained. See also 
\cite{KozDugaveGohmannThermaxFormFactorsXXZ,KozDugaveGohmannThermaxFormFactorsXXZOffTransverseFunctions,KozMailletSlaLowTLimitNLSE} where the analysis of the low-temperature 
limit of so-called thermal form factors in a massless model at finite temperature has been carried out. 

The main issue with the non-integer decay in the volume of individual form factors is that it does not allow one to 
replace the finite-volume form factor expansions by series of multiple integrals. In fact, for a finite spatial and/or temporal separation between the operators
it has been impossible, so far, to write any meaningful form factor series expansion in the thermodynamic limit. Even though intractable in general, form factor series 
expansion for massless models have recently been discovered to be manageable in the limit of  large spatial separations between the operators
building up the correlator from which the expansion originates. Indeed, then, and when the volume is large but finite, the large-distance/time asymptotic behaviour of such series can be extracted 
by means of a variant of the saddle-point method. The evaluation of the leading contribution to the correlator
is achieved through the evaluation of certain multidimensional sums over the massless excitations of the model. 
After re-summing, one can already take the infinite volume limit, hence accessing to the large-distance asymptotic behaviour of the correlator. 
This approach has been developed in \cite{KozKitMailSlaTerRestrictedSums,KozKitMailSlaTerRestrictedSumsEdgeAndLongTime,KozKitMailTerMultiRestrictedSums} 
and culminated in the construction of a direct mapping \cite{KozMailletMicroOriginOfc=1CFTUniversality} between the zero energy excitation sector of a massless model and 
the free boson field theory.

The results which have been reminded above mostly concern integrable models built over rank one algebras, typically $GL(2)$ or its $q$-deformation. 
The correlation functions and form factors of models built over higher rank algebras were much less studied, this despite their relevance to the physics of super-Yang-Mills theories 
\cite{SerbanTheseHdR,FodaJiangKostovSerban3ptFctsinSU3SectorofN=4SYM}
or condensed matter models of mixed particle species \cite{BatchelorGuanLeeFromBetheAnsatzToExperiments,BatchelorFoersterGuanKuhnExactlySolveDModelAndUltraColdFermiGases,LiaoRittnerPaprottaLiPartridgeHuletBaurMuellerExpRealisation1D2componentFermiGasIntegrable}. 
Multiple integral representations for the  form factors of massive quantum field theories
associated with various higher rank algebras were obtained, \textit{e.g.}, in the  works 
\cite{BabujianFoersterKarowskiFormFactorsinSUN,BabujianKarowskiFormFactorsinZNandAN-1Toda,SmirnovFormFactors}. However, so far, only few results were obtained relatively to quantum integrable models
in finite volume. Indeed, although many models underlying to higher rank Lie algebras can still be solved by a variant of the Bethe Ansatz -the so-called nested Bethe Ansatz
\cite{YangFactorizingDiffusionWithPermutations,KulishReshetikhinNestedBAFirstIntroduction,KulishReshetikhinNestedBASomeGeneralisationstoGL(N)Reps}- the structure of these model's Bethe vectors becomes highly intricate hence inducing numerous new technical complications 
to the calculation of correlation functions. 
In fact, until very recently, only a determinant representation for the norm \cite{ReshetikhinNormsSU(3)BetheStates} of Bethe vectors in certain rank $2$ models was known. 
Recently, some progress has been achieved relatively to simplifying the scheme of constructing the Bethe vectors as well as to the calculation of 
their scalar product
\cite{BelliardPakuliakRagoucySlavnovBetheVectorsInGL(3)IntModels,EnriquezKoroshkinPakuliakWeightFctAndDrinfeldCurrent,KhoroshkinPakulianUniversalWeightFctUqGlN,OskinPakuliakSilantevUniversalWeightFctUqGLN,PakuliakRagoucySlavnovBetheVectorsUqGL(N)IntModels}, 
this in the context of spin chain models based on either $GL(N)$ or its q-deformation. 
For models enjoying of an $SU(3)$ symmetry, several expressions -either in terms of determinants or multiple-integrals- were obtained for scalar products between Bethe vectors
\cite{BelliardPakuliakRagoucySlavnovHighestCoeffInSU(3)ScalarProducts,BelliardPakuliakRagoucySlavnovNestedBAScalarProductsSU(3),WheelerMultIntRepScalarProdsInSU(3),WheelerScalarProdsInGenModelWithSU(3)}.
The determinant based representations for the scalar products appeared effective enough so as to lead to determinant based representations for the 
form factors \cite{BelliardPakuliakRagoucySlavnovFormFactorsSU(3)Models,PakuliakRagoucySlavnovFormFactorsWithGL(3)RMatrix}.
Very recently, some progress has been made relatively to characterising the scalar products in models related to a $q$-deformation of a rank $2$
algebra \cite{PakuliakRagoucySlavnovBetheVectorsGL(3)TrigonometricRMatrix,PakuliakRagoucySlavnovScalarProductsGL(3)TrigonometricRMatrix}, 
what ultimately led to a determinant representation for some specific instances of scalar products \cite{SlavnovScalarProductsForGL(3)TrigRMatrix}. 
The aforementioned results could also be generalised to the case of a two-component Bose gas. 
A set of recurrence relations satisfied by the form factors in this model have been obtained in \cite{PozsgayOeiKormosSomeConjFFNestedBA} and some 
special cases of solutions were given. Later, determinant representations for all form factors in the model were obtained in  \cite{PakuliakRagoucySlavnovFFForTwoComponentBoseGas,PakuliakRagoucySlavnovBetheVectorsForGL3QIM,PakuliakRagoucySlavnovFFForGL3QIM}, this on the basis
of preparatory results obtained in \cite{SlavnovNestedBAFor2ComponentBoseGas}. 

Despite the lack of explicit representation for multi-point correlation functions in higher rank models, one can still expect to build on the universality principle so as 
describe some of the properties of the correlators, their large-distance asymptotics in particular. Building on Cardy's \cite{CardyConformalDimensionsFromLowLSpectrum} relations between the $1/L$ corrections to the ground and
excited state's energies on the one hand and critical exponents on the other one, Izergin, Korepin and Reshetikhin \cite{IzerginKorepinReshetikhinConformalDimensionInNestedBAModels} predicted the critical exponents
driving the long-distance asymptotic behaviour of two-point functions in a large class of higher rank massless quantum integrable models. 
The predictions for the critical exponents obtained in \cite{IzerginKorepinReshetikhinConformalDimensionInNestedBAModels} were in fact extending  Cardy's approach in that 
the large-distance regime was not described by an effective $c=1$ conformal field theory but, rather, by a direct sum thereof. This was needed due to 
the different values of Fermi velocities associated with the different branches of massless excitations in such models. 
The $1/L$ corrections to the excitation energies of the Hubbard model were obtained in \cite{EckleWoynarovichFiniteSizeCorrectionsHubbardHalfFilling} for the model
at half-filling, what led to a purely conformal spectrum of low-lying excitations. These results were extended to the case of the Hubbard model away from half-filling
in \cite{WoynaorwiczFiniteSizeCorrectionsHubbardRepulsiveAwayHalfFilling}, and it was apparent that, again, in the general situation for this model
one needs to consider a direct sum of $c=1$ conformal field theories so as to describe the long-distance asymptotics of correlation functions in this model. 
This point of view was developed in \cite{FrahmKorepinCriticalExponents2DHubbardPartI} and some specific limits (strong coupling, magnetic field close to zero or to its 
critical value) of the critical exponents were studied in \cite{FrahmKorepinCriticalExponents2DHubbardPartII}. 
Asymptotics of two-point functions in this model were also predicted on the basis of a generalisation of the Luttinger liquid concept to multi-species excitations in
\cite{SchultzCritExpHubbardByLuttLiquidAnnouncement,SchultzCritExpHubbardByLuttLiquidDvpmtgeneralTheory}.

The aim of the paper is to test the aforementioned predictions through an \textit{ab inicio} calculation. On top of providing a firm ground to the universality predictions, this paper demonstrates
that upon certain reasonable hypothesis on the structure of the spectrum and form factors of local operators -which we explicitly check to be true in the case of the $SU(3)$-invariant XXX magnet-
it is possible to adapt the large-distance form factor summation technique developed in \cite{KozKitMailSlaTerRestrictedSums,KozKitMailSlaTerRestrictedSumsEdgeAndLongTime,KozKitMailTerMultiRestrictedSums,KozMailletMicroOriginOfc=1CFTUniversality}
to the setting of multi-species excitations. Our microscopic approach confirms the predictions of 
\cite{FrahmKorepinCriticalExponents2DHubbardPartI,IzerginKorepinReshetikhinConformalDimensionInNestedBAModels,WoynaorwiczFiniteSizeCorrectionsHubbardRepulsiveAwayHalfFilling,SchultzCritExpHubbardByLuttLiquidAnnouncement,
SchultzCritExpHubbardByLuttLiquidDvpmtgeneralTheory}.
We shall describe the structure of the large-distance asymptotics we obtain on the example of a specific correlator in the $SU(3)$-invariant XXX magnet,
the two-point function $\big< \op{E}_{1+m}^{21} \op{E}_1^{12} \big>$ involving the local operators given by the elementary matrices $E^{ij}$ localised on sites $1$
and $1+m$. The results obtained in the core of this paper demonstrate that the below
large-$m$ expansion holds
\beq
\big< \op{E}_{1+m}^{21} \op{E}_1^{12} \big> \simeq \sul{ \bs{\ell}\in \mathbb{Z}^2 }{} (-1)^{m\ell^{(1)}} \cdot  \ex{ \i m  \big( 2\bs{\ell}+\bs{\mf{n}}_{12}\big)\cdot \bs{p}_{F} }
\cdot \f{  \Big| \mc{F}_{\bs{\ell},\bs{\kappa}_{12}}\big( \op{E}_1^{12} \big)\Big|^2  }{ \big(2\pi m \big)^{ \De_{\bs{\ell},\bs{\kappa}_{12}} }  } 
\enq
with 
\beq
2 \De_{ \bs{\ell},\bs{\kappa}_{12} }  \, = \, \Big(2\bs{\ell}+\bs{\mf{n}}_{12}, \mc{Z}\mc{Z}^{\e{t}} \cdot (2\bs{\ell}+\bs{\mf{n}}_{12})  \Big) \, + \, \big(\bs{\kappa}_{12},\big[\mc{Z}\mc{Z}^{\e{t}}\big]^{-1} \bs{\kappa}_{12} \big) \;. 
\enq
In the above expansion, the sum runs over two-dimensional integer valued vectors $\bs{\ell}=\big(\ell^{(1)},\ell^{(2)} \big)$ which label the possible Umklapp excitations over the model's ground state. 
The vector $\bs{\mf{n}}_{12}$  originates from the fine structure of the class of excited state
connected by the operator  $\op{E}_1^{12}$ to the ground state.
The contribution of each Umklapp excitation produces an oscillatory factor with phase $2\i m \, \bs{\ell} \! \cdot \! \bs{p}_{F} $ in which $\bs{p}_F=\big(p_F^{(1)},p_F^{(2)} \big)$
is built up from the two Fermi momenta associated with the two Fermi zones of the model.   There is an additional oscillatory factor with phase $ \i m \, \bs{\mf{n}}_{12} \! \cdot \! \bs{p}_{F} $ 
associated with the shift in the momentum of the class of excited states connected by the operator $\op{E}_1^{12}$ with the ground state. 
 The quantity $| \mc{F}_{\bs{\ell},\bs{\kappa}_{12}}\big( \op{E}_1^{12} \big)|^2$ represents the properly normalised in the model's volume thermodynamic limit 
of the form factor of the operator $\op{E}_1^{12}$ taken between the ground state and the lowest energy excited state corresponding to an $\bs{\ell}$-Umklapp excitation
connected to the ground state by the operator $\op{E}_1^{12}$. Finally, the critical exponent is given by a scalar product which involves a $2\times 2$ matrix $\mc{Z}$
that can be expressed in terms of the dressed charge matrix of the model.

The paper is organised as follows. In Section \ref{Section General setting} we present the general framework that a model needs to fulfil so that the approach of the paper
becomes applicable. In Section \ref{Section asympt grde distance fct deux points}
we show how the general structure described in Section  \ref{Section General setting} does allow one to access to the large-distance asymptotic behaviour of the 
two-point functions in the model. We then conform the obtained result to the case of the $SU(3)$-invariant XXX magnet and to the Hubbard model away from half-filling. 
In Section \ref{Section SU3 XXX modele et discussions} we build on the determinant representation for the form factors of local operators
in the $SU(3)$-invariant XXX magnet so as to prove that, indeed, this rank $2$ quantum integrable models does fit into the framework described in Section  \ref{Section General setting}. 
The paper contains several appendices were we postpone some of the technical handlings. 
Appendix \ref{Appendix Thermo limit of the model} provides a reminder of the description of the thermodynamic limit of higher rank models. In Section \ref{Appendix Sous section large L expansion EINL}
we briefly recall the derivation of the large-$L$ expansion of the vector counting function associated with nested Bethe Ansatz solvable models. In Section \ref{Appendix Sous section cptmt grd L de moment et energie} we build on these
results so as to obtain the leading large-$L$ expansion of the excitation energies and momenta relatively to the ground state in such models. Finally, in Section \ref{Appendice derivation identite quadratique phase habillee} 
we establish several identities satisfied by solutions to linear integral equations that are of interest to the problem. 
Appendix  \ref{Appendix Proof factorisation kappa deformed determinant}
contains the proof of the smooth-discrete part factorisation of form factors in the $SU(3)$-invariant XXX magnet. 
Also, we have gathered our basic notations in Appendix \ref{sec:notations} so as to ease the reading of this article. 

\section{The general setting}
\label{Section General setting}

In this section we discuss several general features present in a quantum integrable models solvable by the nested Bethe Ansatz. The number of nestings necessary to construct the Bethe
vectors will be denoted by $r-1$, where $r$ corresponds to the rank of the Lie algebra over which the integrable model is constructed. 
We will focus on lattice integrable models, namely those whose Hilbert space $\mf{h}=\mf{h}_1\otimes\cdots \otimes \mf{h}_{L}$ admits a tensor product decomposition into a 
product of so-called local spaces $\mf{h}_{k}$ called the sites of the model. We will assume that these spaces are all isomorphic to some base space $\mf{h}_{k}\simeq \mf{h}_{\e{base}}$. 
The Hamiltonian of the model is assumed to take the form
\beq
\op{H}=\op{H}_{0} \, + \, \sul{k=1}{r} h^{(k)} \op{Q}^{(k)} \;. 
\label{ecriture hamiltonien de Base}
\enq
$\op{H}_0$ represents the base Hamiltonian and $\op{Q}^{(k)}$, $k=1,\dots ,r$ a set of $r$ independent conserved charges $\big[ \op{H}_0, \op{Q}^{(k)} \big]=0$
that take into account the possibility of coupling the base $\op{H}_0$ Hamiltonian with external fields $h^{(1)},\dots, h^{(r)}$. 
The explicit example of the base Hamiltonians $\op{H}_0$ of the $SU(3)$-invariant XXX magnet and of its associated charges $\op{Q}^{(k)}$ 
will be discussed in Sub-section \ref{SousSection application aux modeles specifiques}. 
Here, we only mention that, typically, the auxiliary conserved charges $\op{Q}^{(k)}$ have integer eigenvalues  which are expressible, in a simple way,
in terms of the number of roots arising in the nested Bethe Ansatz equations. 

The base Hilbert space $\mf{h}_{\e{base}}$ is assumed to be endowed with an algebraic basis $\op{O}^{(\a)}$ of operators where the superscript $\a=1,2,\dots$ runs through some finite
or infinite set depending on the dimensionality of $\mf{h}_{\e{base}}$.
These operators can then be raised into operators $\op{O}^{(\a)}_n$ on $\mf{h}$ by tensoring with the identity
\beq
\op{O}^{(\a)}_n \, = \, \underbrace{ \e{id} \otimes \cdots  \otimes \e{id} }_{n-1 \, \e{times} } \otimes \, \op{O}^{(\a)} \otimes \underbrace{ \e{id} \otimes \cdots  \otimes \e{id} }_{L-n \, \e{times} } \;. 
\enq
We will call such operators local since they only act non-trivially on \textit{one} of the spaces appearing in the tensor product decomposition of 
$\mf{h}$. We will assume in the following that the algebraic basis $\op{O}^{(\a)}$ used to build the operators $\op{O}^{(\a)}_n$
is chosen in such a way that the operators  $\op{O}^{(\a)}_n$ only connect eigenstates of the operators $\op{Q}^{(k)}$ having definite eigenvalues.

 \subsection{Bethe equations and basic observables}

The eigenvectors of the Hamiltonian \eqref{ecriture hamiltonien de Base} that are constructed by the nested Bethe Ansatz 
are called Bethe vectors and we denote them by $\ket{\La}$. The Bethe vectors $\ket{\La}$  are parametrised by a collection 
\beq
\La \, = \, \Big\{  \{\la_a^{(k)} \}_{a=1}^{N^{(k)}_{\La} } \, , \, k=1, \dots, r\Big\}
\label{ecriture notation pour ensemble racine Lambda}
\enq
of Bethe roots, split into $r$ families of "roots" $\{\la_a^{(k)} \}_{a=1}^{N^{(k)}_{\La} }$ that we call species. The  notation introduced in \eqref{ecriture notation pour ensemble racine Lambda} 
indicates that the set $\La$ contains $N^{(k)}_{\La}$ roots of species $k$. Note that, when passing from one Bethe vector to another, 
the number of roots of a given species may change. 
The roots $\La$ not only parametrise the Bethe vector $\ket{\La}$  but also all the observables of the model associated with it: the energy, momentum and, more generally, the form factor 
of local operators involving the state $\ket{\La}$. For $\ket{\La}$ to be an eigenvector of the model's Hamiltonian, the roots $\La$ have to satisfy a system of Bethe equations which, in its logarithmic variant,
takes the general form:
\beq
L p_0^{(k)} \big( \la_a^{(k)} \big) \, + \, \sul{  \ell = 1  }{  r } \sul{b=1}{ N^{(k)}_{\La} } \th_{k \ell}\big( \la_a^{(k)}, \la_b^{(\ell)} \big) \; = \;  m_{a;\La}^{(k)}   \,- \, \f{1 + N^{(k)}_{\La} - \mf{n}^{(k)}_{\La} }{ 2 },
\label{ecriture eqn de Bethe log}
\enq
where the shift integer $\mf{n}^{(k)}_{\La}$ can be expressed in terms of the  condition below on the sums of integers 
\beq
w^{(k)}_{\La}=  \sul{ \substack{ \ell=1, \not= k  \\ \ell : \th_{k\ell}\not=0 }  }{ r  } N^{(\ell)}_{\La} + \big(N_{\La}^{(k)}+1 \big)\bs{1}_{\th_{kk} = 0}
\qquad \e{as} \qquad 
\mf{n}^{(k)}_{\La} \;=\; \left\{ \ba{ccc}  0 &  \e{if} & w^{(k)}_{\La} \in 2\mathbb{N}  \vspace{2mm} \\ 
  1 & \e{if}  & w^{(k)}_{\La} \in 2\mathbb{N} +1   \ea \right.  \;. 
\enq
Here $\bs{1}_{\th_{kk} = 0}=1$ if the function $\th_{kk}$ vanishes identically and  $\bs{1}_{\th_{kk} = 0}=0$ otherwise. 
The functions appearing in the \textit{lhs} of \eqref{ecriture eqn de Bethe log} are the bare momenta $p_0^{(k)} $ and the bare phases $\th_{k \ell}$
of the excitations. Throughout this paper, we shall assume that the bare phase is a function of the sole difference of arguments  and that it is symmetric
in its indices
\beq
\th_{k\ell}(\la,\mu)=\th_{\ell k }(\la,\mu)=\vth_{k\ell}(\la-\mu) \;. 
\enq
We shall as well take for granted that  all bare momenta and bare phases are odd functions, namely that 
\beq
p_0^{(k)} \big(- \la \big)  \; = \; - p_0^{(k)} \big( \la \big) \qquad \e{and} \qquad \th_{k\ell}\big( \la, \mu \big) \; = \;   \th_{\ell k}\big(- \mu, -\la \big) \; = \;  - \th_{\ell k}\big( \mu, \la \big)  \;, \qquad \e{for} \quad k,\ell \in \intn{1}{r} \;. 
\label{ecriture pte symmetries moment et phase nus}
\enq
We do stress  that although such an assumption allows for a few technical simplifications, it is not essential at any stage of the handlings that will follow.
Finally, the logarithmic Bethe equations involve integers $m_a^{(k)}\in \mc{I}^{(k)}$ which take value in a model-dependent 
set $\mc{I}^{(k)} \supset \intn{1}{ N^{(k)}_{\La} }$. For models with an unbounded $k^{\e{th}}$ bare momentum -such as the multi-component Bose gas \cite{YangFactorizingDiffusionWithPermutations}-
one has, typically, $\mc{I}^{(k)}=\mathbb{Z}$ while, for models with a bounded $k^{\e{th}}$ bare momentum-such as the $GL(N)$-invariant XXX magnet \cite{KulishReshetikhinNestedBAFirstIntroduction,KulishReshetikhinNestedBASomeGeneralisationstoGL(N)Reps}-, 
one has $\mc{I}^{(k)}=\intn{-M^{(k)}_{-}}{M_{+}^{(k)}}$ for some integers $M^{(k)}_{\pm}$.

 Clearly, a choice of roots $\La$ does fix the value of the integers $m_a^{(k)}$ arising in the \textit{rhs} of the logarithmic Bethe equations \eqref{ecriture eqn de Bethe log}.  
 However, it could happen that the correspondence is not injective, meaning that two distinct collections of Bethe roots could give rise to exactly the same collection of integers.
In the following, we shall however assume that, if one restricts one's attention to a subset of solutions to the Bethe equations called particle-hole excited states, 
then there is a one-to-one correspondence between integers and solutions to the Bethe Ansatz equations. Recently, this property was show to hold, for $L$ large-enough, in the case of the 
XXZ spin-$1/2$ chain \cite{KozProofOfDensityOfBetheRoots} and we believe that the same mechanism will be at play in the more general setting we consider in the present paper.

\subsubsection*{$\bullet$ The ground state}

We shall assume that the model has a non-degenerate ground state\symbolfootnote[3]{Should the mode have a finitely degenerate ground state, there would still be no problem to apply the present setting, although some
of the handlings would become bulkier.}. We shall always denote by 
\beq
\Om=\big\{  \{\om_a^{(k)} \}_{a=1}^{N^{(k)}_{\Om} } , k=1,\dots, r\big\}
\enq
the collection of Bethe roots giving rise to the model's ground state. Furthermore, 
we shall assume that the set of integers associated with the ground state takes the form 
\beq
m_{a;\Om}^{(k)} \; = \; a \; , \quad a=1,\dots, N^{(k)}_{\Om} \quad \e{and} \quad k=1,\dots,r 
\label{ecriture entiers pour fondamental}
\enq
and that, for the ground state, one has $\mf{n}_{\Om}^{(k)}=0$ for $k=1,\dots,r$. The value of the integers  $N^{(k)}_{\Om}$ of $k^{\e{th}}$-species roots for the ground state 
is fixed by the external fields $h^{(k)}$, $k=1,\dots,r$ which couple $\op{H}_0$ to the auxiliary conserved charges $\op{Q}^{(k)}$. 

When taking the thermodynamic limit $L\tend +\infty$ of the model, the integers  $N^{(k)}_{\Om}$, $k=1,\dots,r$ will be all assumed to grow with $L$ in such a way that 
$\lim_{L\tend +\infty} \big( \tf{ N^{(k)}_{\Om} }{ L } \big)=D^{(k)}$, with $D^{(k)} >0$, finite and fixed once for all. 
Furthermore, these integers will be such that the two endpoints $M_{-/+}^{(k)}$ of the set $\mc{I}^{(k)}$ are such that  
\beq
M_{-}^{(k)} \tend +\infty \quad \e{and} \quad M_{+}^{(k)}- N^{(k)}_{\Om} \tend +\infty \quad \e{as} \quad  L\tend +\infty \;. 
\enq

 In this paper we shall build on the assumption that the ground state roots have the densification property, meaning that 
\beq
\f{1}{L} \sul{ a=1 }{ N^{(k)}_{\Om} } f\big( \om^{(k)}_a \big) \; \limit{L}{+\infty}\;  \Int{-q^{(k)} }{q^{(k)} } \!\! f(s) \cdot \rho^{(k)}(s) \dd s \;. 
\enq
In other words, the $k^{\e{th}}$ species ground state Bethe roots will form, when $L\tend +\infty$, a dense distribution on a finite interval $\intff{ - q^{(k)} }{ q^{(k)} }$
with density $\rho^{(k)}$. The interval $\intff{ - q^{(k)} }{ q^{(k)} }$ will be called the Fermi zone associated with the $k^{\e{th}}$ species roots, or 
$k^{\e{th}}$ Fermi zone for short. The specific value of the endpoints $q^{(k)}$ of this interval is fixed by the values taken by the densities $D^{(a)}$, $a=1,\dots,r$.

We do stress that for the XXZ spin-$1/2$ chain which is a rank one ($r=1$) model
one can prove \cite{Yang-YangXXZproofofBetheHypothesis} that, indeed, the choice of integers \eqref{ecriture entiers pour fondamental} does give rise to the roots parametrising the ground state
in the sector with $N_{\Om}^{(1)}$ particles, and that, in this sector, the ground state is non-degenerate. 
The key feature is to first, investigate the model at specific values of its coupling constant where it reduces to free fermions and, thus, where 
the logarithmic Bethe equations become explicitly solvable giving rise to an explicit formula for the energies of the eigenstates. 
The second step consists then in using a continuity argument adjoined with a non-degeneracy of the ground state.
Finally, the densification property of the Bethe roots for the ground state can be established by studying the large-$L$
behaviour of solutions to a non-linear integral equation \cite{KozProofOfDensityOfBetheRoots} of Destri-deVega type.

Unfortunately, such reasoning cannot, in general, be reproduced for higher rank models as the Bethe equations do not seem to enjoy the presence of a free fermion point. 
Some arguments were given in the literature regarding to the fact that the $ c \tend 0^+ $  limit of a multi-species boson gas interacting through two-body Dirac delta functions of strength $ c$
has its ground state given by \eqref{ecriture entiers pour fondamental}. It has been conjectured that such a description holds, in fact, for all $c$ \cite{SutherlandCBAForNSpeciesBosons,YangFactorizingDiffusionWithPermutations}. 
It is as well important to keep in mind that there are several examples 
of higher rank models which are believed \textit{not} to follow our assumptions on the ground state roots \eqref{ecriture entiers pour fondamental}.
For instance, it appears \cite{SutherlandCBAForHigerRankSpinChain} that the ground state of a spin chain built up of a mixture of $r-1$ species  of fermions and one specie of bosons 
is obtained by solving logarithmic Bethe equations with $m_{a;\Om}^{(1)}=a$, $a=1,\dots,N^{(k)}_{\Om}$ and $N^{(k)}_{\Om}$ growing to infinity as described above, this for  $k=1,\dots r-1$ but with $N^{(r)}_{\Om}=0$. 
\textit{A priori}, our setting \textit{does not} directly include these models, although we do trust that it can be appropriately modified so as to encompass these cases as well. 

\subsubsection*{ $\bullet$ The excited states }
 
Throughout this paper, we shall only focus on the particle-hole excited states, hence waving-off the bound states. Since, \textit{in fine} our aim is to study the
asymptotic behaviour of two-point functions, this does not constitute an important limitation in that the bound states are expected to contribute solely
to corrections that are exponentially small  in the distance of separation between the operators. 

Recall that $N^{(k)}_{\Om}$ denotes the number of $k^{\e{th}}$ species roots building up the ground state.  The excited states having a finite, in respect to the thermodynamic limit $L\tend +\infty$, excitation energy relatively to the 
ground state are realized in terms of the Bethe vectors $\ket{\La}$ built out of $N^{(k)}_{\La}$ roots of species $k$, $k=1,\dots, r$, such that 
\beq
N^{(k)}_{\La} \; = \; N^{(k)}_{\Om} \, + \,  \kappa^{(k)} \; 
\enq
with $\kappa^{(k)}$ being some integers, depending on the excitation of interest, but which are kept \textit{bounded} in  $L$. 

In the language of the logarithmic Bethe equations, a particle-hole excitation corresponds to a solution to \eqref{ecriture eqn de Bethe log}
associated with an "almost" contiguous distribution of integers, namely:
\beq
m_{a;\La}^{(k)} \; = \; a  \qquad \e{for} \qquad a \in \intn{ 1 }{ N^{(k)}_{\La} }\setminus\{h_1^{(k)},\dots, h_{n^{(k)}}^{(k)} \}
\qquad \e{and}  \qquad 
m_{ h_{b}^{(k)} ; \La } ^{(k)} \; = \; p_{b}^{(k)}  \qquad \e{for} \quad b=1,\dots, n^{(k)} \;.  
\label{ecriture param entiers nk par part trous}
\enq
Here 
\beq
h_1^{(k)}<\dots< h_{n^{(k)}}^{(k)} \quad , \;\; h_a^{(k)} \in \intn{ 1 }{ N^{(k)}_{\La} } \qquad \e{and} \qquad 
p_1^{(k)}<\dots< p_{n^{(k)}}^{(k)} \quad , \;\; p_a^{(k)} \in \mc{I}^{(k)}\setminus \intn{ 1 }{ N^{(k)}_{\La} } 
\enq
are integers labelling, on a microscopic level, the particles and holes building up the excitation. Finally, the numbers $n^{(k)}$, with $k=1,\dots, r$, count the number of 
particle-hole excitations in the $k^{\e{th}}$ species sector.

Owing to the presumed property that the ground state's shifts all vanish $\mf{n}_{\Om}^{(k)}=0$, one can express the shifts for the excited state solely in terms of 
 the $\kappa$'s. Indeed, defining the vector 
\beq
v^{(k)}_{\La} =  \sul{ \substack{ \ell=1, \not= k  \\ \ell : \th_{k\ell}\not=0 }  }{ r  } \kappa^{(\ell)}  \, + \,  \kappa ^{(k)}\bs{1}_{\th_{kk}=0}
\qquad \e{one}\, \e{has} \quad 
\mf{n}_{\La}^{(k)} \; = \; \left\{ \ba{ccc} 1 & \e{if} & v^{(k)}_{\La} \in 2\mathbb{Z}+1 \\ 
					0 & \e{if} & v^{(k)}_{\La} \in 2\mathbb{Z} 		\ea \right. \;. 
\label{definition vecteur shift n}
\enq
The relationship between the $\kappa^{(k)}$'s and the $\mf{n}^{(k)}_{\La}$'s entails that 
\beq
\bs{\kappa}\cdot \bs{\mf{n}} \; \equiv \; \sul{k=1}{r} \kappa^{(k)} \mf{n}^{(k)}_{\La} \in 2\mathbb{Z} \;. 
\label{propriete parite PS vecteur n et kappa}
\enq
Indeed, $\bs{\kappa}\cdot \bs{\mf{n}} \in 2\mathbb{Z}$ is equivalent to $\bs{\kappa}\cdot \bs{v}_{\La} \in 2\mathbb{Z}$. However, owing to $\th_{k\ell}(\la,\mu)=\th_{\ell k}(-\mu,-\la)$,
one can recast the last scalar product as
\beq
\bs{\kappa}\cdot \bs{v}_{\La} \; = \; \sul{ \substack{ k=1  \\ k : \th_{kk}=0} }{r}\kappa^{(k)}\big(\kappa^{(k)}+1\big) \, + \, 
2 \sul{  \substack{ k>\ell \\ \ell : \th_{k\ell} \not=0} }{}\kappa^{(k)} \kappa^{(\ell)} \;\; \in 2\mathbb{Z} \;. 
\enq

\subsubsection*{ $\bullet$  The energy and momentum}

 When $L$ is finite, the momentum and energy of an excited state  $\ket{\La}$ relatively to the ground state  take the form:
\beqa
\mc{P}^{(\e{ex})}_{\La;\Om} \; = \; \mc{P}_{\La}\, - \, \mc{P}_{\Om} & = &
 \sul{k=1}{r} \bigg\{ \sul{a=1}{ N^{(k)}_{\La} } p^{(k)}_{0}\big(\la_{a}^{(k)} \big) \; - \; \sul{a=1}{N_{\Om}^{(k)}} p^{(k)}_{0}\big( \om_{a}^{(k)} \big)  \bigg\}   \\
\mc{E}^{(\e{ex})}_{\La;\Om} \; = \; \mc{E}_{\La}\, - \, \mc{E}_{\Om} & = &
 \sul{k=1}{r} \bigg\{ \sul{a=1}{ N^{(k)}_{\La} } \veps^{(k)}_{0}\big(\la_{a}^{(k)} \big) \; - \; \sul{a=1}{N_{\Om}^{(k)}} \veps^{(k)}_{0}\big( \om_{a}^{(k)} \big)  \bigg\}  \;. 
\eeqa
We remind that $\Om=\big\{  \{\om_a^{(k)} \}_{ a = 1 }^{  N^{(k)}_{\Om}  } \big\}_{k=1}^{r}$ is the set of Bethe roots associated with the ground state. 
The function $p_0^{(k)}$, resp. $\veps_0^{(k)}$, corresponds to the bare momentum, resp. bare energy, of an excitation in the $k^{\e{th}}$-species sector. 
They are both model dependent. Furthermore, the functions $\veps_0^{(k)}$ depend explicitly on the external magnetic fields $h^{(1)},\dots, h^{(r)}$.

\subsubsection*{ $\bullet$  Linear integral equation describing the thermodynamic limit}

The observables in thermodynamic limit of the model are characterised in terms of a collection of special functions defined as solutions
to auxiliary linear integral equations. We shall present the solutions which will be of interest to our study. 
However, first, we need to introduce a few notation. 

By $\bs{D}$, resp. $\bs{f}(\la)$, we shall denote $r$-dimensional vectors, resp. vector-valued functions, \textit{e.g.}:
\beq
\bs{f}(\om) \; = \; \left( \ba{cc} f^{  (1)}(\om)  \\ \vdots \\  f^{  (r)}(\om) \ea \right) \; \;  ,\quad 
\bs{D} \; = \; \left( \ba{cc} D^{(1)} \\ \vdots \\   D^{(r)} \ea \right) \; \; , \quad 
\bs{\mf{n}} \; = \; \left( \ba{cc} \mf{n}^{(1)}_{\La}  \\ \vdots \\  \mf{n}^{(r)}_{\La} \ea \right)  \quad \e{and}\quad 
\bs{\kappa} \; = \; \left( \ba{cc} \kappa^{(1)}  \\ \vdots \\  \kappa^{(r)} \ea \right) \; . 
\enq
Further, we introduce an integral operator $\op{K}$ acting on vector valued functions $\bs{f}(\la)$ as
\beq
\Big[ \Big( \e{id} \, + \, \op{K} \Big)[\bs{f}](\om) \Big]^{(k)} \; = \; \bs{f}^{(k)}(\om) \; + \; \sul{\ell=1}{r} \Int{ -q^{(\ell)} }{ q^{(\ell)} } \Dp{\mu} \th_{k\ell}(\om,\mu) \cdot \bs{f}^{(\ell)}(\mu) \cdot \dd \mu \;. 
\enq
Note that the integral kernel of $\op{K}$ is built out of the first order derivatives of the bare phases $\th_{ks}$ arising in \eqref{ecriture eqn de Bethe log}. 
 We shall assume that the operator $\e{id} \, + \, \op{K} $ is invertible and denote its inverse by 
$\e{id}-\op{R}$. This inverse acts on vector valued functions as 
\beq
\Big[ \Big( \e{id} \, - \, \op{R} \Big)[\bs{f}](\om) \Big]^{(k)} \; = \; \bs{f}^{(k)}(\om) \; - \; \sul{\ell=1}{r} \Int{ -q^{(\ell)} }{ q^{(\ell)} } R_{k\ell}(\om,\mu) \cdot \bs{f}^{(\ell)}(\mu) \cdot \dd \mu \;. 
\enq
The domains of integrations
$\intff{ -q^{(k)} }{ q^{(k)} }$ correspond to the $k^{\e{th}}$ species Fermi zone. The endpoint $q^{(k)}$ of this zone should be chosen in such a way 
that the configuration of Bethe roots condensing on $\intff{ -q^{(k)} }{ q^{(k)} }$ does indeed realise the minimum of the energy. 
This requirement can be rephrased in terms of the dressed energies $\veps^{(k)}$ which are the components of the vector dressed energy defined as the solution to the linear integral equation:
\beq
\Big( \e{id} \, + \, \op{K} \Big)[\bs{\veps}](\om) = \bs{\veps}_0(\om) \qquad \e{with}\quad \big[ \bs{\veps}_0(\om)  \big]^{(k)} \; = \; \veps^{(k)}_0(\om)\;. 
\label{definition energie habillee vectorielle}
\enq
The endpoints $q^{(k)}$ are chosen in such a way that the associated dressed energies vanish on their respective Fermi boundaries $\veps^{(k)}\big( q^{(k)} \big) = 0$ and, furthermore, satisfy
\beq
\veps^{(k)}(\om) \, < \,  0  \quad \e{for} \quad  \om \in \intoo{ - q^{(k)} }{  q^{(k)} } \qquad \e{and} \qquad \veps^{(k)}(\om) \, > \,  0  \quad \e{for} \quad  \om \in \R \setminus \intff{ - q^{(k)} }{  q^{(k)} } \;. 
\label{ecriture pte fond dressed energy}
\enq
Thus, it  is in fact the equation $\veps^{(k)}\big( q^{(k)} \big) = 0$ that ought to be taken as the definition of $q^{(k)}$ associated with the model's ground state in the presence of the 
external fields $h^{(k)}$.

The other functions of interest  are the vector dressed momentum $\bs{p}$, the vector dressed phase $\bs{\Phi}_s$ associated with species $s$ and the dressed charge matrix $\bs{Z}$.
They are defined as the solutions to the 
linear integral equations:
\beqa
\Big( \e{id} \, + \, \op{K} \Big)[\bs{p}](\om) & = &  \bs{p}_0(\om)  \; + \; \sul{s=1}{r} \f{D^{(s)}}{2} \Big( \bs{\Xi}_s\big(\om,q^{(s)} \big)\, + \, \bs{\Xi}_s\big(\om,-q^{(s)} \big)\Big)    
\label{definition vecteur moment habille}\\
\Big( \e{id} \, + \, \op{K} \Big)[\bs{\Phi}_s(*,z)](\om) & = & \bs{\Xi}_s(\om,z)   \qquad \e{where} \qquad  \Big( \bs{\Xi}_s(\om,z) \Big)^{(k)} \; = \;  \th_{ks}\big(\om,z \big)      \label{definition vector dressed phase}   \\
\Big( \e{id} \, + \, \op{K} \Big)[ \bs{Z}](\om) & = & I_r   \label{definition matrix dressed charge}
\eeqa
where $I_r$  is the $r\times r$ identity matrix and the $*$ indicates the argument of the vector function on which the matrix integral operator $ \e{id} \, + \, \op{K} $ acts. 
Also, the action of $\op{K}$ on matrix valued functions is defined column-wise. By construction, the vector dressed momentum satisfies $ 2 p^{(k)}\big(q^{(k)}\big)= D^{(k)}$. 
The symmetry properties \eqref{ecriture pte symmetries moment et phase nus} then imply that $p$
is an odd vector function $\bs{p}(\la)=-\bs{p}(-\la)$ while $\bs{Z}$ is an even matrix function $\bs{Z}(\la)=\bs{Z}(-\la)$.

\subsubsection*{ $\bullet$ The counting function and thermodynamic limit of eigenvalues}

It is convenient to characterise the roots $\La$ of a particle-hole excited state in terms of its collection of counting functions  
\beq
\wh{\xi}_{\La}^{\, (k)}(\om) \; = \;  p_0^{(k)} \big( \om \big) \, + \, \f{1}{L}\sul{\ell= 1}{ r } \sul{b=1}{ N^{(k)}_{\La} } \th_{k\ell}\big( \om, \la_b^{(\ell)} \big) \, + \, \f{ N^{(k)}_{\La} +1 - \mf{n}^{(k)}_{\La}}{ 2 L }   
\quad , \quad k=1,\dots, r\;. 
\label{definition counting function}
\enq
These counting functions are, by construction, such that, when evaluated at the Bethe roots belonging to $\La$, it holds
\beq
\wh{\xi}_{\La}^{\, (k)}\big( \la_a^{(k)} \big) \; = \; \frac{m_{a;\La}^{(k)} }{ L} \;. 
\enq
We shall lay our analysis on the hypothesis that, for $L$ large enough, the counting functions associated with particle-hole excited states are all strictly increasing: $\big( \wh{\xi}_{\La}^{(k)} \big)^{\prime}(\om) >0$ for $\om \in \R$. 
This being taken for granted, it is useful to introduce a set of auxiliary background roots $ \big\{ \hat{\la}_a^{\,(k)} \big\} $ which are defined as
\beq
\wh{\xi}_{\La}^{\, (k)}\big( \hat{\la}_a^{\,(k)} \big) \; = \; \frac{a}{ L} \;  \qquad \e{for} \quad a\in 
\Big[\!\Big[ - \big[ -\lim_{\om\tend-\infty}\wh{\xi}_{\La}^{\, (k)}(\om) \big] \, ; \,  \lim_{\om\tend+\infty}\wh{\xi}_{\La}^{\, (k)}(\om)   \Big]\!\Big]  \; 
\enq
where $[x]$ stands for the integer part of $x$. Then, the Bethe roots can be recast as
\beq
\Big\{ \la_a^{(k)} \Big\}_{ a=1 }^{ N^{(k)} } \; = \; \bigg\{ \Big\{ \hat{\la}_a^{\,(k)} \Big\}_{a=1}^{ N^{(k)} }\setminus \Big\{ \hat{\la}_{h_a^{(k)}}^{\, (k)} \Big\}_{a=1}^{ n^{(k)} } \bigg\} \,
\cup  \, \Big\{ \hat{\la}_{p_a^{(k)}}^{\, (k)} \Big\}_{a=1}^{ n^{(k)} } \;. 
\enq
The pieces of information gathered above allow one to write down a non-linear integral equation satisfied by the counting function which, in its turn, allows
one to access to the large-$L$ asymptotic expansion of the latter \cite{DestriDeVegaAsymptoticAnalysisCountingFunctionAndFiniteSizeCorrectionsinTBAFirstpaper,DeVegaWoynarowichFiniteSizeCorrections6VertexNLIEmethod}.
We refer to Appendix \ref{Appendix Sous section large L expansion EINL} for a short derivation of the first few terms of this asymptotic expansion 
which, in fact, contain all the information that is needed for the study of the thermodynamic limit of the model. 
The conclusion of Appendix \ref{Appendix Sous section large L expansion EINL} is that the counting function admits the large-$L$ asymptotic expansion:
\beq
\wh{ \bs{\xi} }_{\La}(\om) \; = \; \bs{p}(\om) \, + \; \f{ \bs{D} }{2}  \; + \; \sul{s=1}{r} \f{ \kappa_s  }{ L } \bs{\Phi}_s(\om,q_s) \; + \; 
\f{1}{L} \sul{s=1}{r} \sul{a=1}{n^{(s)}} \bigg[ \bs{\Phi}_s\Big( \om , \mu_{p_a^{(s)}}^{(s)} \Big) \, - \,  \bs{\Phi}_s\Big( \om , \mu_{h_a^{(s)}}^{(s)} \Big) \bigg]
\; + \; \bs{Z}(\om)\cdot \f{ \bs{\kappa} -\bs{\mf{n}} }{2 L } \; + \; \e{O}\Big( \f{ 1 }{ L^2 } \Big)
\label{ecriture DA fct comptage}
\enq
where the remainder is to be understood entry-wise. Some explanations are in order. The parameters $\mu_{a} ^{(k)}$ are defined as the unique solutions to the equations 
\beq
p^{(k)}\big( \mu_{a}^{(k)} \big) \, = \, \f{ a }{ L }
\enq
in which $\bs{p}$ is to be interpreted as the thermodynamic vector counting function. They correspond to the leading large-$L$ behaviour of the $k^{\e{th}}$-species Bethe roots 
in that $\hat{\la}^{(k)}_a-\mu^{(k)}_a=\e{O}\big( L^{-1} \big)$. In fact, $ \mu^{ (k)} \, \!\!\!\! _{h_a^{(k)}}$, resp. $ \mu^{   (k)} \, \!\!\!\! _{p_a^{(k)}} $ 
should be thought of as the macroscopic counterparts of the integers $p_a^{(k)}$, resp. $h_a^{(k)}$. 
It will appear convenient, in the following, to associate, with each excited state,
the set of macroscopic particle-hole rapidities that arise in the parametrisation of the state:
\beq
\mf{R}_{\La}\; = \; \bigg\{ \Big\{ \mu_{h_a^{(k)}}^{(k)} \Big\}_{a=1}^{ n^{(k)} } \, ;   \, \Big\{ \mu_{p_a^{(k)}}^{(k)} \Big\}_{a=1}^{ n^{(k)} } \; \; , k=1,\dots, r \bigg\} \;. 
\enq

We stress that one of the hypothesis used in the derivation of the large-$L$ expansion of the counting function is that  $| D^{(k)}-\tf{ N^{(k)}_{\Om} }{ L } |=\e{O}\big( L^{-2}\big)$. 
The latter is, in fact, a constraint on how the thermodynamic limit of the model ought to be taken. 
The prescription $| D^{(k)}-\tf{ N^{(k)}_{\Om} }{ L } |=\e{O}\big( L^{-2}\big)$ ensures that the model's low-energy spectrum has the structure of a direct sum of conformal field
theories. Should one rather consider the general case, then one would get oscillatory in $L$  corrections of the order $\e{O}\big( L^{-1}\big)$ to the ground and excited states energies as discussed in 
\cite{WoynaorwiczFiniteSizeCorrections,WoynaorwiczFiniteSizeCorrectionsHubbardRepulsiveAwayHalfFilling}.

It follows from \eqref{ecriture DA fct comptage} that the Bethe roots belonging to the $k^{\e{th}}$ species condense, in the thermodynamic limit, on the interval $\intff{ - q^{(k)} }{ q^{(k)} }$
with a density $\rho^{(k)}=\big( p^{(k)} \big)^{\prime}$. However, the distribution of roots posses a small number of microscopic gaps at the rapidities of the holes 
$ \hat{\la}^{ \, (k)} \, \!\!\!\! _{h_a^{(k)}}$. There are also additional roots at the rapidities $ \hat{\la}^{\, (k)}\, \!\!\!\! _{p_a^{(k)}}$ of the particles. 

%Note that the value of the endpoints of the Fermi zones is determined by the equations
%
%
%
%\beq
%
%\lim_{L\tend + \infty} \bigg\{ \wh{ \bs{\xi} }_{\Om}\big( q^{(k)} \big) \bigg\} \; = \; D^{(k)}  \qquad  viz. \quad p^{(k)}\big( q^{(k)} \big) \, \equiv \, p_{F}^{(k)}= \f{ D^{(k)} }{2}\;. 
%
%\enq
%
%
%

The precise control on the large-$L$ behaviour of the counting function associated with a particle-hole excited state $\La$ allows for the characterisation of the excitation energies
and momenta. In the large-$L$ limit, these  are expressed in terms of the dressed energies $\veps^{(k)}$ \eqref{definition energie habillee vectorielle} and momenta $p^{(k)}$  \eqref{definition vecteur moment habille} 
of the particles and holes as 
\beqa
\mc{P}^{(\e{ex})}_{\La;\Om} &  =  & \sul{k=1}{r} \bigg\{ \mf{n}^{(k)}_{\La} p^{(k)}\big( q^{(k)} \big)  
					      \, + \, \sul{a=1}{n^{(k)} } \Big[  p^{(k)}\Big( \mu_{p_a^{(k)}} ^{(k)} \Big) \, - \,  p^{(k)}\Big(\mu_{h_a^{(k)}}^{(k)} \Big) \Big] \; \bigg\}   \; + \;\e{O}\Big(\f{1}{L} \Big) 
\label{Ecriture dressed momentum at large L}
					      \\
\mc{E}^{(\e{ex})}_{\La;\Om} & =  & \sul{k=1}{r}  \sul{a=1}{n^{(k)} } \Big[\veps^{(k)}\Big( \mu_{p_a^{(k)}} ^{(k)} \Big) \, - \,  \veps^{(k)}\Big(\mu_{h_a^{(k)}}^{(k)} \Big)  \Big] \; + \;\e{O}\Big(\f{1}{L} \Big)   \;. 
\label{Ecriture dressed energy at large L}
\eeqa
We refer to Appendix \ref{Appendix Sous section cptmt grd L de moment et energie} for more detail on the derivation of these formulae.

\subsubsection*{$\bullet$ The $\bs{\ell}$-critical classes}

Among all excited states there singles out one specific class of states: the one corresponding to all those excited states whose excitation energy $\mc{E}^{(\e{ex})}_{\La;\Om}$ vanishes in the thermodynamic limit. 
In virtue of the sign properties of the dressed energy \eqref{ecriture pte fond dressed energy} and of the large-$L$ behaviour of the excitation energy \eqref{Ecriture dressed energy at large L},
this can only happen if the particle-hole rapidities of the $k^{\e{th}}$ species collapse, when $L\tend +\infty$, on the Fermi boundaries $\pm q^{(k)}$. Such a combination of collapsing rapidities
is obtained from any collection $\La$ of Bethe roots whose associated particle-hole integers can be put in the form 
\beq
\Big\{ h_a^{(k)} \Big\}_1^{n^{(k)}} \; = \; \Big\{ N^{(k)}_{\Om} + 1 + \kappa^{(k)}  -  h_a^{(k;+)} \Big\}_1^{ n^{(k)}_{h;+} } \cup 
\Big\{  h_a^{(k;-) } \Big\}_1^{ n^{(k)}_{h;-}  }
\label{ecriture entiers trous pour etat critique}
\enq
and
\beq
\Big\{ p_a^{(k)} \Big\}_1^{n^{(k)}} \; = \; \Big\{ N^{(k)}_{\Om} +  \kappa^{(k)}  +  p_a^{(k;+)} \Big\}_1^{ n^{(k)}_{p;+} } \cup 
\Big\{  1-p_a^{(k;-) } \Big\}_1^{ n^{(k)}_{p;-}  } \;,  
\label{ecriture entiers particules pour etat critique}
\enq
with $k=1,\dots,r$. The re-centred particle $p_a^{(k;\pm)}$ and hole $h_a^{(k;\pm)}$ integers are assumed to be such that 
\beq
\lim_{L \tend +\infty}  \bigg\{ \f{1}{L} \cdot \sul{ a=1 }{ n^{(k)}_{p;\pm} } p_a^{(k;\pm)}  \bigg\}\; = \; 
\lim_{L \tend +\infty} \bigg\{  \f{1}{L} \cdot \sul{ a=1 }{ n^{(k)}_{h;\pm} } h_a^{(k;\pm)} \bigg\}\; = \; 0  \;. 
\enq
The integers $  n^{ (k) }_{ p/h ; \pm }$ correspond to the number of particle   $ n^{ (k) }_{ p ; \pm }$ and hole $ n^{ (k) }_{ h ; \pm } $ 
excitations in the $k^{\e{th}}$ species sector that collapse on the right ($+$) and left ($-$) boundaries of the $k^{\e{th}}$ species Fermi zone. These numbers are such that 
\beq
n^{(k)}_{h;+} + n^{(k)}_{h;-} \; = \;  n^{(k)}  \; = \; n^{(k)}_{p;+} + n^{(k)}_{p;-}  \;. 
\enq
States whose particle/hole integers are of the form \eqref{ecriture entiers trous pour etat critique}-\eqref{ecriture entiers particules pour etat critique}
will be called critical. One can gather the critical states into sub-classes depending on the value taken by the thermodynamic limit of the 
excitation momentum attached to these states. Introducing the integer
\beq
\ell^{(k)} \; = \; n^{(k)}_{p;+}  \, - \, n^{(k)}_{h;+}  \; = \;  n^{(k)}_{h;-}  - n^{(k)}_{p;-} 
\enq
one finds that the excitation momentum takes the form 
\beq
\mc{P}^{(\e{ex})}_{\La;\Om} \; = \; \sul{k=1}{r} \Bigg[ \big( 2\ell^{(k)} + \mf{n}^{(k)}_{\La} \big) p_{F}^{(k)}\; + \; \f{2 \pi}{ L }  \bigg\{ \sul{ a=1 }{ n_{p;+}^{(k)}  } \big( p_{a;+}^{(k)}-1\big) \, + \, \sul{ a=1 }{ n_{h;+}^{(k)}  }  h_{a;+}^{(k)}  \bigg\} 
\, - \, \f{2 \pi}{ L }\bigg\{ \sul{ a=1 }{ n_{p;-}^{(k)}  } \big( p_{a;-}^{(k)}-1\big) \, + \, \sul{ a=1 }{ n_{h;-}^{(k)}  }  h_{a;-}^{(k)}  \bigg\}  \Bigg] \; + \; \cdots 
\enq
In this expansion, $p_F^{(k)} \, = \, p^{(k)}\big( q^{(k)} \big)$ stands for the Fermi momentum  associated with the $k^{\e{th}}$ species. 
The dots $\dots$ refer to terms depending on the $p_{a;\pm}^{(k)},h_{a;\pm}^{(k)}$ but preceded by a $L^{-2}$ prefactor, or to terms of the order 
$1/L$ but which are \textit{independent} of the re-centred particle-hole integers.

Since excited states belonging to $\bs{\ell}$-classes are such that all of their associated particle-hole rapidities  collapse on the respective left or right 
Fermi boundaries, the thermodynamic limit of the set of macroscopic rapidities $\mf{R}_{\La}$ attached to any such states actually reduces to
\beq
\mf{R}_{\La} \hookrightarrow  \bigg\{  \Big\{ q^{(k)}  \Big\}_1^{ n^{(k)}_{p;+} }\bigcup \Big\{ -q^{(k)}  \Big\}_1^{ n^{(k)}_{p;-} }
	  \; ; \;  \Big\{ q^{(k)}  \Big\}_1^{ n^{(k)}_{h;+} }\bigcup \Big\{ -q^{(k)}  \Big\}_1^{ n^{(k)}_{h;-} } \; , \; \; k=1,\dots , r  \bigg\}\;. 
\enq
Relatively to the $k^{\e{th}}$ species, there will be $|\ell^{(k)}|$ particles collapsing at  $q^{(k)} \e{sgn}\big( \ell^{(k)} \big)$ and  $|\ell^{(k)}|$ holes collapsing at  $-q^{(k)} \e{sgn}\big( \ell^{(k)} \big)$
plus a certain amount of particle-hole excitations, equal in number, on each of the Fermi boundaries.  
Most, if not all, pertinent observables associated with the model do not actually "see" the effects 
of the "swarm" of particle-holes excitations, equal in number, attached to each of the boundaries of the Fermi zones but solely keep 
track of the difference in the number $\ell^{(k)} $of particle-hole excitations on the right endpoints of the $k^{\e{th}}$-species Fermi zone. 

A good example of such a mechanism is the reduction occurring in the thermodynamic limit $ \bs{F}_{\La, \Om} $ of the finite volume $\wh{ \bs{F} }_{\La, \Om}$ vector shift function of the state $ \La $ relatively to the ground state $\Om$ defined as:
\beq
\wh{ \bs{F} }_{\La, \Om}(\om) \; = \;  L \cdot \Big( \, \wh{\bs{\xi}}_{\Om}(\om) \; - \; \wh{\bs{\xi}}_{\La}(\om) \Big)  \qquad \e{and} \qquad
\bs{F}_{\La, \Om}(\om) \; = \; \lim_{L\tend + \infty} \Big\{ \wh{ \bs{F} }_{\La, \Om}(\om) \Big\} \;. 
\label{definition fonction de comptage vectorielle}
\enq
Indeed, when focusing on an excited state belonging to the $\bs{\ell}$-critical class, one has 
$\bs{F}_{\La, \Om} \hookrightarrow \bs{F}_{ \bs{\ell};\bs{\kappa} } $ where 
\beq
\bs{F}_{ \bs{\ell};\bs{\kappa} }(\om)  \; = \; \; - \; \bs{Z}(\om)\cdot \f{ \bs{\kappa} - \bs{\mf{n}} }{2} \; + \; \sul{s=1}{r} \bigg\{ \ell^{(s)} \bs{\Phi}_s\Big( \om , -q^{(s)} \Big) 
\, - \, \big( \ell^{(s)} +\kappa^{(s)} \big) \bs{\Phi}_s\Big( \om , q^{(s)} \Big) \bigg\} \;. 
\label{definition fct shift F ell kappa}
\enq

It appears convenient to introduce the dressed phase matrix by $\Phi_{ks}(\la,\mu)=\bs{\Phi}_s^{(k)}(\la,\mu)$. 
Straightforward handlings of the matrix linear integral equations allow one to express the dressed charge matrix in terms of the dressed phase matrix
as 
\beq
\bs{Z}_{ks}(\om) \; = \; \de_{ks} \, + \, \Phi_{ks}\big(\om,-q^{(s)}\big)\, - \, \Phi_{ks}\big(\om,q^{(s)}\big) \;. 
\label{expression explicite dressed charge matrix}
\enq
A slightly less obvious identity is established in Appendix \ref{Appendice derivation identite quadratique phase habillee} and relates to a closed expression for the inverse of the 
matrix $\mc{Z}_{sk}=Z_{ks}\big(q^{(k)}\big)$:
\beq
\big[ \mc{Z}^{-1}\big]_{ks} \; = \; \de_{ks} \, - \, \Phi_{ks}\big(q^{(k)},-q^{(s)}\big)\, - \, \Phi_{ks}\big(q^{(k)},q^{(s)}\big) 
\label{expression explicite inverse dressed charge matrix}
\enq
The above two identities  \eqref{expression explicite dressed charge matrix} and \eqref{expression explicite inverse dressed charge matrix} allow one to recast the specific combinations of the shift function 
\beq
\de^{(k); + }_{ \bs{\ell} ; \bs{\kappa} } \; = \;   F_{ \bs{\ell}; \bs{\kappa} }^{(k)} \big( q^{(k)} \big) + \ell^{(k)} +\kappa^{(k)} \qquad \e{and} \qquad
\de^{(k); - }_{ \bs{\ell} ; \bs{\kappa} } \; = \;   F_{\bs{\ell}; \bs{\kappa}}^{(k)} \big( -q^{(k)} \big)+ \ell^{(k)}    \;,
\label{definition exposants delta pm}
\enq
or in compact form
\beq
\de^{(k); \pm }_{ \bs{\ell} ; \bs{\kappa} } \; = \;  \big[ \mc{Z}^{\e{t}} \cdot \big( \bs{\ell} + \f{\bs{\mf{n}}}{2} \big) \big]^{(k)} \; \pm  \; \f{1}{2}\big[ \mc{Z}^{-1}  \cdot \bs{\kappa} \big]^{(k)}  \;. 
\label{ecriture formula explicite de pm k}
\enq
We remind that $\mc{Z}_{sk}=\bs{Z}_{ks}\big(q^{(k)}\big)$ and that $^{\e{t}}$ stands for the matrix transposition.

It will appear useful in the following to introduce the so-called fundamental representative of an $\bs{\ell}$-critical class. It is an excited state whose 
particle-hole integers are packed as tightly as possible. More precisely, given $\bs{\ell}$, such a state corresponds to the configuration 
\beq
\ba{ccc}  \Big\{   \{ p_{a;+}^{(k)} = a \}_1^{\ell^{(k)}} \} \, ;  \, \{ \emptyset \}  \Big\} \bigcup  \Big\{   \{ \emptyset \}   \, ; \,  \{ h_{a;-}^{(k)} = a \}_1^{\ell^{(k)}} \} \Big\} & \e{if} & \ell^{(k)}\geq 0  \vspace{2mm} \\
 \Big\{   \{ \emptyset \}   \, ; \,  \{ h_{a;+}^{(k)} = a \}_1^{-\ell^{(k)}} \} \Big\} \bigcup \Big\{   \{ p_{a;-}^{(k)} = a \}_1^{-\ell^{(k)}} \} \, ;  \, \{ \emptyset \}  \Big\}    & \e{if} & \ell^{(k)} <  0  
\ea  \;. 
\label{definition config fond pour etat classe ell}
\enq

\subsection{The form factors of local operators}

We are now finally in position to discuss the structure of the form factors of local operators.
We are going to state a conjecture relative to the universal form taken by the form factors in nested Bethe Ansatz solvable models. 
As we shall demonstrate in Section \ref{Section asympt grde distance fct deux points}, this universal form is responsible for the universality of the critical behaviour of the 
correlation functions in these models. 

The conjecture has been demonstrated to hold for numerous rank $1$ models, \textit{c}.\textit{f}. \cite{KozKitMailSlaTerEffectiveFormFactorsForXXZ,KozKitMailSlaTerThermoLimPartHoleFormFactorsForXXZ}.
In Section \ref{Section SU3 XXX modele et discussions} of the present paper, we shall demonstrate that it holds as well for the $SU(3)$-invariant XXX magnet. 
The proof heavily builds on the results of \cite{BelliardPakuliakRagoucySlavnovNestedBAScalarProductsSU(3),PakuliakRagoucySlavnovAllFFInSU3ModelsByALimit} which provide
determinant based representations for the form factors of local operators in this model. We do however trust that the conjecture does hold for all models solvable by the nested Bethe Ansatz.

\subsubsection*{ $\bullet$ The main conjecture }

\begin{conj}
\label{Conjecture resultat pple de analyse}

Let $\op{O}^{(\a)}_1$ be a local operator acting on the first site of the model and such that it connects an $\bs{N}$-particle state solely with 
a $\bs{N}+\bs{\kappa}_{\a}$ one, with $\bs{\kappa}^{t}_{\a} \, = \, \big(\kappa^{(1)}_{\a},\dots, \kappa^{(r)}_{\a} \big) $. 
Then the matrix elements of $\op{O}^{(\a)}_1$ taken between two Bethe vectors $\ket{\La}$ and $\ket{\Ups}$ take the form 
\beq
 \Bigg| \f{ \bra{  \Ups}  \op{O}^{ (\a) }_1 \ket{ \La }  }
 {  \norm{ \La } \cdot \norm{ \Ups  }     }  \Bigg|^2 
  \; = \; \wh{\mc{A} }_{  \op{O}^{ (\a) }  }\big( \Ups ; \La \big)	 \cdot 
\pl{k=1}{r}  \bigg\{ \mc{D}\Big( \{ \nu_a^{(k)}\}_{a=1}^{N_{\Ups}^{(k)}}     \mid   \{ \la_a^{(k)} \}_{a=1}^{ N^{(k)}_{\La} }   \Big) \Big[ \wh{\xi}_{\Ups}^{\,(k) },  \wh{\xi}_{ \La }^{\, (k)} \Big]    \bigg\} \;  \qquad 
where  \qquad \left\{ \ba{cc}  \La \; = \;  \Big\{ \la_a^{(k)} \Big\}_{ a = 1 }^{ N^{(k)}_{\La} }   \vspace{3mm} \\
									\Ups \; = \; \Big\{ \nu_a^{(k)} \Big\}_{ a = 1 }^{ N^{(k)}_{\Ups} }   \ea \right. 
\label{conjecture factorization facteur de forme op loc}
\enq
and the vector integers $\bs{N}_{\Ups}$ and  $\bs{N}_{\La}$ satisfy
\beq
N_{\Ups}^{(k)} \; = \; N_{\La}^{(k)} \, + \, \kappa^{(k)}_{\a} \;. 
\enq
The decomposition of the form factor is split in two parts: on the one hand the so-called "smooth" part $\wh{\mc{A} }_{ \op{O}^{ (\a) } } $ and, on the other hand, the product of the so-called discrete parts associated with the $k^{\e{th}}$
species of roots. 
\begin{itemize}

\item The factor $\wh{\mc{A} }_{ \op{O}^{ (\bs{\a}) } } \big(\La; \Ups \big)$ represents the "smooth" part of the form factor in the sense that its thermodynamic limit 
is solely described in terms of the thermodynamic limit of the rapidities of the multi-species particles and holes
that build up the excited state:
\beq
\wh{\mc{A} }_{  \op{O}^{ (\a) }  }\big(  \Ups ; \La \big) \; = \; \mc{A} _{  \op{O}^{ (\a) }  }\bigg( \mf{R}_{\Ups} \, ; \, \mf{R}_{\La} \bigg) \cdot \bigg( 1 \, + \,  \e{O}\Big( \f{1}{L} \Big) \bigg) \;. 
\label{ecriture partie lisse reduction}
\enq
Furthermore, the dependence on the particle-hole rapidities contained in the sets $ \mf{R}_{\Ups} $ and $ \mf{R}_{\La} $ is smooth. 
The set function  $ \mc{A} _{  \op{O}^{ (\bs{\a}) }  }$ enjoys, furthermore, particle-hole reduction properties, namely it holds
\beq
 \mc{A} _{  \op{O}^{ (\a) }  }\bigg( \mf{R}_{\Ups} \, ; \, \mf{R}_{\La} \bigg)_{\mid \mu_{ h_a^{(k)} }^{(k)} =   \mu_{ p_b^{(k)} }^{(k)} } \; = \; 
  \mc{A} _{  \op{O}^{ (\a) }  }\bigg( \check{\mf{R}}_{\Ups} \, ; \, \mf{R}_{\La} \bigg) \;,  
\enq
where $\check{\mf{R}}_{\La}$ is the set obtained from $\mf{R}_{\La}$ by deleting from the set of  $k^{\e{th}}$ species particle-hole rapidities 
the two rapidities:  $\mu_{ h_a^{(k)} }^{(k)}$ and $\mu_{ p_b^{(k)} }^{(k)} $.

\item 
The factor $\mc{D}$ represents the universal part of the form factor. It is operator independent in the sense that it only depends on the two-collections of roots $\La, \Ups$ parametrising the states
connected by the operator. It represents the "discrete" part of the form factor in the sense that its large-$L$ behaviour not only depends on the 
macroscopic momenta $\mf{R}_{\La}$ and $ \mf{R}_{\Ups}$ associated with the two states but also has an explicit dependence on microscopic data of the excited state,
namely the particle-holes integers. $\mc{D}$ contains all the universal part of the structure of a form factor and reads 
\beq
\mc{D}\Big( \{ \nu_a \}_{a=1}^{N_{\nu}}    \mid  \{ \la_a \}_{a=1}^{N_{\la}}   \Big) \Big[ \wh{\xi}_{\mu}, \wh{\xi}_{\la} \Big] 
\; = \; 
\pl{a=1}{ N_{\la} } \bigg\{ \f{ \sin^2\big[ \pi \wh{F}_{\nu;\la}(\la_a) \big] }{ \pi L \wh{\xi}_{\la}^{\prime}(\la_a) }  \bigg\} \; \cdot \; 
\pl{a=1}{ N_{\nu} } \bigg\{  \f{1 }{ \pi L \wh{\xi}_{\nu}^{\prime}(\nu_a) } \bigg\}
\cdot 
\f{ \pl{a<b}{N_{\nu} }  (\nu_a - \nu_b)^2 \cdot  \pl{a<b}{N_{\la} }  (\la_a - \la_b)^2 }
{  \pl{a=1}{N_{\nu}} \pl{b=1}{N_{\la} } (\nu_a -\la_b)^2 }
\label{definition fonction D espece locale}
\enq
where we made use of the shorthand notation
\beq
\wh{F}_{\nu;\la} \; = \; L \cdot \big( \wh{\xi}_{\la}- \wh{\xi}_{\nu} \big)  \;. 
\label{definition fct shift auxiliaire pour partie discrete}
\enq
Here, $\wh{\xi}_{\nu}$ (resp. $\wh{\xi}_{\la}$) are the counting functions associated with the sets of 
parameters $\nu$ (resp. $\la$).

\end{itemize}

\end{conj}

Several remarks are in order. The smooth part is definitely non-generic. It strongly varies from one model to another and also depends non trivially on the operator. 
This can be explicitly seen for those examples of rank one models where the smooth/discrete part decomposition of the form factors has been obtained. 
Also the dependence of the smooth part on the operator is manifest on the level of the expressions found in Section \ref{Section SU3 XXX modele et discussions}
for the form factors of the local operators\symbolfootnote[2]{Here $\op{E}^{ij}$ are the elementary $3 \times 3$ matrices, \textit{viz}. $\op{E}^{ij}_{k\ell}=\de_{ik} \de_{j\ell}$.  } 
$\op{E}_1^{22}, \op{E}_1^{21}$ and $\op{E}_1^{23}$ associated with the 
 $SU(3)$ invariant XXX magnet studied there. However, in what concerns the analysis of the large-distance asymptotic behaviour of the correlation functions, 
only the very broad properties of the smooth part matter, namely the reduction property \eqref{ecriture partie lisse reduction}.

The full discrete part is realized as a product over the discrete parts associated with each species $k$. Such a product decomposition means that the species do not interact on the level
of the discrete parts. The discrete part attached to the $k^{\e{th}}$ species is a function of the $k^{\e{th}}$-component $\wh{ \bs{F} }_{\Ups, \La}^{\, (k)} $ of the vector 
shift function of the state $ \Ups $ relatively to the state $\La$, \textit{c.f.} \eqref{definition fonction de comptage vectorielle} and \eqref{definition fonction D espece locale}-\eqref{definition fct shift auxiliaire pour partie discrete}.
In fact, to the leading order in $L$, the large-$L$ behaviour of the discrete part solely  depends on the thermodynamic limit of the latter
\beq
 \bs{F}_{\Ups, \La}(\om) \; = \; \lim_{L\tend + \infty} \Big\{ \wh{ \bs{F} }_{\Ups,\La}(\om) \Big\} \;. 
\enq

Since it will be of no use for the analysis that will follow, we shall \textit{not} discuss here the form taken by the large-$L$ asymptotic behaviour of
the discrete part in the case when the "in" and "out" states involved in the form factor correspond to general particle-hole excitations. There is
no problem to obtain the large-$L$ behaviour even in such a general setting and we refer the interested reader to \cite{KozKitMailSlaTerThermoLimPartHoleFormFactorsForXXZ} for the corresponding formulae.
However, we will now present the large-$L$ behaviour when one of the states is given by the ground state while the other one corresponds to a particle-hole excitation belonging to an $\bs{\ell}$-class.

\subsubsection*{$\bullet$ Large-$L$ behaviour for $\bs{\ell}$-critical states}

Let $\Ups$ be a particle-hole excited state belonging to the $\bs{\ell}$-critical class as described previously and let $\Om$
stands for the ground state Bethe roots. Then the discrete part associated with the $k^{\e{th}}$-species admits the large-$L$ behaviour 
\bem
\mc{D}\Big( \{ \nu_a^{(k)}\}_{a=1}^{N^{(k)}_{\Ups} }     \mid   
		\{ \om_a^{(k)}\}_{a=1}^{N^{(k)}_{\Om}}   \Big) \Big[ \wh{\xi}_{\Ups}^{\,(k)},  \wh{\xi}_{\Om}^{\,(k)} \Big]  \; \sim \; 
\f{ \mc{D}_{ \bs{\ell} ; \bs{\kappa} }^{(k)}  }{ L^{ \De^{ (k);+ }_{ \bs{\ell} ; \bs{\kappa} }  + \De^{ (k); - }_{ \bs{\ell} ; \bs{\kappa} }  } }
\cdot  \f{ G^2\big( 1 \, + \,   \de^{(k);+}_{ \bs{\ell} ; \bs{\kappa} } -\ell^{(k)}  \big) \, G^2\big(   1 \, - \,   \de^{ (k);- }_{ \bs{\ell} ; \bs{\kappa} }  + \ell^{(k)} \big) }
{   G^2\big( 1 \, + \,  \de^{ (k);+ }_{ \bs{\ell} ; \bs{\kappa} }  \big) \,  G^2\big( 1 \, - \, \de^{ (k);- }_{ \bs{\ell} ; \bs{\kappa} } \big) }
  \\ 
\times \mc{R}_{  n^{(k)}_{p;+} ;  n^{(k)}_{h;+}   } 
\bigg(  \big\{   p_a^{(k;+)} \big\}_1^{ n^{(k)}_{p;+} } \; ; \; \big\{   h_a^{(k;+)} \big\}_1^{ n^{(k)}_{h;+} } 
\mid  \de^{(k);+}_{ \bs{\ell} ; \bs{\kappa} } -\ell^{(k)} \bigg)  \; \cdot \; 
\mc{R}_{  n^{(k)}_{p;-} ;  n^{(k)}_{h;-}   } 
\bigg(  \big\{   p_a^{(k;-)} \big\}_1^{ n^{(k)}_{p;-} } \; ; \; \big\{   h_a^{(k;-)} \big\}_1^{ n^{(k)}_{h;-} } 
\mid -   \de^{(k);-}_{ \bs{\ell} ; \bs{\kappa} } + \ell^{(k)}  \bigg) \; . 
\end{multline}
There are several ingredients in these asymptotics:
\begin{itemize}

\item $\bs{\kappa}$ stands for the vector of $k^{\e{th}}$-species roots number discrepancies between the out and in states $$\kappa^{(k)}=N_{\Ups}^{(k)}-N_{\Om}^{(k)} \;. $$ 

\item $ \mc{D}_{ \bs{\ell} ; \bs{\kappa} }^{(k)}$ is a numerical prefactor that \textit{only} depends on the vector integers $\bs{\ell}$  labelling the critical class of interest and
on the vector integers $\bs{\kappa}$ characterising the pseudo-particle changing nature of the operator. 
\item $\mc{R}_{n_p,n_h}$ represents the microscopic contribution of the swarm of particles and holes living on the left or right Fermi boundary. It is expressed as
\bem
\mc{R}_{n_p;n_h}  \Big( \{ p_a \}_1^{ n_p } \; ; \; \{ h_a \}_1^{ n_h }  \mid \de  \Big)  \;  = \; 
\bigg(\f{\sin[\pi \de] }{ \pi }  \bigg)^{ 2 n_h} 
				 \cdot \f{ \pl{a<b}{n_p} (p_a-p_b)^2 \cdot \pl{a<b}{n_h} (h_a-h_b)^2 }
{ \pl{a=1}{n_p} \pl{b=1}{n_h} (p_a+h_b-1)^2 }    \\ 
\times  \pl{a=1}{n_p} \bigg\{ \f{  \Ga^2 (p_a+\de)  }{  \Ga^2 (p_a ) } \bigg\}
\cdot \pl{a=1}{n_h} \bigg\{   \f{  \Ga^2 (h_a-\de)  }{  \Ga^2 (h_a ) } \bigg\}  \;. 
\nonumber
\end{multline}
\item $G$ stands for the Barnes function. It is a normalisation constant which is chosen such that it counterbalances the contribution of the left/right factors $\mc{R}_{n_p,n_h}$
in the case when the particle-holes on that boundary are chosen to be the fundamental representatives of the $\bs{\ell}$-class, \textit{c.f.} \eqref{definition config fond pour etat classe ell}. 
\item The $k^{\e{th}}$ species discrete part decays algebraically in the volume with an exponent $\De^{(k);-}_{ \bs{\ell} ; \bs{\kappa} }+\De^{(k);+}_{ \bs{\ell} ; \bs{\kappa} }$ 
that is generically non-rational. The collection of the exponents $\De^{(k);\pm}_{ \bs{\ell} ; \bs{\kappa} }$ provides one with the scaling dimensions associated with the specific operator. 
These exponents are given as squares 
\beq
  \De^{(k);\pm}_{ \bs{\ell} ; \bs{\kappa} }  \; =\; \Big(   \de^{(k);\pm}_{ \bs{\ell} ; \bs{\kappa} }  \Big)^2
\label{definition scaling dimension espece k}
\enq
of the specific combinations of the values taken by the shift function $F_{ \bs{\ell}; \bs{\kappa} }^{(k)}$, \textit{c.f.} \eqref{definition fct shift F ell kappa}, 
on the right ($+$) or left ($-$) Fermi boundaries, as defined in \eqref{definition exposants delta pm}. 

\end{itemize}

\vspace{4mm}

Furthermore, when restricted to the $\bs{\ell}$-critical class, due to the reduction properties \eqref{ecriture partie lisse reduction}, the smooth part goes to a constant solely depending on the integers 
$\bs{\ell}$ and $\bs{\kappa}_{\a}$ and on the operator $\op{O}^{(\a)}$:
\beq
\wh{\mc{A} }_{\op{O}^{ (\a) } }  \big( \Ups; \Om \big)	 \simeq  \mc{A} _{ \bs{\ell} ; \bs{\kappa}_{\a} } \big( \op{O}^{ (\a) } \big) \;. 
\enq
For further convenience, it is useful to absorb all these constants in a unique term
\beq
\big|  \mc{F}_{ \bs{\ell} ; \bs{\kappa}_{\a}} \big( \op{O}^{ (\a) } \big) \big|^2  \; = \; 
\mc{A} _{ \bs{\ell} ; \bs{\kappa}_{\a} } \big( \op{O}^{ (\a) } \big)   \cdot \pl{k=1}{s}  \bigg\{ \mc{D}_{ \bs{\ell} ; \bs{\kappa}_{\a} }^{(k)}  \bigg\} \;. 
\enq
%
%
%

%%%%%%%%%%%%%%%%%%%%%%%%%%%%%%%%%%%%%%%%%%%%%%%%%%%%%%%%%%%
%%%%%%%%%%%%%%%%%%%%%%%%%%%%%%%%%%%%%%%%%%%%%%%%%%%%%%%%%%%
%%%%%%%%%%%%%%%%%%%%%%%%%%%%%%%%%%%%%%%%%%%%%%%%%

%%%%%%%%%%%%%%%%%%%%%%%%%%%%%%%%%%%%%%%%%%%%%%%%%%%%%%%%%%%
%%%%%%%%%%%%%%%%%%%%%%%%%%%%%%%%%%%%%%%%%%%%%%%%%
%%%%%%%%%%%%%%%%%%%%%%%%%%%%%%%%%%%%%%%%%%%%%%%%%%%%%%%%%%%
%%%%%%%%%%%%%%%%%%%%%%%%%%%%%%%%%%%%%%%%%%%%%%%%%

\section{Large-distance asymptotic behaviour of two-point functions and applications}
\label{Section asympt grde distance fct deux points}

\subsection{The large-distance asymptotics in the general setting}

In this section we argue that provided the setting of the previous section holds, the zero-temperature two-point functions exhibit the large-distance asymptotics
\beq
\f{  \bra{\Om} \big[ \op{O}^{(\a)}_{m+1} \big]^{\dagger} \cdot \op{O}^{(\a)}_1 \ket{\Om}  }{ \braket{\Om}{\Om} } \; \simeq  \;  \sul{ \bs{\ell} \in \mathbb{Z}^r }{}  
\ex{\i  m \big(2 \bs{\ell}+\bs{\mf{n}}_{\a}\big)\cdot \bs{p}_F  }   \cdot  
\f{ (-1)^{ \f{m}{2}  (\bs{\kappa}, 2\bs{\ell}+\bs{\mf{n}}_{\a} ) }   }{  
\big( 2 \pi m \big)^{ \De_{ \bs{\ell} ; \bs{\kappa}_{\a} }  }   } \cdot \big|  \mc{F}_{ \bs{\ell} ; \bs{\kappa}_{\a}} \big( \op{O}^{ (\a) } \big)  \big|^2 \cdot \big(1+\e{o}(1)\big) \;. 
\label{ecriture DA fct deux points}
\enq
In the above expansion, the vector $\bs{\mf{n}}_{\a}$ is defined in terms of $\bs{\kappa}_{\a}$ as is \eqref{definition vecteur shift n} while 
\beq
2 \De_{ \bs{\ell} ; \bs{\kappa}_{\a} } \, = \, \Big(2 \bs{\ell}+\bs{\mf{n}}_{\a}, \mc{Z}\mc{Z}^{\e{t}} \cdot (2 \bs{\ell}+\bs{\mf{n}}_{\a}) \Big) 
\; + \; \Big(  \bs{\kappa}_{\a}, \big[\mc{Z}\mc{Z}^{\e{t}}\big]^{-1} \bs{\kappa}_{\a} \Big)
\enq
Note that the asymptotics are indeed real valued  since \eqref{propriete parite PS vecteur n et kappa} ensures that $(\bs{\kappa},  \bs{\mf{n}}_{\a} ) \in 2\mathbb{Z}$
while the form factors $ \big|  \mc{F}_{ \bs{\ell} ; \bs{\kappa}_{\a}} \big( \op{O}^{ (\a) } \big)  \big|^2$ and $ \big|  \mc{F}_{ -\bs{\ell}-\bs{\mf{n}}_{\a} ; \bs{\kappa}_{\a}} \big( \op{O}^{ (\a) } \big)  \big|^2$
coincide owing to the symmetries of the Bethe equations.
Finally, we insist that the asymptotic expansion \eqref{ecriture DA fct deux points} provides ones with the leading large-$m$ asymptotic behaviour of each of the oscillating 
with the distance harmonics present in the large-$m$ expansion of the two-point function.  

\vspace{3mm}

In order to derive the result, we follow the strategy of \cite{KozKitMailSlaTerRestrictedSums}. 
We start by writing down the form factor expansion of a two-point function:
\beq
\f{  \bra{\Om} \big[ \op{O}^{(\a)}_{m+1} \big]^{\dagger} \cdot \op{O}^{(\a) }_1 \ket{\Om}  }{ \braket{ \Om }{ \Om } } \; = \; 
\sul{ \{ \Ups \}  }{}   \ex{ \i m \mc{P}_{\Ups; \Om}^{(\e{ex})}  } \cdot 
\Bigg| \f{ \bra{  \Ups}  \op{O}^{(\a)}_1 \ket{ \Om }  }
 {  \norm{ \Ups } \cdot \norm{ \Om  }     }  \Bigg|^2 \;. 
\enq
The sum runs over all the eigenstates of the model characterised by $N^{(k)}_{\Ups}=N^{(k)}_{\Om}+\kappa^{(\a)}$.
Owing to \eqref{definition vecteur shift n}, the integer shift $\bs{\mf{n}}_{\Ups}=\bs{\mf{n}}_{\a}$ will be constant for all such excited states.
In the following, we shall restrict the summation to the part of the spectrum realized in terms of particle-hole excitations. 
It is expected that all other types of excited states (\textit{viz}. the bound states) only generate exponentially small contributions to the large-distance asymptotics of the
correlator. This fact is rather well supported by calculations carried out on the two-point functions 
in integrable models subordinated to Lie algebras of rank 1 (for instance the XXZ spin 1/2 chain). In the long-time and large-distance regime, this property 
quite probably breaks down. In that case, in addition to the terms that will follows from the present analysis,
the contributions of the bound states should be added. These, however, go beyond the scope of the present analysis.

By analogy to the case of rank 1 models, we assume that in the large-$m$ regime the sums over particle-hole 
excitations will localise around the edges of their respective Fermi zones, \textit{viz}. the leading contribution to the large-distance asymptotics
will issue from excitations belonging to the  $\bs{\ell}$-classes. Also then, we assume that it is enough to take into account solely the leading large-$L$ behaviour of the involved form factor. 
This leads to 
\bem
\f{  \bra{\Om} \big[ \op{O}^{(\a)}_{m+1} \big]^{\dagger} \cdot \mc{O}^{(\a) }_1 \ket{\Om}  }{ \braket{ \Om }{ \Om } }\; \simeq  \; 
\pl{k=1}{r} \sul{ \ell^{(k)} \in \mathbb{Z} }{}  \pl{\eps_k= \pm }{}
\sul{ \substack{ n^{(k)}_{p;\eps_k}, n^{(k)}_{h;\eps_k} \\ n^{(k)}_{p;\eps_k}- n^{(k)}_{h;\eps_k} = \eps_k \ell_k  } }{}
\sul{ p_{1}^{(k;\eps_k)}< \dots <  p^{(k;\eps_k)}_{n^{(k)}_{p;\eps_k} } }{}
\sul{ h_{1}^{(k;\eps_k)}< \dots <  h^{(k;\eps_k)}_{n^{(k)}_{h;\eps_k} } }{}
\ex{ \i   (2\bs{\ell}+\bs{\mf{n}}_{\a}) \cdot \bs{p}_F m }   \\
\times 
\f{ \big|   \mc{F}_{ \bs{\ell} ; \bs{\kappa}_{\a}} \big( \op{O}^{ (\a) } \big)  \big|^2 }
{ \pl{k=1}{r} \Big\{ L^{\De^{(k);+}_{ \bs{\ell} ; \bs{\kappa}_{\a} }  + \De^{(k);-}_{ \bs{\ell} ; \bs{\kappa}_{\a} }   } \Big\} }
\cdot \pl{k=1}{r} \pl{ \eps_k= \pm }{} \bigg\{ \pl{a=1}{n_{p}^{(k;\eps_k)}} \ex{ \eps_k 2 \i \pi  \f{m}{L} (p_a^{(k;\eps_k)}-1)  } 
\pl{a=1}{ n_{h}^{(k;\eps_k)} } \ex{ \eps_k 2 \i \pi  \f{m}{L} h_a^{(k;\eps_k)}  } \bigg\}
\\
 \times \pl{k=1}{r} \pl{\eps=\pm }{} \Bigg\{ 
\f{  G^2 \Big(    1 \, + \, \eps \de^{(k); \eps }_{ \bs{\ell} ; \bs{\kappa}_{\a} }-\eps \ell^{(k)}    \Big) }{
						     G^2 \Big( 1 \, + \, \eps  \de^{(k); \eps }_{ \bs{\ell} ; \bs{\kappa}_{\a} }    \Big) }    \; \cdot \; 
\mc{R}_{  n^{(k)}_{p; \eps} ;  n^{(k)}_{h; \eps}   } 
\bigg(  \big\{   p_a^{(k; \eps)} \big\}_1^{ n^{(k)}_{p; \eps} } \; ; \; \big\{   h_a^{(k; \eps)} \big\}_1^{ n^{(k)}_{h; \eps} } 
\mid  \eps \de^{(k);  \eps }_{ \bs{\ell} ; \bs{\kappa}_{\a} }- \eps \ell^{(k)} \bigg)     \Bigg\} \;. 
\end{multline}
After reorganising the sums, the large-$m$ behaviour of the two-point function takes the form 
\bem
\f{  \bra{\Om}\big[ \op{O}^{(\a)}_{m+1} \big]^{\dagger} \cdot \op{O}^{(\a) }_1 \ket{\Om}  }{ \braket{ \Om }{ \Om } }  \\
\; \simeq  \;  
 \sul{ \bs{\ell} \in \mathbb{Z}^r }{}  
\ex{ \i  \bs{p}_F\cdot (2\bs{\ell}+\bs{\mf{n}}_{\a}) m } \cdot  \big|   \mc{F}_{ \bs{\ell} ; \bs{\kappa}_{\a}} \big( \op{O}^{ (\a) } \big)  \big|^2  \cdot 
\pl{k=1}{s} \Bigg\{ \f{
 \mc{L}_{\ell^{(k)}}\Big(  \de^{(k); + }_{ \bs{\ell} ; \bs{\kappa}_{\a} }-\ell^{(k)}  ; m \Big) \cdot  \mc{L}_{-\ell^{(k)}}\Big(  \de^{(k); - }_{ \bs{\ell} ; \bs{\kappa}_{\a} } + \ell^{(k)}   ; -m \Big) }
   { L^{\De^{(k);+}_{ \bs{\ell} ; \bs{\kappa}_{\a} }  + \De^{(k);-}_{ \bs{\ell} ; \bs{\kappa}_{\a} }   } }  \Bigg\}  \;. 
\end{multline}
The function $\mc{L}_{\ell}(\nu; x)$ correspond to so-called restricted sums and can be computed in closed form \cite{KozKitMailSlaTerRestrictedSums}:
\bem
\mc{L}_{\ell}(\nu; x) \; = \; \f{ G^2 \big(  1 \, + \,\nu  \big)  }{ G^2\big( 1 \, + \,  \ell \,+ \,  \nu   \big) } 
\times \sul{ \substack{ n_{p}, n_{h} \\ n_{p}- n_{h} =  \ell  } }{}
\sul{ p_{1} < \dots <  p_{ n_{p} } }{} \sul{ h_{1}< \dots <  h_{n_{h} } }{}
 \pl{a=1}{ n_{p} } \bigg\{\ex{  2 \i \pi  \f{x}{L} (p_a-1)  }  \bigg\}
\pl{a=1}{ n_{h} } \bigg\{ \ex{  2 \i \pi  \f{x}{L} h_a  } \bigg\} \\ 
\times  \mc{R}_{  n_{p} ;  n_{h}   } 
\bigg(  \big\{   p_a \big\}_1^{ n_{p} } \; ; \; \big\{   h_a \big\}_1^{ n_{h} } 
\mid \nu  \bigg)  
\; = \; \f{ \ex{ \i \f{ \pi x}{L} \ell (\ell-1) }  }{  \Big( 1 - \ex{ \f{ 2 \i \pi x }{ L } }  \Big)^{ (\nu +\ell)^2}  } \;. 
\end{multline}
After inserting the expression for $\mc{L}_{\ell}(\nu; x) $ one can already take the $L\tend +\infty$ limit what yields
\beq
\f{  \bra{\Om} \big[ \op{O}^{(\a)}_{m+1} \big]^{\dagger} \cdot \op{O}^{(\a)}_1 \ket{\Om}  }{ \braket{\Om}{\Om} } \; \simeq  \;  \sul{ \bs{\ell} \in \mathbb{Z}^r }{}  
\ex{ \i  (2\bs{\ell}+\bs{\mf{n}}_{\a}) \cdot \bs{p}_F m  }   \cdot  
\f{  \big|  \mc{F}_{ \bs{\ell} ; \bs{\kappa}_{\a}} \big( \op{O}^{ (\a) } \big)  \big|^2  }{  \big( -2\i \pi m \big)^{ \De^{ + }_{ \bs{\ell} ; \bs{\kappa}_{\a} }  }  
\big( 2\i \pi m \big)^{ \De^{ - }_{ \bs{\ell} ; \bs{\kappa}_{\a} }  }   } \big(1+\e{o}(1)\big) \;. 
\enq
where the scaling dimensions driving the large-distance asymptotics are obtained by summing-up over the scaling dimensions 
\eqref{definition scaling dimension espece k} attached to each species 
\beq
 \De^{ \pm }_{ \bs{\ell} ; \bs{\kappa}_{\a} } \; = \; \sul{k=1}{r}  \De^{ (k) ; \pm }_{ \bs{\ell} ; \bs{\kappa}_{\a} }    \;. 
\enq
It then solely remains to observe that 
\beq
 \big( -2\i \pi m \big)^{ -\De^{ + }_{ \bs{\ell} ; \bs{\kappa}_{\a} }  }  \big( 2\i \pi m \big)^{ -\De^{ - }_{ \bs{\ell} ; \bs{\kappa}_{\a} }  } \; = \; 
\f{ \ex{\i\f{\pi}{2} \big(  \De^{ + }_{ \bs{\ell} ; \bs{\kappa}_{\a} }  -  \De^{ - }_{ \bs{\ell} ; \bs{\kappa}_{\a} }  \big)  } }
  { \big(  2  \pi m \big)^{ \De^{ + }_{ \bs{\ell} ; \bs{\kappa}_{\a} } + \De^{ - }_{ \bs{\ell} ; \bs{\kappa}_{\a} } }  }
\enq
and that, due to \eqref{ecriture formula explicite de pm k}
\beq
\De^{ + }_{ \bs{\ell} ; \bs{\kappa}_{\a} } + \De^{ - }_{ \bs{\ell} ; \bs{\kappa}_{\a} }  \; = \;  \De_{ \bs{\ell} ; \bs{\kappa}_{\a} } 
\quad \e{and} \quad 
 \De^{ + }_{ \bs{\ell} ; \bs{\kappa}_{\a} }  -  \De^{ - }_{ \bs{\ell} ; \bs{\kappa}_{\a} } \; = \; (\bs{\kappa}, 2\bs{\ell}+\bs{\mf{n}}_{\a} ) \;. 
\enq
This entails the claim. \qed

\subsection{Application to specific models}
\label{Section application aux modeles specifiques}

\subsubsection{Application to the SU(3) invariant XXX magnet}
\label{SousSection application aux modeles specifiques}

The $SU(3)$ invariant $XXX$ magnet refers to the bare Hamiltonian
\beq
\op{H}_{0;XXX} \; = \; \sul{a=1}{L} \mc{P}_{a a+1}
\enq
acting on the Hilbert space  $\mf{h}_{XXX}=\bigotimes_{a=1}^L\mf{h}_a^{XXX}$ with $\mf{h}_{a}^{XXX}=\Cx^3$, $a=1,\dots,L$. Above, $\mc{P}_{ab}$ stands for the permutation operator on $\mf{h}_{a}^{XXX}\otimes \mf{h}_{b}^{XXX}$, \textit{viz}.
$\mc{P}_{ab}=\sum_{k,j}{}\op{E}_{a}^{kj}\otimes\op{E}_{b}^{jk}$. There are two conserved charges 
\beq
\op{Q}^{(1)}_{XXX}\, = \, \sul{a=1}{L}  \op{E}_{a}^{11}- \op{E}_{a}^{22} \qquad \e{and} \qquad \op{Q}^{(2)}_{XXX}\, = \, \sul{a=1}{L} \op{E}_{a}^{22}- \op{E}_{a}^{33} \;. 
\enq

The quantum integrability of the model is ensured by the existence of a local Lax matrix $\mc{L}_{ab}(\la)=R_{ab}\big(\la-\i\tf{c}{2}\big)$, where  $ R_{ab}(\la) \; = \;  I_9 \; + \; \tf{ \i c \mc{P}_{ab} }{ \la } $,
out of which one builds the twisted monodromy matrix 
\beq
\op{T}_{0;1,\dots, L}^{(\be)}(\la) \; = \; \ex{\be E_{0}^{22}} \cdot \mc{L}_{01}(\la) \cdots \mc{L}_{0L}(\la) \;. 
\enq
This monodromy matrix is a $3\times 3$ matrix on the auxiliary space $\mf{h}_{0}^{XXX}$ whose entries are operators on $\mf{h}_{XXX}$. It is sometimes convenient to represent it as
\beq
\op{T}_{0;1,\dots, L}^{(\be)}(\la) \; = \; \sul{i,j=1}{3} \op{T}_{ij}^{(\be)}(\la) \otimes \op{E}_0^{ij} 
\enq
and we shall also write $\op{T}_{ij}(\la) = \op{T}_{ij}^{(0)}(\la) $. 

The bare Hamiltonian is then reconstructed out of the logarithmic derivative of the transfer matrix at zero twist
\beq
\op{H}_{0;XXX} \; = \; \i c \f{ \Dp{} }{ \Dp{}z} \ln\Big\{ \big(z-\i\tf{c}{2}\big)^L \cdot \e{tr}_0\big[ \op{T}_{0;1,\dots, L}^{(0)}( z ) \big]  \Big\}_{\mid z= \i \tf{c}{2}} \;. 
\enq
 For later convenience, it is useful to discuss the construction of the eigenvectors associated with the $\be$-twisted transfer matrix.  
This model is built over a rank $2$ Lie-algebra: the eigenstates of the  $\be$-twisted transfer matrix are parametrised in terms of a collection $\La_{\be}$ of 
two species of Bethe roots $\big\{ \la_a^{(1)} \big\}_1^{N^{(1)}_{\La}}$ and $\big\{ \la_a^{(2)} \big\}_1^{N^{(2)}_{\La}}$. These solve the system of $\be$-twisted Bethe Ansatz equations
\beqa
\bigg( \f{  \la_k^{(1)} +\i \tf{c}{2}  }{ -\la_k^{(1)}  + \i \tf{c}{2}  } \bigg)^L  & = & \ex{\be} (-1)^{N^{(1)}_{\La}-1} 
\pl{ a=1 }{ N^{(1)}_{\La} }\bigg\{ \f{ \la_k^{(1)} - \la_a^{(1)} + \i c    }{ \la_a^{(1)} - \la_k^{(1)} + \i c }  \bigg\}
\cdot \pl{ a=1 }{ N^{(2)}_{\La} } \bigg\{ \f{ \la_a^{(2)} - \la_k^{(1)} + \i \tf{c}{2}    }{ \la_a^{(2)} - \la_k^{(1)} - \i \tf{c}{2}  }  \bigg\} \label{eqn Bethe type 1}\\
1 & = & \ex{-\be} (-1)^{N^{(2)}_{\La}-1} 
\pl{a=1}{ N^{(2)}_{\La} }\bigg\{ \f{ \la_k^{(2)} - \la_a^{(2)} + \i c    }{ \la_a^{(2)} - \la_k^{(2)} + \i c }  \bigg\}
\cdot \pl{a=1}{ N^{(1)}_{\La} } \bigg\{ \f{ \la_k^{(2)} - \la_a^{(1)} + \i \tf{c}{2}    }{ \la_k^{(2)} - \la_a^{(1)} - \i \tf{c}{2}  }  \bigg\} \label{eqn Bethe type 2} \; . 
\eeqa
Here and in the following, whenever we shall write $\La_{\be}$, it will be understood that the Bethe roots solve the $\be$-twisted Bethe equations. 
When the subscript $\be$ is dropped, namely when writing $\La$, it will mean that the Bethe roots solve the Bethe equations at $\be=0$. Finally, given $\La$, by $\La_{\be}$
we mean the solution to the $\be$-twisted Bethe equations which is a smooth in $\be$ small enough deformation of the roots $\La$. 

The eigenvalues of the $\be$-twisted transfer matrix $\e{tr}_0\big[ \op{T}_{0;1,\dots, L}^{(\be)}( z ) \big] $ associated with the eigenstate $\ket{  \La_{\be} } $
 take the form:
\beq
\tau_{\be}\big( z \mid  \La_{\be} \big) \; = \; \bigg( \f{ z +\i \tf{c}{2}  }{ z  - \i \tf{c}{2}  } \bigg)^L f\Big( \ov{\la}^{(1)},z\Big) 
\; + \; \ex{\be} f\Big(z ,  \ov{\la}^{(1)} \Big) \cdot k\Big( \ov{\la}^{(2)},z \Big)  \; + \; k\Big( \ov{\la}^{(2)} , z \Big)
\enq
where we have introduced the functions
\beq
f(x,y) \; = \; \f{x - y + \i c}{ x - y}  \quad , \quad k(x,y) \; = \; \f{ x - y + \i \tf{c}{2} }{ x - y - \i \tf{c}{2} } \;. 
\enq
Above and in the following, we agree upon the convention that whenever parameters belonging to a set appear with a bar, then the
product over all representatives of the set should be taken, \textit{e.g.}:
\beq
f\Big(\om ,  \ov{\la}^{(k)} \Big) \; = \; \pl{ a=1 }{ N^{(k)}_{\La} } f\Big(\om ,  \la^{(k)}_a \Big) \;. 
\enq
The shift functions associated with this model take the form 
\beqa
\wh{\xi}_{\La_{\be} }^{(1)}(\om) & = & p_{0;XXX}^{(1)}(\om) \, -\i \f{\be}{2\pi L}   \; + \; \f{1}{ L } \sul{a=1}{N^{(1)}_{\La}} \vth_1\big( \om- \la^{(1)}_a \big)
\; - \; \f{1}{ L } \sul{a=1}{N^{(2)}_{\La}}  \vth_2\big( \om- \la^{(2)}_a \big)   \; + \; \f{N^{(1)}_{\La} -\mf{n}^{(2)}_{\La;XXX} +1 }{ 2L }\\
\wh{\xi}_{\La_{\be} }^{(2)}(\om) & = & \, \i \f{\be}{2\pi L}   \; + \; \f{1}{ L } \sul{a=1}{N^{(2)}_{\La}} \vth_1\big( \om- \la^{(2)}_a \big)
\; - \; \f{1}{ L } \sul{a=1}{N^{(1)}_{\La}}  \vth_2\big( \om- \la^{(1)}_a \big)   \; + \; \f{N^{(2)}_{\La}-\mf{n}^{(1)}_{\La;XXX}  +1 }{ 2L } \;.
\eeqa
Above, $\mf{n}^{(a)}_{\La;XXX}=1$ if $N^{(a)}_{\La}$ is odd and $\mf{n}^{(a)}_{\La}=0$ if $N^{(a)}_{\La}$ is even. Also we have introduced
\beq
\vth_n(\om) \; = \; \f{1}{2 \i \pi } \ln \bigg( \f{ \tf{\i c}{n} +\om}{ \tf{\i c}{n} -\om  } \bigg) \qquad \e{and}\qquad 
p_{0;XXX}^{(1)}(\om) \; = \; \f{ \i }{ 2 \pi } \ln \bigg( \f{ \tf{\i c}{2} +\om}{ \tf{\i c}{2} -\om  } \bigg) \; . 
\enq
Note that $p_{0;XXX}^{(1)}(\om)$ is strictly increasing while the $\vth_k(\om)$ are strictly decreasing. Both functions are odd. 
From the above, one deduces that the bare phase matrix takes the form 
\beq
\left( \ba{cc} \th_{11}(\la,\mu) & \th_{12}(\la,\mu) \\ 
\th_{21}(\la,\mu) & \th_{22}(\la,\mu) \ea \right) \; = \; \left( \ba{cc} \vth_{1}(\la-\mu) & -\vth_{2}(\la-\mu) \\ 
									  -\vth_{2}(\la-\mu) & \vth_{1}(\la-\mu) \ea \right)
\enq
and thus does indeed satisfy to the general hypothesis stated earlier on. 
The equations defining the ground state of the model take the form
\beq
\wh{\xi}_{\Om}^{(k)}\big( \om_{a}^{(k)} \big) \, = \, a \quad \e{with} \quad  a=1,\dots, N_{\Om}^{(k)} \quad \e{and} \quad k=1,2 \;. 
\enq
Finally, we remind that for this model the shift functions are defined as 
\beq
\wh{F}^{(a)}_{\Ups,\La}(z)\, =\, L \Big( \wh{\xi}_{\La}^{(k)}\big( z \big)  \, - \,  \wh{\xi}_{\Ups}^{(k)}\big( z \big)   \Big) \;. 
\enq

The critical exponents arising in the large-distance asymptotics of two-point functions in the model will then be given by the dressed charge matrix
for this specific model. It solely remains to list the vector integers $\bs{\kappa}_{\a}$ associated with the elementary operators of the model and the vector $\bs{\mf{n}}_{\a}$
attached to the class of excited states arising in the form factor expansion of the two-point functions:
\beq
\ba{|c|c|c|c|c|}
\hline 
\e{operator} \;  \op{O}^{(\a)}  & \op{E}^{aa}_{1} & \op{E}^{12}_{1} & \op{E}^{13}_{1} & \op{E}^{23}_{1}    \\ \hline
\e{vector} \; \bs{\kappa}_{\a} & 	(0,0)  		 &  (-1,0) 	&     (1,-1)    &  (0,1)		\\ \hline
\e{vector} \; \bs{\mf{n}}_{\a} & 	(0,0)  		 &  (0,1) 	&     (1,1)    &  (1,0)		\\ \hline
\ea
\enq
The vectors associated with the other elementary operators can be obtained by hermitian conjugation.  
The asymptotic behaviour of the two-point functions $\big< \op{E}^{ba}_{1+m}\op{E}_{1}^{ab}\big>$
is then readily deduced from \eqref{ecriture DA fct deux points} by picking the appropriate vectors $\bs{\kappa}_{ab}$ and $\bs{\mf{n}}_{ab}$
from the above table.

\subsubsection{The Hubbard model}

The one-dimensional Hubbard model is a rank 2 model of particular interest that has been extensively studied since the seminal calculation of its spectrum 
by Lieb and Wu through nested Bethe Ansatz methods. We refer to \cite{EsslerFrahmGohmanKlumperKorepinOneDimensionalHubbardModel} for a thorough discussion
of the model. Despite the numerous developments relative to the model, not much is still known relatively to the exact expression for its correlation functions. 
In fact, in the present state of the art, solely the norm of the Bethe state was conjectured in \cite{GohmannKorepinConjectureNormBetheStateHubbard}. 
As mentioned in the introduction, the large-distance asymptotic behaviour of two-point functions in this model was obtained 
on the basis of the study of the $1/L$ corrections to the ground and excited state's energies by means of conformal field theoretic \cite{FrahmKorepinCriticalExponents2DHubbardPartI} or Luttinger liquid based
\cite{SchultzCritExpHubbardByLuttLiquidDvpmtgeneralTheory} reasonings.

The bare Hamiltonian of the Hubbard model is defined in terms of the set of fermionic operators
\beq
\Big\{ \op{c}_{j,a}^{\dagger},\op{c}_{k,b}^{\dagger} \Big\} \, = \, \Big\{ \op{c}_{j,a}^{\dagger},\op{c}_{k,b}^{\dagger} \Big\}\, = \; 0 \qquad 
\Big\{ \op{c}_{j,a},\op{c}_{k,b}^{\dagger} \Big\} \, = \, \de_{a,b}\de_{j,k} \qquad \op{n}_{k,a}=\op{c}_{k,a}^{\dagger},\op{c}_{k,a}
\enq
and takes the form 
\beq
\op{H}_{0;HB}\,= \, -\sul{k=1}{L} \sul{a=\ua, \da }{} \Big\{ \op{c}_{j,a}^{\dagger}\op{c}_{j+1,a} \, + \, \op{c}_{j+1,a}^{\dagger}\op{c}_{j,a}  \Big\}\, + \, 2c \sul{k=1}{L} \op{n}_{k,\ua}\op{n}_{k,\da} \;. 
\enq
The model has two conserved charges 
\beq
\op{Q}^{(1)}_{HB} \; = \; \sul{k=1}{L} \Big( \op{n}_{k,\ua}+ \op{n}_{k,\da} \Big) \qquad \e{and} \qquad 
\op{Q}^{(2)}_{HB} \; = \; \f{1}{2} \sul{k=1}{L} \Big( \op{n}_{k,\ua} - \op{n}_{k,\da} \Big)  
\enq
which have the interpretation of the total number of particle and the total longitudinal spin operators. 
The eigenvectors $\ket{\La}$ of $\op{H}_{0;HB}$ are parametrised 
by two species of Bethe roots $\big\{ \la_a^{(1)} \big\}_1^{N^{(1)}_{\La}}$ and $\big\{ \la_a^{(2)} \big\}_1^{N^{(2)}_{\La}}$. The conserved global charges
act on the Bethe vectors as
\beq
\op{Q}^{(1)}_{HB} \; = \; N_{\La}^{(1)} \ket{\La}  \qquad \e{and} \qquad   \op{Q}^{(2)}_{HB} \; = \;\f{1}{2} \Big( N_{\La}^{(1)} -2 N_{\La}^{(2)} \Big) \ket{\La}    \;. 
\enq
The Bethe equations for the Hubbard model take the form 
\beqa
\bigg(  \i \la_k^{(1)} + \sqrt{ 1- \big( \la_k^{(1)}\big)^2   }  \bigg)^L  & = &
						      \pl{ a=1 }{ N^{(2)}_{\La} } \bigg\{ \f{ \la_a^{(2)} - \la_k^{(1)} - \i \tf{c}{2}    }{ \la_a^{(2)} - \la_k^{(1)} + \i \tf{c}{2}  }  \bigg\} \\
1 & = &   (-1)^{N^{(2)}_{\La}-1} 
\pl{a=1}{ N^{(2)}_{\La} }\bigg\{ \f{ \la_a^{(2)} - \la_k^{(2)} + \i c    }{ \la_k^{(2)} - \la_a^{(2)} + \i c }  \bigg\}
\cdot \pl{a=1}{ N^{(1)}_{\La} } \bigg\{ \f{ \la_k^{(2)} - \la_a^{(1)} + \i \tf{c}{2}    }{ \la_k^{(2)} - \la_a^{(1)} - \i \tf{c}{2}  }  \bigg\}  \; . 
\eeqa
They this give rise to the below shift functions:
\beqa
\wh{\xi}_{\La }^{(1)}(\om) & = & p_{0;HB}^{(1)}(\om) \; - \; \f{1}{ L } \sul{a=1}{N^{(2)}_{\La}}  \vth_2\big( \om- \la^{(2)}_a \big)   \; + \; \f{N^{(1)}_{\La} -\mf{n}^{(1)}_{\La;HB} +1 }{ 2L }\\
\wh{\xi}_{\La }^{(2)}(\om) & = & \,  \f{1}{ L } \sul{a=1}{N^{(2)}_{\La}} \vth_1\big( \om- \la^{(2)}_a \big)
\; - \; \f{1}{ L } \sul{a=1}{N^{(1)}_{\La}}  \vth_2\big( \om- \la^{(1)}_a \big)   \; + \; \f{N^{(2)}_{\La}-\mf{n}^{(2)}_{\La;HB}  +1 }{ 2L } \;.
\eeqa
Above, the integer shift is defined as 
\beq
\mf{n}^{(1)}_{\La;HB}\, = \,  \left\{ \ba{ccc} 1 & \e{if} & N^{(1)}_{\La}+N_{\La}^{(2)}+1 \in 2\mathbb{Z}+1 \\ 
						0 & \e{if} & N^{(1)}_{\La}+N_{\La}^{(2)}+1 \in 2\mathbb{Z} \ea \right. \qquad \e{and} \qquad 
\mf{n}^{(1)}_{\La;HB}\, = \,  \left\{ \ba{ccc} 1 & \e{if} & N^{(1)}_{\La} \in 2\mathbb{Z}+1 \\ 
						0 & \e{if} & N^{(1)}_{\La} \in 2\mathbb{Z} \ea \right. \;. 
\enq
Also, the phase functions are defined as for the XXX chain while the bare momentum takes the form 
\beq
p_{0;HB}^{(1)}(\om) \; = \; \f{ - \i }{ 2 \pi } \ln \Big( \i \om +\sqrt{1-\om^2 } \Big) \; . 
\enq
Thus, in the Hubbard model the bare phase matrix takes the form 
\beq
\left( \ba{cc} \th_{11}(\la,\mu) & \th_{12}(\la,\mu) \\ 
\th_{21}(\la,\mu) & \th_{22}(\la,\mu) \ea \right) \; = \; \left( \ba{cc}0 & -\vth_{2}(\la-\mu) \\ 
									  -\vth_{2}(\la-\mu) & \vth_{1}(\la-\mu) \ea \right)
\enq
and hence does indeed satisfy to the general hypothesis stated earlier on. 
The equations defining the ground state of the model take the form
\beq
\wh{\xi}_{\Om}^{(k)}\big( \om_{a}^{(k)} \big) \, = \, a \quad \e{with} \quad  a=1,\dots, N_{\Om}^{(k)} \quad \e{and} \quad k=1,2 \;. 
\enq
Therefore, in order to characterise the large-distance asymptotics of two-point functions it remains to list the vector integers $\bs{\kappa}_{\a}$ associated with the elementary operators of the model and the vector $\bs{\mf{n}}_{\a}$
attached to the class of excited states arising in the form factor expansion of the two-point functions:
\beq
\ba{|c|c|c|c|c|c|}
\hline 
\e{operator} \;  \op{O}^{(\a)}  & \op{S}^{z}_{1} & \op{S}^{+}_{1} & \op{n}_{1;\ua/\da} & \op{c}^{\dagger}_{1;\ua}  &   \op{c}^{\dagger}_{1;\da}    \\ \hline
\e{vector} \; \bs{\kappa}_{\a} & 	(0,0)  		 &  (0,-1) 	&     (0,0)    &  (1,-1)	& (1,1)	\\ \hline
\e{vector} \; \bs{\mf{n}}_{\a} & 	(0,0)  		 &  (1,0) 	&     (0,0)    &  (0,1)		& (0,1) \\ \hline
\ea
\enq
The vectors associated with the other elementary operators can be obtained by hermitian conjugation.  Note that, on  top of the fermionic operators, we have also introduced 
the local spin operators:
\beq
\op{S}^z_1\; = \; \f{1}{2} \Big( \op{n}_{1,\ua} \, - \,  \op{n}_{1,\da} \Big) \qquad \e{and} \qquad 
\op{S}^+_1\; = \; \op{c}_{1,\ua}^{\dagger}  \op{c}_{1,\da} \;. 
\enq

The large-distance expansions obtained for the Hubbard model within our approach do confirm the conformal field theoretic predictions for this model, see \textit{e.g.} \cite{EsslerFrahmGohmanKlumperKorepinOneDimensionalHubbardModel}.

\section{The SU(3) invariant XXX magnet as a check of the form factor's structure}
\label{Section SU3 XXX modele et discussions}

In this section, we follow the setting and notation introduced in Section  \ref{SousSection application aux modeles specifiques}.

\subsection{The $\be$-twisted scalar products}

The authors of \cite{BelliardPakuliakRagoucySlavnovNestedBAScalarProductsSU(3)} introduced a function 
\beq
\mc{S}_{\be}\big( \Ups_{\be} \mid  \La \big) 	\; = \; \braket{\Ups_{\be} }{ \La} 
\label{ecriture fonction generatrice S kappa}
\enq
called the $\be$-twisted scalar product. This function depends on two collections of Bethe roots 
\beq
\Ups_{\be}\; = \; \Big\{  \{ \mu_{a}^{(1)} \}_1^{ N^{(1)}_{\Ups} } , \{ \mu_{a}^{(2)} \}_1^{ N^{(2)}_{\Ups} } \Big\}  \qquad \e{and} \qquad 
\La\; = \; \Big\{  \{ \la_{a}^{(1)} \}_1^{ N^{(1)}_{\La} } , \{ \la_{a}^{(2)} \}_1^{ N^{(2)}_{\La} } \Big\}
\enq
which solve, respectively, the $\be$-twisted Bethe equations of the model \eqref{eqn Bethe type 1}-\eqref{eqn Bethe type 2} and those at $\be=0$. At $\be=0$, it gives rise to the scalar product between the (un-normalised)-state 
parametrised by $\Ups$ and the one parametrised by $\La$.  Further, the function $\mc{S}_{\be}\big( \Ups_{\be} \mid  \La \big) $
corresponds \cite{BelliardPakuliakRagoucySlavnovNestedBAScalarProductsSU(3)} to the generating function of the form factors of the $\op{T}_{22}(z)$ entry of the monodromy matrix in that
\beq
\bra{  \Ups   } \op{T}_{22}(z) \ket{ \La	  } 
\; = \;\f{ \Dp{} }{  \Dp{}\be} \bigg\{ \Big[ \tau_{\be}\big( z \mid \Ups_{\be}  \big)   -   \tau_{0}\big( z \mid \La  \big)  \Big]
 \cdot \mc{S}_{\be}\big( \Ups_{\be}  \mid  \La \big) \bigg\}_{\mid \be=0}  \;. 		 
\label{ecriture relation valeur moyenne T22 et PS kappa deforme}
\enq

\begin{prop}
\label{Proposition reecriture scalar products}
The generating function \eqref{ecriture fonction generatrice S kappa} admits the representation
\bem
 \mc{S}_{\be}\big( \Ups_{\be}  \mid  \La \big)\; = \; 
 (1-\ex{\be} ) \cdot \ex{ N_{\La}^{(2)} \be} \cdot \pl{ \substack{ a=1 \\ \mu_{a}^{(2)} \not= \th}  }{ N_{\Ups_{\be}}^{(2)} } \Big( 1- \ex{2 \i \pi L \wh{F}^{(2)}_{\Ups_{\be};\La}\big(\mu_a^{(2)}\big) } \Big)
\pl{a=1}{ N_{\La}^{(1)} } \Big( 1- \ex{ 2 \i \pi L \wh{F}^{(1)}_{\Ups_{\be};\La}\big(\la_a^{(1)}\big) } \Big)  \\
\times \f{ h\big( \ov{\mu}^{(2)}, \th  \big) }{   h\big( \ov{\la}^{(2)}, \th  \big) }
\cdot k\big( \ov{\mu}^{(2)}, \ov{\mu}^{(1)}\cup \ov{\la}^{(1)}  \big) \cdot
 k\big( \ov{\la}^{(2)}, \ov{\la}^{(1)}  \big) \cdot  f\big( \ov{\la}^{(2)}, \ov{\mu}^{(2)}  \big) 
f\big( \ov{\la}^{(1)}, \ov{\mu}^{(1)}  \big) 
\cdot \det\Big[ \e{id} \; + \; \wh{ \op{U} }_{ \th} ( \Ups_{\be} ; \La )  \Big]
\end{multline}
in which $\th$ can be taken to be equal to any of the roots $\mu_{1}^{(2)},\dots, \mu_{N_{\Ups_{\be}}^{(2)}}^{(2)}$.  
This representation involves the function 
\beq
h(x,y) \; =  \; \f{x-y + \i c}{ \i c} 
\enq
and depends on an auxiliary parameter $\th$ which is arbitrary provided that it belongs to the set $\{ \mu_a^{(2)} \}_1^{ N_{\Ups}^{(2)} } $. 
The representation also involves the Fredholm determinant of the operator $\e{id}   +  \wh{ \op{U} }_{ \th} ( \Ups_{\be} \mid \La )  $, with $\wh{ \op{U} }_{ \th} ( \Ups_{\be} \mid \La )$ 
an integral operator on $L^2\big( \msc{C}_{ \Ups_{\be} , \La } \big) $, with 
\beq
\msc{C}_{ \Ups_{\be} , \La } \; = \; \Ga\Big( \{ \la^{(1)}_a\}_1^{N_{\La}^{(1)}} \Big) \bigcup  \Ga\Big( \{ \mu^{(2)}_a\}_1^{ N_{ \Ups_{\be} }^{(2)}} \Big)
\label{definition contour C Upsilon et Lambda}
\enq
being a small counter-clockwise loop around the indicated above Bethe roots. The operator  $\wh{\op{U}}_{\th}$  admits the following block decomposition relatively to the above 
partitioning of the contour $\msc{C}_{ \Ups_{\be} , \La }$:
\beq
\wh{\op{U}}_{\th} ( \Ups_{\be} ; \La )   \; = \; 
\left( \ba{cc}  \wh{\op{U}}^{(11)}_{\th} ( \Ups_{\be} ; \La )   &  \wh{\op{U}}^{(12)}_{\th} ( \Ups_{\be} ; \La ) 	\vspace{2mm} \\
\wh{\op{U}}^{(21)}_{\th} ( \Ups_{\be} ; \La )  & \wh{\op{U}}^{(22)}_{\th} ( \Ups_{\be} ; \La ) 	  \ea \right) \;. 
\enq
The integral kernels of the operators in the block decompositions take the form
\bem
\wh{U}^{(11)}_{\th}( \Ups_{\be} ; \La ) \big(z , z^{\prime} \big)  \; = \; 
\f{ (2 \i \pi)^{-1 } }{ 1-\ex{ 2 \i \pi \wh{F}^{(1)}_{\Ups_{\be};\La}\big(z^{\prime}\big) }  }
\f{ f\big(z^{\prime},\ov{\la}^{(1)}\big) }{  f\big( z^{\prime} ,\ov{\mu}^{(1)}\big) }
\Bigg\{  \bigg[ \f{ 1 }{ \th  - z + \i \tf{c}{2} } \; - \;   \f{ 1 }{z^{\prime} - z +  \i c }   \bigg] \\
\; - \; \bigg[ \f{ \ex{-\be} }{ z  - z^{\prime} + \i c } \cdot
 \f{ k\big(\ov{\la}^{(2)}, z^{\prime}  \big) }{ k\big(\ov{\mu}^{(2)}, z^{\prime} \big) }  
\; + \;   \f{ 1 }{\th - z - \i \tf{c}{2} }  
		\cdot \f{ k\big( \th, \ov{\mu}^{(1)}  \big) }{ k\big(\th ,\ov{\la}^{(1)} \big) }   \; \bigg] \Bigg\}
\end{multline}
\beq
\wh{U}^{(21)}_{\th}( \Ups_{\be} ; \La ) \big( z , z^{\prime} \big)  \; = \; 
\f{ (2 \i \pi)^{-1 } }{ 1 - \ex{ 2 \i \pi  \wh{F}^{(1)}_{\Ups_{\be};\La}\big(z^{\prime}\big) }  }
\f{ f\big(z^{\prime},\ov{\la}^{(1)}\big) }{  f\big( z^{\prime} ,\ov{\mu}^{(1)}\big) }
\Bigg\{  k(z, z^{\prime})  \cdot \f{ k\big(\ov{\la}^{(2)}, z^{\prime}  \big) }{ k\big(\ov{\mu}^{(2)}, z^{\prime} \big) }  
\; - \;\f{ \ex{\be} }{ f(\th, z)   } \cdot \f{ k\big( \th, \ov{\mu}^{(1)}  \big) }{ k\big(\th ,\ov{\la}^{(1)} \big) } 
\; + \; \f{\i c}{z^{\prime} - z + \i \tf{c}{2}}  \; + \; \f{ \i c}{z - \th + \i c} \Bigg\} 
\enq
in what concerns the first column while the second column reads
\bem
\wh{U}^{(12)}_{\th}( \Ups_{\be} ; \La ) \big( z , z^{\prime} \big)  \; = \; 
\f{  \ex{\be}  \cdot (2 \i \pi)^{-1 } }{ 1 - \ex{2 \i \pi  \wh{F}^{(2)}_{\Ups_{\be};\La}(z^{\prime} ) } }
\f{ f\big(\ov{\mu}^{(2)},z^{\prime}\big) }{  f\big( \ov{\la}^{(2)},z^{\prime}\big) } \cdot
\Bigg\{  \f{1}{z^{\prime} - z - \i \tf{c}{2}} \cdot 
\f{ k\big( z^{\prime}, \ov{\mu}^{(1)}  \big) }{ k\big(z^{\prime} ,\ov{\la}^{(1)} \big) }  \\
\; - \;\f{1}{\th - z - \i \tf{c}{2}} \cdot \f{ k\big( \th, \ov{\mu}^{(1)}  \big) }{ k\big(\th ,\ov{\la}^{(1)} \big) }  \; + \; \f{1}{ \th  -  z  + \i  \tf{c}{2} }  \; -\; \f{1}{ z^{\prime} - z + \i \tf{c}{2} } \Bigg\} 
\end{multline}
\beq
\wh{U}^{(22)}_{\th}( \Ups_{\be} ; \La ) \big( z , z^{\prime} \big)  \; = \; 
\f{  (2 \i \pi)^{-1 } }{ 1 - \ex{2 \i\pi  \wh{F}^{(2)}_{\Ups_{\be};\La}(z^{\prime} ) } }
\f{ f\big(\ov{\mu}^{(2)},z^{\prime}\big) }{  f\big( \ov{\la}^{(2)},z^{\prime}\big) } \cdot
\Bigg\{  \f{ \ex{\be} }{ f\big(z^{\prime}, z \big) } 
			\cdot \f{ k\big( z^{\prime}, \ov{\mu}^{(1)}  \big) }{ k\big(z^{\prime} ,\ov{\la}^{(1)} \big) } 
\; - \;\f{ \ex{\be} }{f\big(\th,z) } \cdot \f{ k\big( \th, \ov{\mu}^{(1)}  \big) }{ k\big(\th ,\ov{\la}^{(1)} \big) } 
\; + \; \f{ \i c}{ z - \th + \i c}  \; -\; \f{ \i c}{z - z^{\prime} + \i c} \Bigg\} 
\enq
Note that in writing the four integral kernels we have dropped writing out their dependence on the Bethe roots explicitly since there
is no possibility of confusion. Finally, the vector shift function $\bs{F}_{\Ups_{\be},\La}$ is as defined in \eqref{definition fonction de comptage vectorielle}.

\end{prop}

%Finally, the representation involves the counting functions whose exponents take the form 
%
%
%
%\beqa
%
%\ex{2 \i \pi \wh{F}_1(\om)} & = & \ex{-\be} \cdot
%
% \f{ h\big( \ov{\mu}^{(1)}, \om \big)  \cdot h\big(\om,  \ov{\la}^{(1)} \big)  \cdot k\big( \ov{\la}^{(2)}, \om \big) }
%
%{ h\big(\om ,  \ov{\mu}^{(1)} \big)  \cdot h\big( \ov{\la}^{(1)}, \om  \big) \cdot k\big( \ov{\mu}^{(2)}, \om \big) } \\
%
%
%\ex{2 \i \pi \wh{F}_2(\om)} & = & \ex{\be} \cdot
%
% \f{ h\big( \ov{\mu}^{(2)}, \om \big)  \cdot h\big(\om,  \ov{\la}^{(2)} \big)  \cdot k\big(\om ,  \ov{\mu}^{(1)} \big) }
%
%{ h\big( \ov{\la}^{(2)}, \om  \big)  \cdot h\big( \om , \ov{\mu}^{(2)}  \big) \cdot k\big( \om , \ov{\la}^{(1)} \big) }
%
%\eeqa
%
%
%

\begin{cor} \cite{BelliardPakuliakRagoucySlavnovNestedBAScalarProductsSU(3),ReshetikhinNormsSU(3)BetheStates}
 The norms of an eigenstate of the $SU(3)$-invariant XXX spin chain admit the determinant representation 
\bem
\mc{S}_{\be}\big( \Ups_{\be} \mid  \Ups_{\be} \big)   \; = \; 
\big\{ k\big( \ov{\mu}^{(2)}, \ov{\mu}^{(1)} \big) \big\}^{3}	\cdot \sla{f}\big( \ov{\mu}^{(1)}, \ov{\mu}^{(1)} \big)  \cdot  
\sla{f}\big( \ov{\mu}^{(2)}, \ov{\mu}^{(2)} \big)		 \\
\times  \big( 2\i\pi L c \big)^{N_{\Ups}^{(1)}+N_{\Ups_{\be}}^{(2)}} \cdot 
\Big( \wh{\xi}^{\,(1)}_{\Ups_{\be}} \Big)^{\prime} \big( \ov{\mu}^{(1)} \big) \cdot \Big( \wh{\xi}^{\,(2)}_{\Ups_{\be}} \Big)^{\prime} \big( \ov{\mu}^{(2)} \big)  \cdot 
\det \Big[ I_{| \Ups_{\be} |} \; + \; \mc{K}_{ \Ups_{\be} } \Big] \;. 
\end{multline}
Above, $I_n$ is the $n$-dimensional identity matrix and $|\Ups_{\be}|=\# \Ups_{\be}=N_{\Ups_{\be}}^{(1)}+N_{\Ups_{\be}}^{(2)}$. 
Finally, given any set of Bethe roots $\La$, the matrix $\mc{K}_{ \La }$ admits the block decomposition 
\beq
\mc{K}_{ \La  } \; = \; 
\left( \ba{cc} \wh{K}_{\La }^{(11)} \big( \la^{(1)}_j,  \la_k^{(1)} \big)  &  \wh{K}_{\La }^{(12)}\big( \la^{(1)}_j, \la_k^{(2)} \big) \vspace{3mm} \\
\wh{K}_{\La }^{(21)}\big( \la^{(2)}_j, \la_k^{(1)} \big)  &  \wh{K}_{\La }^{(22)}\big( \la^{(2)}_j, \la_k^{(2)} \big) \ea \right) 
\qquad with \qquad \wh{K}_{\La }^{(a1)}(\la,\mu) \; = \; \f{ K_a(\la-\mu) }{ L \,  \Big( \wh{\xi}^{\,(1)}_{\La} \Big)^{\prime}(\mu) }
\label{definition matrice K Lambda et de ses entrees}
\enq
\beq
 \wh{K}_{\La }^{(12)}(\la,\mu) \; = \; \f{ K_2(\la-\mu) }{ L \,  \Big( \wh{\xi}^{\,(2)}_{\La} \Big)^{\prime}(\mu) } \qquad , \qquad 
 \wh{K}_{\La }^{(22)}(\la,\mu) \; = \; \f{ K_1(\la-\mu) }{ L \,  \Big( \wh{\xi}^{\,(2)}_{\La} \Big)^{\prime}(\mu) } 
\enq
and where the two difference kernels read
\beq
K_1(\la) \; = \;  \f{  2c   }{ 2\pi \big( \la^2 + c^2 \big) } \hspace{1cm} and \hspace{1cm}
K_2(\la) \; = \;  \f{  c   }{ 2\pi \big(\la^2 + \tf{c^2}{4}\big) } \;. 
\enq

\end{cor}

Note that one has $\vth_{n}^{\prime}=-K_n$.

\begin{cor}
Let $\La$, $\Ups$ be any two collections of Bethe roots such that $N_{\La}^{(a)}=N_{\Ups}^{(a)}$, $a=1,2$, and assume that $\be$ is purely imaginary.  Then, the normalised $\be$-twisted scalar product admits the factorisation 
\beq
\bigg| \, \f{ \mc{S}_{\be}\big( \Ups_{\be} \mid  \La  \big)  }
   { || \Ups_{\be}  || \cdot || \La ||   } \, \bigg|^2 \; = \; |1-\ex{\be}  |^2 \cdot 
\Big( \wh{\mc{W} } \cdot \wh{ \mc{R} }^{(22)} \cdot \wh{ \mc{D} } \Big)\big(\Ups_{\be} ;  \La \big) \;.  						
\label{ecriture factorization PS kappa twiste}
\enq
The three functions appearing in such a decomposition take the form 
\beq
\wh{ \mc{W} }\big(\Ups ;  \La \big) \; = \; W\Big( \{\mu^{(1)}_a\}_1^{N^{(1)}_{\Ups}} ; \{\la^{(1)}_a\}_1^{N^{(2)}_{\La}} \Big) \cdot 
W \Big( \{\mu^{(2)}_a\}_1^{N^{(2)}_{\Ups}} ; \{\la^{(2)}_a\}_1^{N^{(2)}_{\La}} \Big) \cdot
\f{ k\big( \ov{\mu}^{(2)}, \ov{\la}^{(1)} \big)  \cdot k\big( \ov{\la}^{(1)}, \ov{\mu}^{(2)} \big)}
	{ | k\big( \ov{\mu}^{(2)}, \ov{\mu}^{(1)} \big)  \cdot k\big( \ov{\la}^{(2)}, \ov{\la}^{(1)} \big) |   }
\label{definition fonction W hat des ensembles}
\enq
where we agree upon 
\beq
W ( \{ y_a \}_1^N ; \{ z_a \}_1^{N^{\prime}} ) \; = \; \f{ h\big(\ov{y}, \ov{z} \big) h\big(\ov{z}, \ov{y} \big) }
	{  h\big(\ov{y}, \ov{y} \big) h\big(\ov{z}, \ov{z} \big) } \;. 
\enq
Furthermore, we have set 
\bem
\wh{\mc{R}}^{(22)}\big(\Ups_{\be} ;  \La \big) \; = \; \pl{ a=1 }{ M } \bigg( \f{ \sin\big[ \pi  \wh{F}_2\big(\mu^{(2)}_a\big) \big] }
			{  \sin\big[ \pi  \wh{F}_2\big(\la^{(2)}_a\big) \big]  } \bigg)^2 \cdot 
\f{ 1 }{ 4 \sin^2\big[ \pi \wh{F}_2\big(\th \big)  \big] }	
\cdot  \bigg| \f{ h\big( \ov{\mu}^{(2)}, \th \big)  }
		{ h\big( \ov{\la}^{(2)}, \th \big)    } \bigg|^2		 \\
\times \f{  \big| \det\Big[ \e{id} \; + \; \wh{ \op{U} }_{\th}\big(\Ups_{\be};\La\big) \Big]   \big|^2 }
{  \det \Big[ I_{ |\Ups_{\be}| } +  \mc{K}_{ \Ups_{\be} } \Big]  \cdot 
			\det \Big[ I_{ |\La| } +  \mc{K}_{\La} \Big] }
\label{definition fonction R hat des ensembles}
\end{multline}
where $\th \in \{ \mu_a^{(2)} \}_1^{ N_{\Ups}^{(2)} }$ is arbitrary. Finally, 
\beq
\wh{\mc{D}}\big(\Ups ;  \La \big) \; = \; \mc{D}\Big( \{ \mu_a^{(1)}\}_{a=1}^{ N_{\Ups}^{(1)} }     \mid   \{ \la_a^{(1)}\}_{a=1}^{  N_{\La}^{(1)} }   \Big)\big[ \wh{\xi}^{(1)}_{\Ups}, \wh{\xi}^{(1)}_{\La} \big]  
\cdot \mc{D}\Big( \{ \mu_a^{(2)}\}_{a=1}^{  N_{\Ups}^{(2)} }     \mid   \{ \la_a^{(2)}\}_{a=1}^{  N_{\La}^{(2)} }    \Big)\big[ \wh{\xi}^{(2)}_{\Ups}, \wh{\xi}^{(2)}_{\La} \big]    \;, 
\label{definition fonction D hat des ensembles}
\enq
where we remind that the functions $\mc{D}$ are as defined in \eqref{definition fonction D espece locale}. 

\end{cor}

\subsection{Form factors of local operators}

In this section, we discuss the form taken by the form factors of three local operators, namely the operators $E_{1}^{22}$, $E_{1}^{23}$ and $E_{1}^{21}$. 
The moduli squared of the form factors of the diagonal operator $E_{1}^{22}$ follow directly from \eqref{ecriture relation valeur moyenne T22 et PS kappa deforme}. 
The form factors of the diagonal operators $E_{1}^{23}$ and $E_{1}^{21}$ can be deduced from those of $E_{1}^{22}$ by using certain identities that were discovered in \cite{PakuliakRagoucySlavnovAllFFInSU3ModelsByALimit}
and which relate  to each other the various form factors of the model. We will show that the decomposition for the $\be$-twisted scalar products given in the previous section 
does indeed lead to the thermodynamic limit of form factors for this model as claimed in Conjecture \ref{conjecture factorization facteur de forme op loc}.

First, we remind some of the results of \cite{PakuliakRagoucySlavnovAllFFInSU3ModelsByALimit}.

\begin{lemme}\label{lem:E21-E23}
Let $\Ups$ and $\La$ be two collections of Bethe roots. Then it holds
\beq
 \bigg|  \f{ \bra{\Ups} \op{E}^{21}_1 \ket{\La}  }{  ||  \Ups ||  \cdot || \La ||    } \bigg|^2 \; = \; \big( L \, + \, N_{\La}^{(2)}  \, - \,2 N_{\La}^{(1)} \big)
\lim_{\om \tend +\infty} \Bigg\{  \bigg|  \f{ \bra{\Ups} \op{E}^{22}_1 \ket{\La^{(1)}_{\om}  }  }{  ||  \Ups ||  \cdot || \La^{(1)}_{\om} ||    } \bigg|^2 \Bigg\}
\label{formule reconstruction E21 via E22} 
\enq
as well as
\beq
\bigg|  \f{ \bra{\Ups} \op{E}^{23}_1 \ket{\La}  }{  ||  \Ups ||  \cdot || \La ||    } \bigg|^2 \; = \; \big(  N_{\Ups}^{(1)}  \, - \,2 N_{\Ups}^{(2)} \big)
\lim_{\om \tend +\infty} \Bigg\{  \bigg|  \f{ \bra{ \Ups^{ (2) }_{\om} } \op{E}^{22}_1 \ket{\La  }  }{  ||  \Ups^{ (2) }_{\om} ||  \cdot || \La ||    } \bigg|^2 \Bigg\} \;. 
\label{formule reconstruction E23 via E22}
\enq
Above, we have introduced the following convention for the "augmented" sets of Bethe roots
\beq
 \La^{(1)}_{\om} \; = \; \Big\{   \{ \la^{(1)}_a \}_1^{ N_{\La}^{(1)}+1 }  \; ; \;  \{ \la^{(2)}_a \}_1^{ N_{\La}^{(2)} }  \Big\} \qquad and \qquad 
 \Ups^{(2)}_{\om} \; = \; \Big\{   \{ \mu^{(1)}_a \}_1^{ N_{\Ups}^{(1)} }  \; ; \;  \{ \mu^{(2)}_a \}_1^{ N_{\Ups}^{(2)}+1 }  \Big\}
\enq
where respectively, $ \la_{ N_{\La}^{(1)}+1 }^{(1)}=\om$ or  $ \mu_{ N_{\Ups}^{(2)}+1 }^{(2)}=\om$.

\end{lemme}

\Proof 

Recall that the matrix elements of local operators are reconstructed as \cite{MailletTerrasGeneralsolutionInverseProblem}:
\beq
\op{E}_1^{k \ell } \; = \; \lim_{z \tend \i \tf{c}{2} } \bigg\{ \op{T}_{\ell k}(z) \cdot  \Big\{  \e{tr}_{0}\big[ \op{T}^{(0)}_{0;1,\dots,L}(z) \big]\Big\}^{-1} \bigg\} \;.  
\label{ecriture formule IP}
\enq
The action of the inverse transfer matrix is readily computed on the Bethe eigenvectors. Then, it solely remains to use the  following  identities (see \cite{PakuliakRagoucySlavnovAllFFInSU3ModelsByALimit} for more details):
\beq
\bra{\Ups } \op{T}_{32}(z) \ket{ \La } \; = \; -\lim_{\om \tend + \infty}\Big\{   \f{ \om }{ \i c } \bra{\Ups_{\om}^{(2)} } \op{T}_{22}(z) \ket{ \La }  \Big\} 
\quad \e{and} \quad
\bra{\Ups } \op{T}_{12}(z) \ket{ \La } \; = \; \lim_{\om \tend + \infty}\Big\{   \f{ \om }{ \i c } \bra{\Ups } \op{T}_{22}(z) \ket{ \La_{\om}^{(1)} }  \Big\} 
\enq
as well as 
\beq
\norm{ \La }^2 \; = \; \lim_{\om \tend + \infty}\bigg\{  \f{  \om^2 \cdot  \norm{ \La_{\om}^{(1)}  }^2 /c^2  }{ L + N_{\La}^{(2)} - 2 N_{\La}^{(1)}  } \bigg\} \quad \e{and} \quad
\norm{ \Ups }^2 \; = \; \lim_{\om \tend + \infty}\bigg\{  \f{  \om^2 \cdot  \norm{ \Ups_{\om}^{(2)}  }^2 /c^2  }{  N_{\Ups}^{(1)} - 2 N_{\Ups}^{(2)}  } \bigg\} 
\enq
and use that $ \tau_{0}\big(z \mid \Ups \big)  = \lim_{\om\tend +\infty} \Big\{  \tau_{0}\big(z \mid \Ups^{(a)}_{\om} \big)  \Big\} $. \qed

\begin{lemme}\label{lem:Ejk}
Let $\Ups$ and $\La$ be two collections of Bethe roots. Then it holds
\beq
 \bigg|  \f{ \bra{\Ups} \op{E}^{12}_1 \ket{\La}  }{  ||  \Ups ||  \cdot || \La ||    } \bigg|^2 \; = \;  \bigg|  \f{ \bra{\La} \op{E}^{21}_1 \ket{\Ups}  }{  ||  \Ups ||  \cdot || \La ||    } \bigg|^2 \,,\
\qquad
\bigg|  \f{ \bra{\Ups} \op{E}^{32}_1 \ket{\La}  }{  ||  \Ups ||  \cdot || \La ||    } \bigg|^2 \; = \;  \bigg|  \f{ \bra{\La} \op{E}^{23}_1 \ket{\Ups}  }{  ||  \Ups ||  \cdot || \La ||    } \bigg|^2 
\label{formule reconstruction Ejk via Ekj} 
\enq
and
\beq
 \bigg|  \f{ \bra{\Ups} \op{E}^{13}_1 \ket{\La}  }{  ||  \Ups ||  \cdot || \La ||    } \bigg|^2 \; = \;  \bigg|  \f{ \bra{\La} \op{E}^{31}_1 \ket{\Ups}  }{  ||  \Ups ||  \cdot || \La ||    } \bigg|^2 \;.
\label{formule reconstruction E13 via E31} 
\enq
Also, one has 
\beq
 \bigg|  \f{ \bra{\Ups} \op{E}^{31}_1 \ket{\La}  }{  ||  \Ups ||  \cdot || \La ||    } \bigg|^2 \; = \; \ \big(  N_{\Ups}^{(1)}  \, - \,2 N_{\Ups}^{(2)} \big)
\lim_{\om \tend +\infty} \Bigg\{  \bigg|  \f{ \bra{ \Ups^{ (2) }_{\om} } \op{E}^{12}_1 \ket{\La  }  }{  ||  \Ups^{ (2) }_{\om} ||  \cdot || \La ||    } \bigg|^2 \Bigg\} .
\label{formule reconstruction E31 via E12} 
\enq
Finally, in what concerns the diagonal amplitudes, one has:
\beq
 \bigg|  \f{ \bra{\Ups} \op{E}^{11}_1-\op{E}^{22}_1 \ket{\La}  }{  ||  \Ups ||  \cdot || \La ||    } \bigg|^2 \; = \; \big( L \, + \, N_{\La}^{(2)}  \, - \,2 N_{\La}^{(1)} \big)
\lim_{\om \tend +\infty} \Bigg\{  \bigg|  \f{ \bra{\Ups} \op{E}^{12}_1 \ket{\La^{(1)}_{\om}  }  }{  ||  \Ups ||  \cdot || \La^{(1)}_{\om} ||    } \bigg|^2 \Bigg\}
\label{formule reconstruction E11 via E12} 
\enq
\beq
 \bigg|  \f{ \bra{\Ups} \op{E}^{22}_1-\op{E}^{33}_1 \ket{\La}  }{  ||  \Ups ||  \cdot || \La ||    } \bigg|^2 \; = \; \ \big(  N_{\Ups}^{(1)}  \, - \,2 N_{\Ups}^{(2)} \big)
\lim_{\om \tend +\infty} \Bigg\{  \bigg|  \f{ \bra{ \Ups^{ (2) }_{\om} } \op{E}^{32}_1 \ket{\La  }  }{  ||  \Ups^{ (2) }_{\om} ||  \cdot || \La ||    } \bigg|^2 \Bigg\} \;.
\label{formule reconstruction E33 via E32} 
\enq

\end{lemme}
\Proof

Relations \eqref{formule reconstruction E31 via E12}, \eqref{formule reconstruction E11 via E12}  and \eqref{formule reconstruction E33 via E32} are obtained using the limits
\beqa
\bra{\Ups } \op{T}_{31}(z)\ket{ \La } &=& - \lim_{\om \tend + \infty}\Big\{   \f{ \om }{ \i c } \bra{\Ups_{\om}^{(2)} } \op{T}_{21}(z) \ket{ \La }  \Big\} \\
\bra{\Ups } \op{T}_{11}(z)-\op{T}_{22}(z) \ket{ \La } &=& \lim_{\om \tend + \infty}\Big\{   \f{ \om }{ \i c } \bra{\Ups } \op{T}_{12}(z) \ket{ \La_{\om}^{(1)} }  \Big\}  \\
\bra{\Ups } \op{T}_{22}(z)-\op{T}_{33}(z) \ket{ \La } &=&  \lim_{\om \tend + \infty}\Big\{   \f{ \om }{ \i c } \bra{\Ups_{\om}^{(2)} } \op{T}_{23}(z) \ket{ \La }  \Big\} 
\eeqa
Then, to establish relations \eqref{formule reconstruction Ejk via Ekj}  and \eqref{formule reconstruction E13 via E31}, we use the antimorphism $\psi$ that acts as $\psi(T_{jk}(z))=T_{kj}(z)$ and $\psi(\ket{\La})=\bra{\La}$, 
see \cite{PakuliakRagoucySlavnovAllFFInSU3ModelsByALimit} for details.

\qed

In the following, owing to Lemma \ref{lem:Ejk}, a good deal of amplitudes can be deduced from the three that are described in Lemma \ref{lem:E21-E23}. We shall therefore only focus our attention the latter.

In order to state the formulae for the moduli squared of the form factors, we need to introduce some notations.

\begin{prop}
\label{Proposition factorisation lisse discrete pour quelques facteurs de forme}

Let $\La$ and $\Ups$ denote two sets of Bethe roots. Then,
the amplitudes factorise as follows. 

For the diagonal operator 
\beq
 \bigg|  \f{ \bra{\Ups} \op{E}_{1}^{22} \ket{\La}  }{  ||  \Ups ||  \cdot || \La ||    } \bigg|^2 \; = \; 
 -\f{1}{2}\f{\Dp{}^2 }{ \Dp{}\be^2  }\bigg\{ \wh{\mc{D}}\big( \Ups_{\be} , \La \big)  \cdot \wh{\mc{A}}^{(22)}\big( \Ups_{\be} , \La \big) \bigg\}_{ \mid \be=0}
\label{ecriture decomposition lisse discrete E22}
\enq
 the function $\wh{\mc{D}}\big( \Ups , \La \big)$ being defined as in \eqref{definition fonction D hat des ensembles}, while
\beq
\wh{\mc{A}}^{(22)}\big( \Ups_{\be} , \La \big)  \; = \; \Big|  \f{ \tau_0\big( \i\tf{c}{2} \mid \Ups_{\be}\big) }{  \tau_0\big( \i\tf{c}{2} \mid \La\big) } -  1   \Big|^{2}
\cdot \;  \big| 1-\ex{ \be }\big|^2 \cdot  \wh{ \mc{W} }\big(\Ups_{\be} ;  \La \big) \cdot \wh{\mc{R}}^{(22)}\big(\Ups_{\be} ;  \La \big)
\label{ecriture amplitude pour FF type 22}
\enq
where the two functions $\wh{ \mc{W} }$ and $\wh{\mc{R}}^{(22)}$ are, respectively, defined in \eqref{definition fonction W hat des ensembles} and 
\eqref{definition fonction R hat des ensembles}. 

For the off-diagonal terms, one has
\beq
 \bigg|  \f{ \bra{\Ups} \op{E}_{1}^{21} \ket{\La}  }{  ||  \Ups ||  \cdot || \La ||    } \bigg|^2 \; = \; \wh{\mc{D}}\big( \Ups , \La \big)  \cdot \wh{\mc{A}}^{(21)}\big( \Ups , \La \big)\,,
\quad\mbox{and}\quad
 \bigg|  \f{ \bra{\Ups} \op{E}_{1}^{23} \ket{\La}  }{  ||  \Ups ||  \cdot || \La ||    } \bigg|^2 \; = \; \wh{\mc{D}}\big( \Ups , \La \big)  \cdot \wh{\mc{A}}^{(23)}\big( \Ups , \La \big)\,,
\label{ecriture decomposition lisse discrete E21 et E23}
\enq
with
\beqa
\wh{\mc{A}}^{(21)}\big( \Ups , \La \big) &=& \Big|  \f{ \tau_0\big( \i\tf{c}{2}\mid \Ups\big) }{  \tau_0\big( \i\tf{c}{2}\mid \La\big) } -  1   \Big|^{2} \cdot \wh{ \mc{W} }\big(\Ups ;  \La \big) \cdot \wh{\mc{R}}^{(21)}\big(\Ups ;  \La \big)\,,
\label{ecriture amplitude pour FF type 21}
\\
\wh{\mc{A}}^{(23)}\big( \Ups , \La \big) &=& \Big|  \f{ \tau_0\big( \i\tf{c}{2}\mid \Ups\big) }{  \tau_0\big( \i\tf{c}{2}\mid \La\big) } -  1   \Big|^{2} \cdot \wh{ \mc{W} }\big(\Ups ;  \La \big) \cdot \wh{\mc{R}}^{(23)}\big(\Ups ;  \La \big)\,,
\label{ecriture amplitude pour FF type 23}
\eeqa
where 
\beqa
\wh{\mc{R}}^{(23)}\big(\Ups ;  \La \big) &=& \f{1}{8c} \f{ \sin^2\Big[ \pi \wh{F}_{\Ups;\La}^{(2)} \big( \ov{\mu}^{(2)} \big) \Big]  }{ \sin^2\Big[ \pi \wh{F}_{\Ups;\La}^{(2)} \big( \ov{\la}^{(2)} \big) \Big] }
\cdot \f{  \big| \det\Big[ \e{id} \; + \; \wh{ \op{U} }_{\infty}\big(\Ups;\La\big) \Big]   \big|^2 }
{  \det \Big[ I_{ |\Ups | } +  \mc{K}_{ \Ups  } \Big]  \cdot 
\det \Big[ I_{ |\La| } +  \mc{K}_{\La} \Big] } \nonumber \\
\wh{\mc{R}}^{(21)}\big(\Ups ;  \La \big) &=& \f{1}{8  c} \f{ \sin^2\Big[ \pi \wh{F}_{\Ups;\La}^{(2)} \big( \ov{\mu}^{(2)} \big) \Big]  }{ \sin^2\Big[ \pi \wh{F}_{\Ups;\La}^{(2)} \big( \ov{\la}^{(2)} \big) \Big] }
\cdot \Bigg| \f{ h\big( \ov{\mu}^{(2)}, \th \big)  }{ h\big( \ov{\la}^{(2)}, \th \big)    } \Bigg|^2 \cdot 
\f{ \big|  1 \, - \,  \tf{ k\big( \th, \ov{\mu}^{(1)}  \big) }{ k\big(\th ,\ov{\la}^{(1)} \big) }  \big|^2 \cdot \big| \det\Big[ \e{id} \; + \; \wh{ \op{V} }_{\th}\big(\Ups;\La\big) \Big]   \big|^2 }
{ \sin^2\Big[ \pi \wh{F}_{\Ups;\La}^{(2)} \big( \th \big) \Big] \cdot  \det \Big[ I_{ |\Ups| } +  \mc{K}_{ \Ups } \Big]  \cdot \det \Big[ I_{ |\La| } +  \mc{K}_{\La} \Big] } \;. \nonumber
\eeqa
The parameter $ \th $ is any of the Bethe roots $\{ \mu_a^{(2)} \}_1^{ N_{\Ups}^{(2)} }$ and the operator $\wh{ \op{V} }_{\th}\big(\Ups;\La\big) $ is an integral operator on the 
contour $\mc{C}_{\Ups;\La}$ defined by \eqref{definition contour C Upsilon et Lambda} which corresponds to a rank one perturbation of $\wh{ \op{U} }_{\th}\big(\Ups;\La\big) $:
\beq
\wh{ \op{V} }_{\th}^{(ab)}\big(\Ups;\La\big)(z,z^{\prime} ) \; = \; \wh{ \op{U} }_{\th}^{(ab)}\big(\Ups;\La\big)(z,z^{\prime} ) \; - \; \wh{G}_{\th;L}^{(a)}\big(\Ups;\La\big)(z)\cdot \wh{G}_{\th;R}^{(b)}\big(\Ups;\La\big)( z^{\prime} )
\enq
where the functions $\wh{G}_{\th;L/R}^{(a)}(\Ups,\La)(z)$ read 
\beqa
\wh{G}_{\th;L}^{(1)}\big(\Ups;\La\big)(z) & = & \f{ \i c }{ 2 } \Bigg\{   \f{ 1 }{ \th  - z + \i \tf{c}{2} } \; - \;    \f{ 1 }{\th - z - \i \tf{c}{2} }  
		\cdot \f{ k\big( \th, \ov{\mu}^{(1)}  \big) }{ k\big(\th ,\ov{\la}^{(1)} \big) }   \; \Bigg\}  \;, \\
\wh{G}_{\th;L}^{(2)}\big(\Ups;\La\big)(z) & = & \f{ \i c }{ 2 } \Bigg\{1 \; - \;\f{ 1 }{ f(\th, z)   } \cdot \f{ k\big( \th, \ov{\mu}^{(1)}  \big) }{ k\big(\th ,\ov{\la}^{(1)} \big) } 
\, + \, \f{ \i c  }{ z-\th +\i c   } \Bigg\}  \;,  \\
\wh{G}_{\th;R}^{(1)}\big(\Ups;\La\big)(z) & = & \f{ ( c \pi)^{-1 } }{ 1-\exp\Big\{ 2 \i \pi \wh{F}^{(1)}_{\Ups;\La}\big(z\big) \Big\}  } \cdot 
\f{ f\big(z,\ov{\la}^{(1)}\big) }{  f\big( z ,\ov{\mu}^{(1)}\big) } \cdot
\Bigg\{  \f{ k\big( \th, \ov{\mu}^{(1)}  \big) }{ k\big(\th ,\ov{\la}^{(1)} \big) }   \, - \,  \f{ k\big(\ov{\la}^{(2)}, z  \big) }{ k\big(\ov{\mu}^{(2)}, z \big) }  \Bigg\}
\cdot \Bigg\{ 1 \, - \,  \f{ k\big( \th, \ov{\mu}^{(1)}  \big) }{ k\big(\th ,\ov{\la}^{(1)} \big) } \Bigg\}^{-1} \hspace{-2mm} ,  \\
\wh{G}_{\th;R}^{(2)}\big(\Ups;\La\big)(z) & = & \f{ ( - c \pi)^{-1 } }{ 1-\exp\Big\{ 2 \i \pi \wh{F}^{(2)}_{\Ups;\La}\big(z\big) \Big\}  }\cdot  \f{ f\big(\ov{\mu}^{(2)},z\big) }{  f\big( \ov{\la}^{(2)},z\big) } \cdot
\Bigg\{ \f{ k\big( z, \ov{\mu}^{(1)}  \big) }{ k\big(z ,\ov{\la}^{(1)} \big) } \, - \,  \f{ k\big( \th, \ov{\mu}^{(1)}  \big) }{ k\big(\th ,\ov{\la}^{(1)} \big) }  \Bigg\}
\cdot \Bigg\{ 1 \, - \,  \f{ k\big( \th, \ov{\mu}^{(1)}  \big) }{ k\big(\th ,\ov{\la}^{(1)} \big) } \Bigg\}^{-1} \hspace{-2mm}.
\eeqa

\end{prop}

\Proof

Formula \eqref{ecriture decomposition lisse discrete E22} is a direct consequence of the reconstruction \eqref{ecriture relation valeur moyenne T22 et PS kappa deforme} and \eqref{ecriture formule IP},
the decomposition of the scalar product \eqref{ecriture factorization PS kappa twiste} and the fact that $\be \in \i \R$. 

\subsubsection*{$\bullet$ The amplitudes for $\op{E}^{21}_1$}

The starting point to establish \eqref{ecriture decomposition lisse discrete E21 et E23}
and \eqref{ecriture amplitude pour FF type 21} is the identity \eqref{formule reconstruction E21 via E22}.  
Since the two sets $\La, \Ups$ are different, $\wh{\mc{R}}^{(22)}\big( \Ups_{\be}, \La \big)$ is regular at $\be=0$, so that the $\be$-derivative has to act on $|1-\ex{\be}|^2$. Thus, 
due to  $ \tau_{0}\big(z \mid \La \big)  = \lim_{\om\tend +\infty} \Big\{  \tau_{0}\big(z \mid \La^{(1)}_{\om} \big)  \Big\} $, one gets 
\beq
 \bigg|  \f{ \bra{\Ups} \op{E}_{1}^{21} \ket{\La}  }{  ||  \Ups ||  \cdot || \La ||    } \bigg|^2 \; = \; \Big|  \f{ \tau_0\big( \i\tf{c}{2} \mid \Ups\big) }{  \tau_0\big(  \i\tf{c}{2} \mid \La\big) } -  1   \Big|^{2}
\lim_{\om \tend + \infty}\bigg\{  \wh{\mc{W}}\big( \Ups , \La^{(1)}_{\om} \big) \cdot \wh{\mc{D}}\big( \Ups , \La^{(1)}_{\om} \big)  \cdot  \wh{\mc{R}}^{(22)}\big( \Ups, \La^{(1)}_{\om} \big) \bigg\} \;. 
\enq
It follows from straightforward algebra that 
\beq
\wh{\mc{W}}\big( \Ups , \La^{(1)}_{\om} \big) \; = \;  \f{ \om^{2} }{ c^2   } \cdot  \wh{\mc{W}}\big( \Ups , \La \big) 
\cdot \Big(1+\e{O}\big( \om^{-1} \big) \Big)\;. 
\enq
Further, observe that, for any fixed $z$ and $a=1,2$, it holds
\beq
\big( \wh{\xi}^{\,(a)}_{\La^{(1)}_{\om}}\big)^{\prime}(z) \; = \; \big( \wh{\xi}^{\,(a)}_{\La}\big)^{\prime}(z)\; + \; \e{O}\big(\om^{-2}\big)
\label{ecriture DA en om des densite om modifiees}
\enq
as well as 
\beq
\wh{F}^{\,(1)}_{\Ups, \La^{(1)}_{\om} }(z) \; = \; \wh{F}^{\,(1)}_{\Ups, \La }(z) \, + \, \underbrace{ \tf{1}{2} \, + \, \vth_1(z-\om) }_{1 \, + \, \e{O}\big( \om^{-1} \big)}
\qquad \e{and} \qquad
\wh{F}^{\,(2)}_{\Ups, \La^{(1)}_{\om} }(z) \; = \; \wh{F}^{\,(2)}_{\Ups, \La }(z) \, + \,  \underbrace{ \tf{1}{2} \, - \, \vth_2(z-\om) }_{ \e{O}\big( \om^{-1}\big) } \, - \, \de_{ n_{\La}^{(1)},0 }  \;. 
\enq
The above also implies that $\wh{F}^{\,(1)}_{\Ups, \La^{(1)}_{\om} }(\om) = \tf{1}{2}  \, - \, \tf{c}{(\pi \om) } \, + \,   \e{O}\big( \om^{-2} \big) $\;. As a consequence, one gets that 
\beq
\wh{\mc{D}}\big( \Ups , \La^{(1)}_{\om} \big) \; = \;  \f{   \wh{\mc{D}}\big( \Ups , \La \big)   }{ \pi L \om^{2} \big( \wh{\xi}^{\,(1)}_{\La^{(1)}_{\om}}\big)^{\prime}(\om)   } 
\cdot \Big(1+\e{O}\big( \om^{-1} \big) \Big)\;. 
\enq
It remains to deal with the limit of $\wh{\mc{R}}^{(22)}\big( \Ups, \La^{(1)}_{\om} \big)$. The $\om\tend +\infty$ behaviour of the pre-factors in $\wh{\mc{R}}^{(22)}\big(\Ups;\La^{(1)}_{\om} \big) $ is easy computed leading to 
\beq
\wh{\mc{R}}^{(22)}\big( \Ups, \La^{(1)}_{\om} \big) \; = \; \f{  \det\big[ I_{|\La | } \, + \,  \mc{K}_{ |\La | }  \big]  }{ \det\big[ I_{|\La^{(1)}_{\om} | } \, + \,  \mc{K}_{ |\La^{(1)}_{\om} | }  \big] }
\cdot \Bigg|  \f{ \det\big[ \e{id} \, + \,  \wh{\op{U}}_{\th}\big(\Ups;\La^{(1)}_{\om} \big)  \big] }{ \det\big[ \e{id} \, + \,  \wh{\op{U}}_{\th}\big(\Ups;\La \big)  \big] } \Bigg|^2_{\mid \be=0} 
\cdot \wh{\mc{R}}^{(22)}\big( \Ups, \La^{(1)}_{\om} \big)  \cdot \Big(1\, + \, \e{O}\big( \om^{-1} \big) \Big) \;. 
\enq
The treatment of the ratio of determinants is a bit more involved since one has to extract out of the determinants the part which 
vanishes when $\om \tend +\infty$. 
We first focus on the norm determinant. One has the $N_{\La}^{(1)}\times 1 \times N_{\La}^{(2)}$ block form decomposition: 
\beq
I_{|\La^{(1)}_{\om} | } \, + \,  \mc{K}_{ |\La^{(1)}_{\om} | } \; = \; \left( \ba{ccc}  
\de_{ab} + \wh{K}_{ \La^{(1)}_{\om} }^{(11)} \big(\la^{(1)}_a, \la^{(1)}_b\big)  & \wh{K}_{ \La^{(1)}_{\om} }^{(11)} \big(\la^{(1)}_a, \om \big) & \wh{K}_{ \La^{(1)}_{\om} }^{(12)}\big(\la^{(1)}_a, \la^{(2)}_b\big) \vspace{1mm}  \\ 
\wh{K}_{ \La^{(1)}_{\om} }^{ (11) } \big( \om , \la^{(1)}_b\big)  & 1+\wh{K}_{ \La^{(1)}_{\om} }^{ (11) } \big(\om, \om \big) & \wh{K}_{ \La^{(1)}_{\om} }^{(12)} \big(\om, \la^{(2)}_b\big)  \vspace{1mm} \\ 
 \wh{K}_{ \La^{(1)}_{\om} }^{(21)} \big(\la^{(2)}_a, \la^{(1)}_b\big)  & \wh{K}_{\La^{(1)}_{\om}}^{(21)}\big(\la^{(2)}_a, \om \big) & \de_{ab} + \wh{K}_{\La^{(1)}_{\om}}^{(2)}\big(\la^{(2)}_a, \la^{(2)}_b\big) \ea  \right)
\enq
where $\wh{K}_{\La}^{(ab)}(y,z)$ has been defined in \eqref{definition matrice K Lambda et de ses entrees}. 
In virtue of \eqref{ecriture DA en om des densite om modifiees}, it follows that for any fixed $y,z$
\beq
\wh{K}_{ \La^{(1)}_{\om} }^{ (ab) }(y,z)\; = \; \wh{K}_{ \La }^{ (ab) }(y,z)\Big( 1 + \e{O}\big(\om^{-1}\big) \Big) \quad \e{while} \quad
\Big| \wh{K}_{\La^{(1)}_{\om}}^{(a1)}(y,\om) \big| + \Big| \wh{K}_{\La^{(1)}_{\om}}^{(1a)}(\om,y) \big|\; = \; \e{O}\big(\om^{-2}\big) \;. 
\enq
Note that, the last bounds follow from the large-$\om$ asymptotics
\beq
\big( \wh{\xi}^{\,(1)}_{\La^{(1)}_{\om}}\big)^{\prime}(\om) \; = \; - \; \f{1}{L} K_{1}(0) \, + \, \f{ c }{ 2 \pi L \om^2 }\big(L \, - \, 2 N_{\La}^{(1)} \, + \, N_{\La}^{(2)} \big)  \; + \; \e{O}\big( \om^{-3} \big)  
\enq
and the fact that $K_1(0)=\tf{1}{(\pi c)}\not=0$. 
These asymptotics also imply that 
\beq
1+\wh{K}_{\La^{(1)}_{\om}}^{(11)}\big(\om, \om \big)\; = \; c \cdot \f{ L \, - \, 2 N_{\La}^{(1)} \, + \, N_{\La}^{(2)}    }{ 2 \pi L \om^2  \big( \wh{\xi}^{\,(a)}_{\La^{(1)}_{\om}}\big)^{\prime}(\om)  }\; \cdot \big( 1+\e{O}(\om^{-2}) \big)\;. 
\enq
In order to extract the $\om^{-2}$ decay out of $\det\big[ I_{|\La^{(1)}_{\om} | } \, + \,  \mc{K}_{ |\La^{(1)}_{\om} | }  \big]$, it is then enough to "kill" the off-diagonal entries
of the column associated with the root $\om$ by making linear combinations of lines. Since the line associated with the root $\om$ is entry-wise $\e{O}\big( \om^{-2} \big)$, 
doing so only modifies the other entries of the matrix by a $\e{O}\big( \om^{-2} \big)$ quantity. Therefore, all-in-all,
\beq
\det\big[ I_{|\La^{(1)}_{\om} | } \, + \,  \mc{K}_{ |\La^{(1)}_{\om} | }  \big] \; = \; \det\big[ I_{|\La | } \, + \,  \mc{K}_{ |\La | }  \big] \cdot 
c \cdot \f{ L \, - \, 2 N_{\La}^{(1)} \, + \, N_{\La}^{(2)}    }{ 2 \pi L \om^2  \big( \wh{\xi}^{\,(1)}_{\La^{(1)}_{\om}}\big)^{\prime}(\om)  }\; \cdot \big( 1+\e{O}(\om^{-2}) \big) \; . 
\enq

The strategy to extract the decaying term out of $\det\big[ \e{id} \, + \,  \op{U}_{\th}\big(\Ups;\La^{(1)}_{\om} \big)  \big]$ is roughly similar. By doing contour deformations so as to 
explicitly evaluate the residue at $\om$, one recasts the determinant one starts with as
\bem
\det\big[ \e{id} \, + \,  \wh{\op{U}}_{\th}\big(\Ups;\La^{(1)}_{\om} \big)  \big] \;=\;  \\
\det\left[ \ba{ccc} \e{id} + \wh{\op{U}}_{\th}^{(11)}\big(\Ups;\La^{(1)}_{\om}\big)(z,y) 
		    & 2\i \pi \e{Res}_{s=\om}\Big(\wh{\op{U}}_{\th}^{(11)}\big(\Ups;\La^{(1)}_{\om}\big)(z,s) \Big) &  \wh{\op{U}}_{\th}^{(12)}\big(\Ups;\La^{(1)}_{\om}\big)(z,y^{\prime} )  \\ 
 \wh{\op{U}}_{\th}^{(11)}\big(\Ups;\La^{(1)}_{\om}\big)(\om,y) 
		    & 1+2\i \pi \e{Res}_{s=\om}\Big(\wh{\op{U}}_{\th}^{(11)}\big(\Ups;\La^{(1)}_{\om}\big)(\om,s) \Big) &  \wh{\op{U}}_{\th}^{(12)}\big(\Ups;\La^{(1)}_{\om}\big)(\om,y^{\prime} ) \\
 \wh{\op{U}}_{\th}^{(21)}\big(\Ups;\La^{(1)}_{\om}\big)(z^{\prime},y) 
		    & 2\i \pi \e{Res}_{s=\om}\Big(\wh{\op{U}}_{\th}^{(21)}\big(\Ups;\La^{(1)}_{\om}\big)(z^{\prime},s) \Big) & \e{id}+  \wh{\op{U}}_{\th}^{(22)}\big(\Ups;\La^{(1)}_{\om}\big)(z^{\prime},y^{\prime} ) \ea  \right] \;. 
\end{multline}
The entries arising in the block determinant correspond to integral kernels where the unprimed variables belong to $\Ga\big( \{ \la^{(1)}_a \}_1^{N_{\La}^{(1)}} \big)$ and the primed ones 
belong to $\Ga\big( \{ \mu^{(2)}_a \}_1^{N_{\Ups}^{(1)}} \big)$. After some algebra, one finds
\beqa
 \wh{\op{U}}_{\th}^{(1a)}\big(\Ups;\La^{(1)}_{\om}\big)(\om,y) & = &  - \f{ \i c }{ 2\om } \wh{G}_{\th;R}^{(a)}\big(\Ups;\La\big)(y) \cdot \Bigg\{ 1 \, - \,  \f{ k\big( \th, \ov{\mu}^{(1)}  \big) }{ k\big(\th ,\ov{\la}^{(1)} \big) } \Bigg\} 
 \cdot \Big( 1+\e{O}\big( \om^{-1} \big) \Big)\\
  2\i \pi \e{Res}_{s=\om}\Big(\wh{\op{U}}_{\th}^{(a1)}\big(\Ups;\La^{(1)}_{\om}\big)(z,s) \Big)  & = &   \wh{G}_{\th;L}^{(a)}\big(\Ups;\La\big)(z)  \cdot \Big( 1+\e{O}\big( \om^{-1} \big) \Big) \;. 
\eeqa
Finally, one has
\beq
1+2\i \pi \e{Res}_{s=\om}\Big(\wh{\op{U}}_{\th}^{(11)}\big(\Ups;\La^{(1)}_{\om}\big)(\om,s) \Big) \; = \;  - \f{ \i c }{ 2\om } \Bigg\{ 1 \, -  \,  \f{ k\big( \th, \ov{\mu}^{(1)}  \big) }{ k\big(\th ,\ov{\la}^{(1)} \big) } \Bigg\} 
 \cdot \Big( 1+\e{O}\big( \om^{-1} \big) \Big) \;. 
\enq
Once these expansions are established, it is enough to factor out the diagonal term associated with the $\om$-column and then "kill" the off-diagonal entries of the $\om$-column by linear combinations. 
Since, for bounded $z,y$ one has
\beq
 \wh{\op{U}}_{\th}^{(ab)}\big(\Ups;\La^{(1)}_{\om}\big)(z,y)  \; = \;  \wh{\op{U}}_{\th}^{(ab)}\big(\Ups;\La\big)(z,y) \cdot \Big( 1+\e{O}\big( \om^{-1} \big) \Big)
\enq
one gets that 
\beq
\det\big[ \e{id} \, + \,  \wh{\op{U}}_{\th}\big(\Ups;\La^{(1)}_{\om} \big)  \big] \;=\;  - \f{ \i c }{ 2\om } \Bigg\{ 1 \, - \,  \f{ k\big( \th, \ov{\mu}^{(1)}  \big) }{ k\big(\th ,\ov{\la}^{(1)} \big) } \Bigg\}  \cdot 
\det\big[ \e{id} \, + \,  \wh{\op{V}}_{\th}\big(\Ups;\La \big)  \big]  
 \cdot \Big( 1+\e{O}\big( \om^{-1} \big) \Big) \;. 
\enq
The above handlings thus leads to 
\beq
\wh{\mc{R}}^{(22)}\big(\Ups;\La^{(1)}_{\om} \big)\; = \; \f{ c^2 \pi  L  \big( \wh{\xi}^{\,(a)}_{\La^{(1)}_{\om}}\big)^{\prime}(\om) }{ L - 2N_{\La}^{ (1) } + N^{ (2) }_{ \La } }
\cdot \wh{\mc{R}}^{(21)}\big(\Ups;\La \big) \cdot \Big( 1\, + \, \e{O}\big( \om^{-1} \big) \Big) \;. 
\enq
Then, it remains to put the various large-$\om$ asymptotics together. 

\subsubsection*{$\bullet$ The amplitudes for $\op{E}^{23}_1$}

Likewise, the starting point to establish \eqref{ecriture decomposition lisse discrete E21 et E23}
and \eqref{ecriture amplitude pour FF type 23} is the identity \eqref{formule reconstruction E23 via E22}. 
The limits of $\wh{D}$ and $\wh{W}$ are treated similarly to the above, leading to 
\beq
\wh{\mc{D}}\big( \Ups^{(2)}_{\om} , \La \big) \; = \;  \f{  \wh{\mc{D}}\big( \Ups , \La \big)   }{ \pi L \om^{2} \big( \wh{\xi}^{\,(2)}_{\Ups^{(2)}_{\om}}\big)^{\prime}(\om)   } 
\cdot \Big(1+\e{O}\big( \om^{-1} \big) \Big) \qquad \e{and} \quad 
\wh{\mc{W}}\big( \Ups^{(2)}_{\om} , \La \big) \; = \;  \f{ \om^{2} }{ c^2   } \cdot  \wh{\mc{W}}\big( \Ups , \La \big) 
\cdot \Big(1+\e{O}\big( \om^{-2} \big) \Big)\;. 
\enq
In order to compute the large-$\om$ behaviour of $\wh{\mc{R}}^{(23)}\big(\Ups^{(2)}_{\om} ;  \La \big) $ it is convenient to set the arbitrary parameter to $\th=\om$. Then, the
Fredholm determinant occurring in the numerator can be recast as
\bem
\det\big[ \e{id} \, + \,  \wh{\op{U}}_{\om}\big(\Ups;\La^{(1)}_{\om} \big)  \big] \;=\; 
\det\left[ \ba{ccc} \e{id} + \wh{\op{U}}_{\om}^{(11)}\big(\Ups;\La^{(1)}_{\om}\big)(z,y) 
		    &\wh{\op{U}}_{\om}^{(12)}\big(\Ups;\La^{(1)}_{\om}\big)(z,y^{\prime} ) & 0   \\ 
 \wh{\op{U}}_{\om}^{(21)}\big(\Ups;\La^{(1)}_{\om}\big)(z^{\prime},y) 
		     & \e{id}+  \wh{\op{U}}_{\om}^{(22)}\big(\Ups;\La^{(1)}_{\om}\big)(z^{\prime},y^{\prime} ) & 0  \\
%
%*
*   & *   & 1  \ea  \right] \\ 
 \; = \; \det\big[ \e{id} \, + \,  \wh{\op{U}}_{\infty}\big(\Ups;\La \big)  \big] \cdot \Big( 1+\e{O}\big( \om^{-1} \big) \Big) \;, 
\end{multline}
where $*$ are some coefficients.   

Taking the $\om\tend+\infty$ out of the norm determinant $\det\big[ I_{|\Ups_{\om}^{(2)}|} \, + \,  \mc{K}_{\Ups_{\om}^{(2)}}  \big]$
is done as previously and one gets 
\beq
\det\big[ I_{|\Ups_{\om}^{(2)}|} \, + \,  \mc{K}_{\Ups_{\om}^{(2)}}  \big] \; = \; 
\f{ c \cdot \big( N^{ (1) }_{ \Ups } - 2N_{\Ups}^{ (2) }  \big) }{ 2\pi L \om^2  \cdot \big( \wh{\xi}^{\,(2)}_{\Ups^{(2)}_{\om}}\big)^{\prime}(\om)  }  
			    \cdot  \det\big[ I_{|\Ups |}   \, + \,    \mc{K}_{\Ups }  \big] \cdot \Big( 1 \, + \, \e{O}\big( \om^{-1} \big) \Big) \; .
\enq
Then, upon extracting the large-$\om$ behaviour of all the pre-factors in $\wh{\mc{R}}^{(22)}\big(\Ups^{(2)}_{\om};\La \big)$ one obtains
\beq
\wh{\mc{R}}^{(22)}\big(\Ups^{(2)}_{\om};\La \big) \; = \; \f{ c^2 \pi  L  \big( \wh{\xi}^{\,(a)}_{\La^{(1)}_{\om}}\big)^{\prime}(\om) }{ N^{ (1) }_{ \Ups } - 2N_{\Ups}^{ (2) }   }
\cdot \wh{\mc{R}}^{(23)}\big(\Ups;\La \big) \cdot \Big( 1\, + \, \e{O}\big( \om^{-1} \big) \Big) \;, 
\enq
so that it solely remains to put all the pieces together. \qed

\begin{cor}

The factorisations given in Proposition \ref{Proposition factorisation lisse discrete pour quelques facteurs de forme} is precisely of the form stated in the main conjecture. 
\end{cor}

\Proof

The statement relatively to the smooth parts  $\wh{\mc{A}}^{(ab)}$ is readily obtained by repeating the handlings of \cite{KozKitMailSlaTerThermoLimPartHoleFormFactorsForXXZ}. 
We refer the reader to that paper for more details. \qed

\section*{Acknowledgements}

K.K.K., E.R.  are supported by CNRS. This work has been partly done within the financing of the grant  ANR grant "DIADEMS" SIMI 1 2010-BLAN-0120-02.
K.K.K. acknowledged support from the  Burgundy region PARI 2013-2014 FABER grant "Structures et asymptotiques d'int\'{e}grales multiples" when part of this work was realized. 
K.K.K. would like to thank the LAPTh for its warm hospitality which allowed this work to take place. 
We warmly thank L. Frappat for his contribution at the early stage of this work.

\appendix

\section{Thermodynamic limit of the model}
\label{Appendix Thermo limit of the model}

\subsection{Non-linear integral equations and large-$L$ expansion}
\label{Appendix Sous section large L expansion EINL}

In this sub-section we establish that the counting function \eqref{definition counting function} associated with a particle-hole excited state admits the large-$L$ asymptotic behaviour 
\beq
\wh{ \xi }_{\La}^{\,(k)}(\om) \; = \; \wh{p}^{(k)}(\om) \, + \; \f{ 1 }{2 L } \big( N_{\Om}^{(k)} + 1  \big) 
\; + \; \f{1}{L } \wh{\Psi}^{(k)}\Big( \om \mid  \big\{ \hat{\la}_{ \wt{p}_a^{(s)}}^{(s)}  \big\}  \;; \; \big\{ \hat{\la}_{ \wt{h}_a^{(s)}}^{(s)}  \big\} \Big)
\; + \; \f{ 1 }{2 L } \Big[\wh{\bs{Z}}(\om)\cdot  (\bs{\kappa} -\bs{\mf{n}}) \Big]^{(k)} \; + \; \e{O}\Big( \f{ 1 }{ L^2 } \Big) \;. 
\label{ecriture DA size fini fct comptage}
\enq
Above, the integers $\wt{h}_a^{(s)}$ and  $\wt{p}_a^{(s)}$ are defined as
\beq
\Big\{ \wt{p}_a^{(s)} \Big\}_1^{\wt{n}_p^{(s)}}  = 
      \left\{ \ba{cc} \big\{ p_a^{(s)}  \big\}_1^{n^{(s)}}  \cup \big\{  N_{\Om}^{(s)}+a \big\}_1^{\kappa^{(s)}}  &  \kappa^{(s)} \geq 0 \vspace{2mm} \\ 
			 \big\{ p_a^{(s)}  \big\}_1^{n^{(s)}} 						& \kappa^{(s)}< 0	\ea     \right. 
\quad   , \quad 
\Big\{ \wt{h}_a^{(s)} \Big\}_1^{\wt{n}_h^{(s)}} \; = \; 
      \left\{ \ba{cc} \big\{ h_a^{(s)}  \big\}_1^{n^{(s)}}  &  \kappa^{(s)} \geq 0 \vspace{2mm} \\ 
			 \big\{ h_a^{(s)}  \big\}_1^{n^{(s)}} 	 \cup \big\{ N_{\Om}^{(s)}+a \big\}_1^{ |\kappa^{(s)}| }					& \kappa^{(s)}< 0	 \ea    \right. \;. 
\nonumber
\enq
The asymptotic expansion \eqref{ecriture DA size fini fct comptage} involves the auxiliary function
\beq
\wh{\Psi}^{(k)}\Big( \om \mid  \big\{ \mu_{ a }^{(s)}  \big\}  \;; \; \big\{ \nu_{ a }^{(s)}  \big\} \Big) \; = \; 
\sul{s=1}{r} \bigg\{ \sul{a=1}{n^{(s)}_{\mu} }  \wh{\Phi}_{ks}\Big( \om , \mu_{ a }^{(s)} \Big) \, - \,  \sul{a=1}{n^{(s)}_{\nu} } \wh{\Phi}_{ks}\Big( \om , \nu_{ a }^{(s)}  \Big) \bigg\}
\enq
which is expressed in terms of the finite-size dressed phase matrix $ \wh{\Phi}_{ks}=\wh{\bs{\Phi}}_s^{(k)}$. 
The finite-size counterparts $\wh{\bs{\Phi}}_s^{(k)}$, $\wh{\bs{p}}$ and $\wh{\bs{Z}}$ of the functions 
introduced in \eqref{definition vecteur moment habille},\eqref{definition vector dressed phase} and \eqref{definition matrix dressed charge} are defined as the solutions to the linear integral equations 
\beqa
\Big( \e{id} \, + \, \wh{\op{K}} \Big)[ \wh{\bs{p}} ](\om) &= & \bs{p}_0(\om)  \; + \; \sul{s=1}{r} \f{N_{\Om}^{(s)} }{2 L } \Big( \Xi_s\big(\om,\wh{q}^{\,(s)} \big)\, + \, \Xi_s\big(\om,-\wh{q}^{\,(s)} \big)\Big)  \vspace{2mm}  \\
\Big( \e{id} \, + \, \wh{\op{K}} \Big)[ \wh{\bs{\Phi}}_s(*,z)](\om) = \bs{\Xi}_s(\om,z)    &\e{and} & \Big( \e{id} \, + \, \wh{\op{K}} \Big)[ \wh{\bs{Z}} ](\om)  =  I_r 
\eeqa
where the finite-size kernel acts as 
\beq
\Big[ \Big( \e{id} \, + \, \wh{\op{K}} \Big)[\bs{f}](\om) \Big]^{(k)} \; = \; \bs{f}^{(k)}(\om) \; + \; \sul{\ell=1}{r} \Int{ -\wh{q}^{\,(\ell)} }{ \wh{q}^{\,(\ell)} } \Dp{\mu} \th_{k\ell}(\om,\mu) \cdot \bs{f}^{(\ell)}(\mu) \cdot \dd \mu \;. 
\enq
Finally, $\wh{q}^{\, (\ell)}$ is defined as the positive solution to 
\beq
\wh{p}^{(k)}\big( \wh{q}^{\,(k)} \big) = \f{N^{(k)}_{\Om} }{2L} \, .
\label{definition endpoint F zone type fini}
\enq
We assume that this solutions does exist and is furthermore unique. 
The condition $D^{(s)}-\tf{ N_{\Om}^{(s)} }{L}=\e{O}\big(L^{-2}\big)$ ensures that $q^{(s)}-\wh{q}^{\,(s)}=\e{O}\big( L^{-2} \big)$ and thus $\wh{p}^{\,(s)}-p^{(s)}=\e{O}\big( L^{-2} \big)$. 
This being settled, the very definition of the roots $\hat{\la}_{a}^{(s)}$ then ensures that
\beq
\hat{\la}_{a}^{(s)}-\mu_a^{(s)} \; = \; \e{O}\Big( L^{-1} \Big) \qquad \e{and} \qquad \mu_{N_{\Om}^{(s)}+p}^{(s)}-\wh{q}^{\,(s)} \; = \;  \e{O}\Big( \f{p}{L} \Big)\;. 
\label{ecriture deviation racines exactes et racines thermo}
\enq
Given these pieces of information, one readily deduces the form \eqref{ecriture DA fct comptage} of the asymptotic expansion presented in the body of the text.

Under the hypothesis that the counting functions are strictly increasing on $\R$, following \cite{DestriDeVegaAsymptoticAnalysisCountingFunctionAndFiniteSizeCorrectionsinTBAFiniteMagField} one obtains 
the non-linear integral equation
\bem
\wh{ \xi }_{\La;\e{sym}}^{\,(k)}(\om) \; = \; p^{(k)}_0(\om) \, + \; \f{ 1 }{2 L } \big( \kappa^{(k)} - n_{\La}^{(k)}  \big) 
\; + \; \f{1}{L } \Theta^{(k)}\Big( \om \mid  \big\{ \hat{\la}_{ \wt{p}_a^{(s)}}^{(s)}  \big\}  \;; \; \big\{ \hat{\la}_{ \wt{h}_a^{(s)}}^{(s)}  \big\} \Big) 
\; - \; \sul{s=1}{r} \Int{ \wh{q}_{-}^{\,(s)} }{ \wh{q}_{+}^{\,(s)} } \Dp{\tau} \th_{ ks}\big(\om,\tau \big)\wh{ \xi }_{\La;\e{sym} }^{\,(s)}(\tau)  \cdot \dd \tau \\
\, + \, \sul{s=1}{r} \f{N_{\Om}^{(s)} }{2 L } \Big( \th_{ks}\big(\om,\wh{q}^{\,(s)}_{+} \big)\, + \, \th_{ks}\big(\om,\wh{q}^{\,(s)}_{-} \big)\Big) 
\; - \; \sul{s=1}{r} \sul{\eps_{s}=\pm}{} \hspace{-3mm}  \Int{  \big( \wh{\xi}_{\La}^{\,(s)}\big)^{-1} \big(\Ga^{(s)}_{\eps_{s}} \big) }{} \hspace{-3mm} \Dp{\tau}\th_{ks}(\la, \tau)  
\ln\bigg[ 1 - \ex{ 2\i\pi \eps_{s} L \wh{\xi}_{\La}^{\,(s)}(\tau) }  \bigg] \cdot \f{ \dd \tau }{2\i\pi L}
\end{multline}
where the endpoints of integration are defined as
\beq
\wh{ \xi }_{\La;\e{sym}}^{\,(k)}\big( \wh{q}^{\,(s)}_{\pm} \big) \, = \, \pm \f{ N_{\Om}^{(s)} }{ 2 L } \qquad \e{with} \qquad 
\wh{ \xi }_{\La;\e{sym}}^{\,(k)}(\om) \, = \, \wh{ \xi }_{\La }^{\,(k)}(\om) \, - \,  \f{1}{2L} \big( N^{(k)}_{\Om} + 1 \big) \;. 
\label{definition q hat pm}
\enq
Note that we have also introduced 
\beq
\Theta^{(k)}\Big( \om \mid  \big\{ \mu_{ a }^{(s)}  \big\}  \;; \; \big\{ \nu_{ a }^{(s)}  \big\} \Big) \; = \; 
\sul{s=1}{r} \bigg\{ \sul{a=1}{n^{(s)}_{\mu} }  \th_{ks}\Big( \om , \mu_{ a }^{(s)} \Big) \, - \,  \sul{a=1}{n^{(s)}_{\nu} } \th_{ks}\Big( \om , \nu_{ a }^{(s)}  \Big) \bigg\} \;. 
\enq
Finally, the integration in the non-linear term runs through the contours depicted in Fig.~\ref{Figure contours pour l'integration EINL}.
\begin{figure}[ht]
\begin{center}

\begin{pspicture}(7,7)

\psline[linestyle=dashed, dash=3pt 2pt]{->}(1,4)(6.7,4)
\psdots(1.5,4)(6,4) 
\rput(1.1,3.5){$\f{1}{2L}$}
\rput(6.9,3.4){$\f{N_{\Om}^{(k)}+\tf{1}{2}}{L}$}

\rput(0.7,5.9){$\f{1}{2L}+\i\a$}
\rput(7.1,6){$\f{N_{\Om}^{(k)}+\tf{1}{2}}{L}+\i\a$}

\rput(0.5,1.8){$\f{1}{2L}-\i\a$}
\rput(6.8,1.6){$\f{N_{\Om}^{(k)}+\tf{1}{2}}{L}-\i\a$}

\psline{-}(1.5,2.5)(1.5,5.5)
\pscurve{-}(1.5,5.5)(1.65,5.9)(2,6)

\psline{-}(2,6)(5.5,6)
\pscurve{-}(5.5,6)(5.85,5.9)(6,5.5)

\psline{-}(6,5.5)(6,2.5)
\pscurve{-}(6,2.5)(5.85,2.1)(5.5,2)

\psline{-}(5.5,2)(2,2)
\pscurve{-}(2,2)(1.65,2.1)(1.5,2.5)

%region p H_i

%\rput(14.5,4.5){$p\pa{H_{III}}$}

%region H_i

\rput(3.5,6.5){ $ \Ga_{+}^{(k)} $ }
\rput(4,1.5){ $ \Ga_{-}^{(k)} $ }

\psline[linewidth=2pt]{->}(5,2)(5.1,2)
\psline[linewidth=2pt]{->}(4,6)(3.9,6)

\end{pspicture}
\caption{ The contours  $\Ga_{+}^{(k)} \cup \Ga_{-}^{(k)}$. \label{Figure contours pour l'integration EINL} }
\end{center}
\end{figure}

\noindent The above can be recast as 
\bem
\Big[ \Big( \e{id} \, + \, \wh{\op{K}} \Big)\big[\wh{ \bs{\xi} }_{\La;\e{sym}}\big](\om) \Big]^{(k)}   \; = \;
p^{(k)}_0(\om) \, + \, \sul{s=1}{r} \f{N_{\Om}^{(s)} }{2 L } \Big( \th_{ks}\big(\om,\wh{q}^{\,(s)} \big)\, + \, \th_{ks}\big(\om,- \wh{q}^{\,(s)} \big)\Big) \\
\, + \; \f{ 1 }{2 L } \big( \kappa^{(k)} - n_{\La}^{(k)}  \big) 
\; + \; \f{1}{L } \Theta^{(k)}\Big( \om \mid  \big\{ \hat{\la}_{ \wt{p}_a^{(s)}}^{(s)}  \big\}  \;; \; \big\{ \hat{\la}_{ \wt{h}_a^{(s)}}^{(s)}  \big\} \Big) 
\, + \, \mc{R}^{(k)}\big[ \bs{\xi}_{\La}\big] \;, 
\end{multline}
where 
\beq
\mc{R}^{(k)}\big[ \bs{\xi}_{\La}\big] \, = \,  - \; \sul{s=1}{r} \sul{\eps_s=\pm}{} \eps_{s} \hspace{-2mm}    \Int{ \eps_{s} \wh{q}^{\,(s)} }{ \wh{q}_{\eps_{s}}^{\,(s)} }
   \hspace{-2mm}   \Dp{\tau} \th_{ ks}\big(\om,\tau \big) \cdot \Big[ \wh{ \xi }_{\La }^{\,(s)}(\tau)  \,- \,  \wh{ \xi }_{\La }^{\,(s)}( \wh{q}_{\eps_{s}}^{\,(s)}  )  \Big]\cdot \dd \tau 
\, - \, \sul{s=1}{r} \sul{\eps_s=\pm}{}\hspace{-3mm}  \Int{ \big(\wh{\xi}_{\La}^{\,(s)}\big)^{-1}\big(\Ga^{(s)}_{\eps_{s}} \big) }{} \hspace{-3mm} \Dp{\tau}\th_{ks}(\la, \tau)  
\ln\bigg[ 1 - \ex{ 2\i\pi \eps_{s} L \wh{\xi}_{\La}^{\,(s)}(\tau) }  \bigg] \cdot \f{ \dd \tau }{2\i\pi L} \;. 
\nonumber
\enq
It then remains to invert the operator. The remainder is easily seen to be a $\e{O}\big( L^{-2} \big)$ by means of an application of Watson's lemma what allows
one to recover the expansion \eqref{ecriture DA size fini fct comptage}. 

Finally, going back to the definition \eqref{definition q hat pm} of $\wh{q}^{\,(s)}_{ \pm}$  and \eqref{definition endpoint F zone type fini} of $\wh{q}^{\, (s)}$,  one can deduce from the asymptotic expansion \eqref{ecriture DA size fini fct comptage} that 
\beq
\wh{q}^{\,(k)}_{\pm} \, \pm \, \wh{q}^{\,(k)} \; = \; \f{-1 }{ L \cdot \Big( \wh{p}^{\,(k)}\Big)^{\prime}\big(\pm \wh{q}^{\,(k)}\big)  } 
\cdot \bigg\{ \wh{\Psi}\Big(\pm \wh{q}^{\,(k)}\mid  \big\{ \hat{\la}_{ \wt{p}_a^{(s)}}^{(s)}  \big\}  \;; \; \big\{ \hat{\la}_{ \wt{h}_a^{(s)}}^{(s)}  \big\} \Big)
\; + \; \f{ 1 }{2 }  \wh{\bs{Z}}\big(\pm \wh{q}^{\,(k)}\big)\cdot  (\bs{\kappa} -\bs{\mf{n}})   \bigg\}^{(k)} \; + \; \e{O}\Big( \f{1}{L^2} \Big) \;. 
\enq

\subsection{Several identities satisfied by the dressed phase}
\label{Appendice derivation identite quadratique phase habillee}

In this subsection we derive various identities that are satisfied by the dressed phase. These will appear useful in later computations. 

\begin{lemme}
The dressed charge and phase matrices satisfy the differential identities
\beq
\Dp{\mu}\Phi_{ks}(\la,\mu)  \, = \, R_{ks}(\la,\mu) \qquad \Dp{\la}\Phi_{ks}(\la,\mu)  \, = \, -R_{ks}(\la,\mu)  
\, + \, \sul{\ell=1}{r}\sul{\eps_{\ell}=\pm }{} \eps_{\ell} R_{k\ell}\big( \la,\eps_{\ell} q^{(\ell)} \big) \cdot   \Phi_{\ell s}\big( \eps_{\ell} q^{(\ell)} ,\mu \big) 
\label{ecriture identites diff pour matrice phase habillee}
\enq
and
\beq
\Dp{\la}Z_{ks}(\la) \, = \, R_{ks}\big(\la,q^{(s)} \big) \, - \, R_{ks}\big(\la,-q^{(s)} \big) 
\, - \, \sul{\ell=1}{r}\sul{\eps_{s}, \eps_{\ell}=\pm }{} \hspace{-2mm} \eps_{s} \eps_{\ell} \, R_{k\ell}\big( \la,\eps_{\ell} q^{(\ell)} \big)  \Phi_{\ell s}\big( \eps_{\ell} q^{(\ell)} ,\eps_{s} q^{(s)}\big) .
\label{ecriture derivee dressed charge matrix}
\enq
Moreover, the dressed charge and the resolvent enjoy reflection properties
\beq
\Phi_{ks}(\la,\mu)\, + \, \Phi_{sk}(\mu,\la)\, = \, \sul{\ell=1}{r}\sul{ \eps_{\ell}=\pm }{}   \eps_{\ell} \Phi_{\ell k}\big( \eps_{\ell} q^{(\ell)},\la \big)  \Phi_{\ell s}\big( \eps_{\ell} q^{(\ell)} , \mu \big) 
\label{ecriture reflection pour matrice de phase habillee}
\enq
and
\beq
R_{k\ell}(\la,\mu) \, =\,  R_{\ell k }(\mu,\la) \quad , \quad \Phi_{ks}(\la,\mu)\, = \, -\Phi_{ks}(-\la,-\mu)\,. 
\label{ecriture identites de reflection pour resolvent et phase habillee}
\enq
\end{lemme}

\Proof 

The two differential identities satisfied by the dressed phase matrix are obtained by differentiating the linear integral equations fulfilled by the matrix entries
and integrating by parts, using the fact that the bare phases are functions of difference of arguments only. The derivative of the dressed charge matrix  \eqref{ecriture derivee dressed charge matrix}
can be computed by means of \eqref{ecriture identites diff pour matrice phase habillee} and \eqref{expression explicite dressed charge matrix}. 
The identity \eqref{ecriture identites de reflection pour resolvent et phase habillee} relative to the resolvent is obtained by a direct inspection of the 
Neumann series representation for the resolvent and by using that $\Dp{\mu}\th_{k\ell}(\la,\mu)=\Dp{\la}\th_{\ell k}(\mu,\la)$. As for the 
second identity, one invokes the integral representation 
\beq
 \Phi_{k\ell}\big( \la, \mu \big) \; = \;  \th_{k\ell}\big( \la, \mu \big) \, - \, \sul{s=1}{r}\Int{ -q^{(s)} }{ q^{(s)} }  R_{ks}(\la,\nu) \th_{s\ell}\big( \nu, \mu \big) \cdot  \dd \nu 
\label{ecriture rep int pour phase habillee}
\enq
and uses the symmetry properties of the bare phase. 
Finally, the identity \eqref{ecriture reflection pour matrice de phase habillee} is obtained by introducing the function 
\beq
f_{k s}(\la,\mu) \, = \, \Phi_{ks}(\la,\mu)\, + \, \Phi_{sk}(\mu,\la)
\enq
and computing its partial $\la$ and $\mu$ derivatives by means of \eqref{ecriture identites diff pour matrice phase habillee}. This shows that 
\beq
f_{k s}(\la,\mu) \, = \, \sul{\ell=1}{r}\sul{ \eps_{\ell}=\pm }{}   \eps_{\ell} \Phi_{\ell k}\big( \eps_{\ell} q^{(\ell)},\la \big)  \Phi_{\ell s}\big( \eps_{\ell} q^{(\ell)} , \mu \big)  \, + \, c_{ks} \;. 
\enq
However, owing to \eqref{ecriture identites de reflection pour resolvent et phase habillee}, it holds $f_{k s}(\la,\mu) \, = \, - f_{k s}(-\la,-\mu)$ what implies that $c_{ks}=0$. \qed

\vspace{2mm}

In the case of rank one models, Slavnov \cite{SlavnovNonlinearIdentityScattPhase} identified the mechanism which gives rise to a quadratic identity satisfied by the dressed
phase. The latter allows one to find a closed expression for the inverse of the dressed charge in terms of the dressed phase. 
We establish a generalisation of this identity below, hence giving rise to the 
\begin{prop}
The inverse of the matrix 
\beq
\mc{Z}_{ k \ell } \, \equiv  \, Z_{k\ell}\big( q^{(k)} \big) \,  = \, \de_{k \ell} + \Phi_{k\ell}\big(q^{(k)}, -q^{(\ell)} \big) \, - \, \Phi_{k\ell}\big(q^{(k)}, q^{(\ell)} \big) 
\enq
takes the form 
\beq
\Big[\mc{Z}^{-1}\Big]_{ k \ell } \; = \;\de_{k \ell} - \Phi_{k\ell}\big(q^{(k)}, -q^{(\ell)} \big) \, - \, \Phi_{k\ell}\big(q^{(k)}, q^{(\ell)} \big) \; .
\label{ecriture formule pour matrice Z moins 1}
\enq
\end{prop}

\Proof

Starting from the integral representation \eqref{ecriture rep int pour phase habillee} for dressed phase matrix   and 
using the symmetry properties of the bare phase $\th_{kp}(\la,\mu) \, = \, \th_{pk}(-\mu,-\la) \, = \, -\th_{pk}(\mu,\la)$, one gets that 
\bem
\de=  \Phi_{k\ell}\big( \la, -\mu \big) \, - \,  \Phi_{\ell k}\big( \mu, -\la \big) \,  + \,  \Phi_{k\ell}\big( \la, \mu \big)    \,  + \,  \Phi_{\ell k}\big( \mu, \la \big)   \\
\; = \;  \sul{s=1}{r}\Int{-q^{(s)} }{ q^{(s)} } \bigg\{ -R_{ks}(\la,\nu ) \th_{sl}(\nu,-\mu)   \, + \,  R_{\ell s}(\mu,\nu ) \th_{sk}(\nu,-\la)   \, -\,  R_{k s}(\la,\nu ) \th_{s \ell}(\nu, \mu) 
 \, -\, R_{\ell s}(\mu,\nu ) \th_{sk}(\nu,\la)  \bigg\} \\
 \; = \; \sul{s=1}{r}\Int{-q^{(s)} }{ q^{(s)} } \bigg\{ -R_{ks}(\la,\nu ) \th_{\ell s}(\mu , -\nu)   \, + \,  R_{\ell s}(\mu, - \nu ) \th_{ks}(\la,\nu)   
\, + \,  R_{k s}(\la,\nu ) \th_{ \ell s }(\mu, \nu)  \, +\, R_{\ell s}(\mu,\nu ) \th_{ks}(\la, \nu)  \bigg\} \, .
\end{multline}
Re-expressing bare phase in terms of the dressed phase and of the resolvent by means of \eqref{ecriture rep int pour phase habillee} one observes that the quadratic terms in the resolvent kernel
cancel out so that one gets 
\bem
\de \; = \; \sul{s=1}{r}\Int{-q^{(s)} }{ q^{(s)} } \bigg\{ -R_{ks}(\la,\nu ) \Phi_{\ell s}(\mu , -\nu)   \, + \,  R_{\ell s}(\mu, - \nu )  \Phi_{ks}(\la,\nu)   
\, + \,  R_{k s}(\la,\nu )  \Phi_{ \ell s }(\mu, \nu)  \, +\, R_{\ell s}(\mu,\nu )  \Phi_{ks}(\la, \nu)  \bigg\} \\
\; = \; \sul{s=1}{r}\Int{-q^{(s)} }{ q^{(s)} } \Dp{\nu}\bigg\{ \Phi_{ks}(\la,\nu ) \Phi_{\ell s}(\mu , \nu)  - \Phi_{ks}(\la,\nu ) \Phi_{\ell s}(\mu , -\nu) \bigg\} \cdot \dd \nu  \\ 
\; = \; \sul{s=1}{r} \sul{\eps_{s}=\pm }{} \eps_s  \bigg\{ \Phi_{ks}\big( \la, \eps_s  q^{(s)} \big) \Phi_{\ell s}\big(\mu , \eps_s  q^{(s)} \big)  
%
%\, - \, \Phi_{ks}\big( \la,-q^{(s)} \big) \Phi_{\ell s}\big(\mu , -q^{(s)} \big) 
%
\, + \, \Phi_{ks}\big( \la, -\eps_s  q^{(s)} \big) \Phi_{\ell s}\big(\mu ,  \eps_s q^{(s)} \big)  \bigg\} 
%\, - \, \Phi_{ks}\big( \la,q^{(s)} \big) \Phi_{\ell s}\big(\mu , -q^{(s)} \big) \;. 
%
%
\end{multline}
where we used the first equation given in \eqref{ecriture identites diff pour matrice phase habillee}. Specialising to $\la=q^{(k)}$ and $\mu=q^{(\ell)}$ one gets the identity
\bem
\Phi_{k\ell}\big( q^{(k)}, -q^{(\ell)} \big) \, - \,  \Phi_{\ell k}\big( q^{(\ell)}, -q^{(k)} \big) \,  + \,  \Phi_{k\ell}\big( q^{(k)}, q^{(\ell)} \big)    \,  + \,  \Phi_{\ell k}\big( q^{(\ell)}, q^{(k)} \big) 
\; = \; 
\sul{s=1}{r} \bigg\{ \Phi_{ks}\big( q^{(k)}, q^{(s)} \big) \Phi_{\ell s}\big(q^{(\ell)} ,  q^{(s)} \big)   \\ 
\, - \, \Phi_{ks}\big( q^{(k)},-q^{(s)} \big) \Phi_{\ell s}\big(q^{(\ell)} , -q^{(s)} \big) 
\, + \, \Phi_{ks}\big( q^{(k)}, -q^{(s)} \big) \Phi_{\ell s}\big(q^{(\ell)} ,  q^{(s)} \big)  \, - \, \Phi_{ks}\big( q^{(k)},q^{(s)} \big) \Phi_{\ell s}\big(q^{(\ell)} , -q^{(s)} \big) \bigg\}
\end{multline}
which is all that one needs to check that the matrix $\mc{M}_{ k \ell} \; = \;\de_{k \ell} - \Phi_{k\ell}\big(q^{(k)}, -q^{(\ell)} \big) \, - \, \Phi_{k\ell}\big(q^{(k)}, q^{(\ell)} \big) $
satisfies $\mc{M} \cdot \mc{Z}= I_{r}$. \qed

\subsection{The large-$L$ expansion of energies and momenta}
\label{Appendix Sous section cptmt grd L de moment et energie}

Given a functions $f^{(k)}$ holomorphic in a neighbourhood of $\intff{-q^{(k)}}{q^{(k)}}$ consider the sums
\beq
S_{\La}\big[ \bs{f}\big] \, = \, \sul{s=1}{r} \sul{a=1}{N_{\La}^{(k)}} f^{(s)}\big( \la_{a}^{(s)} \big) \;. 
\enq
 By repeating the steps outlined in the derivation of the non-linear integral equation satisfied by the counting function one recasts $S_{\La}\big[ \bs{f}\big]$ in the form
\beq
S_{\La}\big[ \bs{f}\big] \; = \; L \sul{s=1}{r} \Int{-\wh{q}^{\,(s)} }{ \wh{q}^{\,(s)} } f^{(s)}(\tau) \Big( \wh{p}^{\,(s)}(\tau) \Big)^{\prime} \dd \tau
\; + \; \mc{S}^{(0)}\big[ \bs{f} \big] \; + \; \de S_{\La}\big[ \bs{f} \big]
\enq
where $\de S_{\La}\big[ \bs{f} \big] = \e{O}\big( L^{-1} \big)$ and
\bem
\mc{S}^{(0)}\big[ \bs{f} \big] \; = \; \sul{s=1}{r} \Int{-\wh{q}^{\,(s)}}{ \wh{q}^{\,(s)} } f^{(s)}(\tau)
\bigg\{   \Dp{\tau}\wh{\Psi}^{(s)}\Big( \tau \mid  \big\{ \hat{\la}_{ \wt{p}_a^{(\ell)}}^{(\ell)}  \big\}  \;; \; \big\{ \hat{\la}_{ \wt{h}_a^{(\ell)}}^{(\ell)}  \big\} \Big)
\; + \; \f{ 1 }{2  } \Big[\wh{\bs{Z}}^{\prime}(\tau)\cdot  (\bs{\kappa} -\bs{\mf{n}}) \Big]^{(s)}  \bigg\}  \cdot \dd \tau \\
\, + \, \sul{s=1}{r} \eps_{s}\Int{\eps_s \wh{q}^{\,(s)} }{ \wh{q}^{\,(s)}_{\eps_s}  }  f^{(s)}(\tau) \Dp{\tau} \wh{p}^{(s)}(\tau) \cdot \dd \tau  
\, + \, \sul{s=1}{r}\bigg\{ \sul{a=1}{\wt{n}^{(s)}_{p} } f^{(s)}\big( \hat{\la}_{ \wt{p}_a^{(s)}}^{(s)}  \big) \, - \,  \sul{a=1}{\wt{n}^{(s)}_{h} } f^{(s)}\big( \hat{\la}_{ \wt{h}_a^{(s)}}^{(s)}  \big) \bigg\}.
\end{multline}
 Owing to $ \big| \wh{q}^{\,(s)} - q^{(s)} \big|= \e{O}\big( L^{-2} \big)$, one can replace $\wh{q}^{\,(s)}$  by its $L\tend +\infty$ limit without altering the form 
of the asymptotic expansion up to $\e{O}\big(L^{-1}\big)$. After some algebra, one recasts $\mc{S}^{(0)}\big[ \bs{f} \big]$ in the form
\beq
\mc{S}^{(0)}\big[ \bs{f} \big] \; = \; - \sul{s=1}{r} \sul{\eps_{s}=\pm }{} \eps_{s} F^{(s)}[\bs{f}]\big( \eps_{s} q^{(s)}\big)\,  \big[ \mc{Z}^{\e{t}}\cdot \f{\bs{\kappa}-\bs{\mf{n}}}{2}  \big]^{(s)}
\; + \;  \sul{s=1}{r}\bigg\{ \sul{a=1}{\wt{n}^{(s)}_{p} } \mc{U}^{(s)}[\bs{f}]\big( \hat{\la}_{ \wt{p}_a^{(s)}}^{(s)}  \big)
\, - \,  \sul{a=1}{\wt{n}^{(s)}_{h} } \mc{U}^{(s)}[\bs{f}]\big( \hat{\la}_{ \wt{h}_a^{(s)}}^{(s)}  \big) \bigg\} \; + \; \e{O}\Big( L^{-1} \Big)
\nonumber
\enq
where 
\beq
 F^{(k)}[\bs{f}]\big(\la\big) \; =\; f^{(k)}(\la)   \, - \, \sul{s=1}{r}\Int{ -q^{(s)} }{ q^{(s)} }  R_{ks}(\la,\nu) f^{(s)}(\nu) \cdot  \dd \nu 
\enq
and
\beq
 \mc{U}^{(s)}[\bs{f}]\big(\la \big) \;=\;  F^{(k)}[\bs{f}]\big(\la\big)\, - \, \sul{\ell=1}{r}  \sul{\eps_{\ell}=\pm}{} \eps_{\ell} F^{(\ell)}[\bs{f}]\big(\eps_{\ell} q^{(\ell)} \big) \Phi_{\ell s}\big(\eps_{\ell} q^{(\ell)},\la \big) \;. 
\enq

In order to obtain the first terms in the large-$L$ expansion of the momenta and energies of the eigenstates one should observe that 
\beq
\sul{\eps_{s}=\pm }{} \eps_{s} F^{(s)}[\bs{p}_0]\big( \eps_{s} q^{(s)}\big)\, =\, 2 \sul{\ell=1}{r} p^{(\ell)}\big( \wh{q}^{(\ell)}\big)\big[ \mc{Z}^{-1} \big]_{s\ell}
\enq
while
\beq
 F^{(k)}[\bs{p}_0]\big(\la\big) \; =\; p^{(k)}(\la) \, - \, \sul{\ell=1}{r} \sul{\eps_{\ell} = \pm }{}   p^{(\ell)}\big( \wh{q}^{(\ell)}\big) \Phi_{k\ell}\big( \la, \eps_{\ell} \wh{q}^{\ell} \big)
\enq
so that, owing to \eqref{ecriture reflection pour matrice de phase habillee}, $ \mc{U}^{(s)}[\bs{p}_0]\big(\la \big) \;=\;p^{(s)}(\la)$. 
The situation with the dressed energy is even simpler since, by construction $\veps\big( \pm q^{(s)} \big)=0$ so that 
$\mc{U}^{(s)}[\bs{\veps}_0]\big(\la \big) \;=\;  F^{(s)}[\bs{\veps}_0]\big(\la\big)=\veps^{(k)}(\la)$. 
In order to obtain \eqref{Ecriture dressed momentum at large L}-\eqref{Ecriture dressed energy at large L}, it remains to invoke \eqref{ecriture deviation racines exactes et racines thermo}. 

%%%%%%%%%%%%%%%%%%%%%%%%%%%%%%%%%%%%%%%%%%%%%%%%%%%%%%%%%%%%%%%%%%%%%%%%%%%%%%%%%%%%%%%%%%%%%%%%%%%%%%%%%%%%%%%%%%%%%%%%%%%%%%%%%%%%%%%%%%%%%%%%%%%%%%%%%%%%%%%%%%%%%%%%%%%%%%%%%%%%%%%%%%%%%%%%%%%%%%%%%%%%%%%%%%%%%%%%%%%%%%%%%%%%%%%%%%%%%%%%
%%%%%%%%%%%%%%%%%%%%%%%%%%%%%%%%%%%%%%%%%%%%%%%%%%%%%%%%%%%%%%%%%%%%%%%%%%%%%%%%%%%%%%%%%%%%%%%%%%%%%%%%%%%%%%%%%%%%%%%%%%%%%%%%%%%%%%%%%%%%%%%%%%%%%%%%%%%%%%%%%%%%%%%%%%%%%%%%%%%%%%%%%%%%%%%%%%%%%%%%%%%%%%%%%%%%%%%%%%%%%%%%%%%%%%%%%%%%%%%%

%%%%%%%%%%%%%%%%%%%%%%%%%%%%%%%%%%%%%%%%%%%%%%%%%%%%%%%%%%%%%%%%%%%%%%%%%%%%%%%%%%%%%%%%%%%%%%%%%%%%%%%%%%%%%%%%%%%%%%%%%%%%%%%%%%%%%%%%%%%%%%%%%%%%%%%%%%%%%%%%%%%%%%%%%%%%%%%%%%%%%%%%%%%%%%%%%%%%%%%%%%%%%%%%%%%%%%%%%%%%%%%%%%%%%%%%%%%%%%%%
%%%%%%%%%%%%%%%%%%%%%%%%%%%%%%%%%%%%%%%%%%%%%%%%%%%%%%%%%%%%%%%%%%%%%%%%%%%%%%%%%%%%%%%%%%%%%%%%%%%%%%%%%%%%%%%%%%%%%%%%%%%%%%%%%%%%%%%%%%%%%%%%%%%%%%%%%%%%%%%%%%%%%%%%%%%%%%%%%%%%%%%%%%%%%%%%%%%%%%%%%%%%%%%%%%%%%%%%%%%%%%%%%%%%%%%%%%%%%%%%

\section{Proof of Proposition \ref{Proposition reecriture scalar products}}
\label{Appendix Proof factorisation kappa deformed determinant}

The work \cite{BelliardPakuliakRagoucySlavnovNestedBAScalarProductsSU(3)}
 provided a determinant based representation for 
the scalar product $\mc{S}_{\be}\big(\Ups_{\be},\La\big)$ that was introduced in \eqref{ecriture fonction generatrice S kappa}. For the intermediate calculations that we ought to carry out below 
it is convenient to agree upon the following re-parametrisation of the sets $\La \, = \, \big\{ \{\la_k^{(a)}\}_{k=1}^{N_{\La}^{(a)}} ,\,a=1,2 \big\}$ and 
$\Ups_{\be}\, = \, \big\{ \{\mu_k^{(a)}\}_{k=1}^{N_{\Ups}^{(a)}},\,a=1,2 \big\}$. 
In what concerns the variables of the first type we set 
\beq
\la_a^{(1)}-\i \f{c}{2} \; = \; u_a^B \qquad \e{and} \qquad \mu_{a}^{(1)} -\i \f{c}{2} \; = \; u_a^C \qquad \e{for} \quad a=1,\dots, N\equiv N_{\La}^{(1)}= N_{\Ups}^{(1)}\; , 
\enq
whereas, in what concerns the variables of the second type,
\beq
\la_a^{(2)} - \i c \; = \; v_a^B \qquad \e{and} \qquad \mu_{a}^{(2)} -\i c \; = \; v_a^C \qquad \e{for} \quad a=1,\dots, M\equiv N_{\La}^{(2)}= N_{\Ups}^{(2)} \; .  
\enq

Within such a re-parametrisation, we are precisely in the normalisation adopted in \cite{BelliardPakuliakRagoucySlavnovNestedBAScalarProductsSU(3)}. The latter is slightly 
more adapted for dealing with the large-size formulae that ought to be handled in the course of the proof. 
We start with the representation 

\bem
\mc{S}_{\be}\big( \{u_a^C \}_1^N , \{ v_{a}^{C} \}_1^M \mid  
		 \{ u_{a}^{B} \}_1^N , \{ v_{a}^{B} \}_1^M	 \big) 		\; = \;  
 f(\ov{v}^{C}, \ov{u}^{C} ) \cdot  f(\ov{v}^{B}, \ov{u}^{B} ) 
\cdot t(\ov{v}^C, \ov{u}^{B} ) \cdot \De_N^{\prime}( \ov{u}^{C}  ) \De_N( \ov{u}^{B}  ) \\
\times \De_M^{\prime}( \ov{v}^{C}  ) \De_M( \ov{v}^{B}  )  \cdot 
h^2(\ov{v}^{C}, \ov{u}^{B} )  
\cdot h(\ov{u}^{B}, \ov{u}^{C} ) \cdot  h(\ov{v}^{B}, \ov{v}^{C} ) 
\cdot \det_{N+M}[\mc{N}]\;. 
\end{multline}
In addition to the functions that were already introduced in the course of the statement of the conclusion of the
proposition, we 	also need to define
\beq
t(x,y) \; = \; \f{ - c^2  }{ (x-y) \cdot (x - y + \i c) }  \; , \quad \De_N(\ov{u}^{C}) \; = \; \pl{j<k}{N} g(u_j^{C}, u_k^{C})
\qquad \e{and} \qquad   \De_N^{\prime} (\ov{u}^{C}) \; = \; \pl{j>k}{N} g(u_j^{C}, u_k^{C}) \;. 
\enq
The function $g$ appearing above reads
\beq
 g(x,y) \; = \; \f{ \i c }{ x - y }   \;. 
\enq
Finally the matrix $\mc{N}$ reads, in its block decomposition subordinate to a splitting in respect to the 
two types of roots $u$ and $v$,
\beq
\mc{N} \; = \; \left( \ba{cc} \ex{\be} t\big(u_k^{B}, u_j^{C}\big) + t\big(u_j^{C},u_k^{B}\big) V_{11}\big(u_k^B\big) & 
				\ex{\be} t\big(v_k^C, u_j^{C}\big) V_{12}\big(v_k^C\big) \\
 t\big(v_j^B, u_k^{B}\big) V_{21}\big(u_k^B\big)  & 
 	t\big(v_j^{B}, v_k^{C}\big) + \kappa t\big(v_k^{C},v_j^{B}\big) V_{22}\big(v_k^C\big) 				\ea \right)  
 \; = \; \left( \ba{cc} \mc{N}^{(11)}\big( u_j^{C},u_k^{B} \big)  &  \mc{N}^{(12)}\big( u_j^{C},v_k^{C} \big) \\
\mc{N}^{(21)}\big( v_j^{B},u_k^{B} \big) &  \mc{N}^{(22)}\big( v_j^{B},v_k^{C} \big)   \ea \right) \;. 
\enq
Above, we have introduced four shorthand notations 
\beqa
 V_{11}( \la )  =  \f{ f\big(\ov{v}^B,\la \big) h\big(\ov{u}^C,\la \big) h\big(\la, \ov{u}^B \big)   }
 { f\big(\ov{v}^C,\la \big) h\big(\ov{u}^B,\la \big) h\big(\la, \ov{u}^C \big)   }    & \qquad &
 V_{12}( \la )  =  \f{ h\big(\la, \ov{u}^C \big) h\big(\ov{v}^C,\la \big)    }
{ h\big(\la, \ov{u}^B \big) h\big(\ov{v}^B,\la \big)  }  \hspace{1cm} \\
 V_{21}( \la )  =  \big[ V_{12}\big( \la \big) \big]^{-1}   & \qquad & 
 V_{22}( \la )  =  \f{ f\big(\la, \ov{u}^C \big) h\big(\ov{v}^C,\la \big) h\big(\la, \ov{v}^B \big)    }
 { f\big(\la, \ov{u}^B \big)  h\big(\ov{v}^B,\la \big) h\big(\la, \ov{v}^C \big)  }  \;. 
\eeqa

Let $A$ be the $(N+M)\times (N+M)$  matrix written in block form 
\beq
A \; = \; \left( \ba{cc} A^{(11)}  &  0_{N\times M} \\
A^{(21)} &  A^{(22)} \ea \right)
\enq
where we agree upon 
\beq
A^{(11)}_{kj} \; = \; \f{1}{u_j^C - u_k^B } \f{ \pl{s=1}{N} \big( u_j^C - u_s^B \big) }
{ \pl{ \substack{ s=1 \\ \not= j} }{N} \big( u_j^C - u_s^C \big)  } 
\qquad , \qquad
A^{(21)}_{kj} \; = \; \f{ \pl{s=1}{N} \big( u_j^C - u_s^B \big) }
{ \pl{ \substack{ s=1 \\ \not= j} }{N} \big( u_j^C - u_s^C \big)  } 
\enq
and finally
\beq
A^{(22)}_{kj} \; = \; \f{ \pl{s=1}{N} \big(v_j^B - v_s^C \big) }
{ \pl{ \substack{ s=1 \\ \not= j} }{N} \big( v_j^B - v_s^B \big)  } 
\cdot \times \left\{ \ba{cc}  \i c \cdot  \big(v_j^B - v_k^C \big)^{-1}  & k \not= M \\
								 1 & k=M  \ea \right. \; . 
\enq
It is readily seen that 
\beq
\underset{N+M}{\det}\; \big[ A \big] \; = \; \big (\i c \big)^{M} \f{ \De_N\big(\ov{u}^C\big) \De_M\big(\ov{v}^B\big) }
			{ \De_N\big(\ov{u}^B\big) \De_M\big(\ov{v}^C\big)  } \cdot 
\f{ \sla{g}\big( \ov{v}^{C}, v_M^C \big) }{ g\big( \ov{v}^{B}, v_M^C \big) }			\;. 
\enq
It is easy to derive, with the help of contour integrals, the sum identities
\bem
\sul{j=1}{N}  \f{ - c^2 }{ \big( x-u_j^C\big) \cdot \big(x-u_j^C  \pm \i c \big) }   \cdot 
\f{ \pl{s=1}{N} \big( u_j^C - u_s^B \big) }
{ \pl{ \substack{ s=1 \\ \not= j} }{N} \big( u_j^C - u_s^C \big)  }  \cdot \f{1}{ \big( u_j^C - u_k^B\big)^{\a}} 
\; = \; \f{ \mp \i c  }{ \big( x-u_k^B \pm \i  c\big)^{\a} } 
\pl{s=1}{N} \bigg\{ \f{   x - u_s^B  \pm \i c }{  x - u_s^C \pm \i c  } \bigg\}  \\
\; \pm \;     
\f{ \i c \de_{\a,1} \de_{x,u_k^B} }{ \pl{s=1}{N} \big(u_k^B - u_s^C \big)  } 
\cdot   \pl{\substack{ s=1 \\ \not= k } }{N} \big(  u_k^{B} - u_s^B \big) 
 \; \pm \; \i c \f{ 1-\de_{x,u_k^B}  }{ \big( x-u_k^B \big)^{\a} } \cdot \pl{s=1}{N} \bigg\{ \f{x-u_s^B}{ x-u_s^C }  \bigg\} \;. 
\label{identite somme pour action A11}
\end{multline}
There, we assume that $\a \in \{0,1\}$ and that the parameters $ u^B_{a}$ and $ u^C_{a}$ are all generic. 
In particular, setting $\a=0$ yields 
\beq
\sul{j=1}{N}  \f{ - c^2 }{ \big( x-u_j^C\big) \cdot \big(x-u_j^C  \pm \i c \big) }  
\f{ \pl{s=1}{N} \big( u_j^C - u_s^B \big) }
{ \pl{ \substack{ s=1 \\ \not= j} }{N} \big( u_j^C - u_s^C \big)  }  \; = \; 
\pm \i c \pl{s=1}{N} \bigg\{  \f{   x  - u_s^B  }{  x - u_s^C  }   \bigg\}
\;  \mp \; \i  c \pl{s=1}{N} \f{   x - u_s^B  \pm \i  c }{  x - u_s^C \pm \i c  }  \;. 
\label{identite somme pour action A21}
\enq
Finally, a similar identity involving the $v$-like variables reads 
\bem
\sul{j=1}{M}  \f{ - c^2 }{ \big( x-v_j^B\big) \cdot \big(x-v_j^B  \pm \i c \big) }   \cdot 
\f{ \pl{s=1}{M} \big( v_j^B - v_s^C \big) }
{ \pl{ \substack{ s=1 \\ \not= j} }{M} \big( v_j^B - v_s^B \big)  }  \cdot \f{1}{ \big( v_j^B - v_k^C\big)^{\a}} 
\; = \; \f{ \mp \i c  }{ \big( x-v_k^C \pm \i c\big)^{\a} } 
\pl{s=1}{M} \bigg\{ \f{   x - v_s^C  \pm \i c }{  x - v_s^B \pm \i c  } \bigg\} \\
\; \pm \;   \f{ \i c \de_{\a,1} \de_{x,v_k^C} }{ \pl{s=1}{M} \big(v_k^C - v_s^B \big)  } 
\cdot   \pl{\substack{ s=1 \\ \not= k } }{M} \big(  v_k^{C} - v_s^C \big) 
 \; \pm \; \i c \f{ 1-\de_{x,v_k^C}  }{ \big( x-v_k^C \big)^{\a} } \cdot \pl{s=1}{ M } \bigg\{ \f{x-v_s^C}{ x-v_s^B }  \bigg\} \;. 
\label{identite somme pour action A22}
\end{multline}

A little algebra yields the identities necessary for computing the matrix products relative to the lines arising in 
the upper block:
\beq
\sul{j=1}{N} A_{kj}^{(11)} \mc{N}^{(11)}\big( u_j^C, u_{\ell}^B \big) \; = \; 
+  \Big[ \ex{\be} - V_{11}\big( u_{\ell}^B \big) \Big]  \de_{k \ell} \cdot 
\f{ g\big(u_{k}^{B},\ov{u}^{C}\big) }{  \sla{g}\big(u_{k}^{B},\ov{u}^{B}\big) }
\; - \; \i c \cdot \f{ h\big(u_{\ell}^{B},\ov{u}^{B}\big) }{ h\big(u_{\ell}^{B},\ov{u}^{C}\big) }  
 \cdot \Bigg\{  \f{ \ex{\be} }{u_{\ell}^B - u_{k}^B + \i c } \; + \;   \f{ 1 }{u_{k}^B - u_{\ell}^{B} + \i c }  
 \cdot \f{ f\big(\ov{v}^{B},u_{\ell}^{B}\big) }{ f\big(\ov{v}^{C},u_{\ell}^{B}\big) }  \Bigg\}
\enq
and similarly 
\beq
\sul{j=1}{N} A_{kj}^{(11)} \mc{N}^{(12)}\big( u_j^C, v_{\ell}^C \big) \; = \; 
\ex{\be} \i c \f{ h(\ov{v}^{C}, v_{\ell}^{C} )  }{ h(\ov{v}^{B}, v_{\ell}^{C} )  } \Bigg\{ 
\f{ 1 }{ \big(v_{\ell}^{C}-u_k^{B} \big) } 
\cdot \f{ f\big(v_{\ell}^{C},\ov{u}^{C}\big) }{ f\big(v_{\ell}^{C},\ov{u}^{B}\big) }  
\; - \; \f{  1}{ \big(v_{\ell}^{C}-u_k^{B} + \i c \big)}  \Bigg\}\;. 
\enq
Above, we agree that in the product $\sla{g}\big(u_{k}^{B},\ov{u}^{B}\big)$ one should omit the index $a=k$ which produces a singular term, namely
\beq
\sla{g}\big(u_{k}^{B},\ov{u}^{B}\big) \, = \, \pl{ \substack{ a=1 \\ \not= k}  }{ N } g\big(u_{k}^{B}, u^{B}_a\big) \;. 
\label{definition produits slashes}
\enq
Very similar manipulation yield the identities necessary for computing the matrix products relative 
to the lines arising in the bottom block: 
\beq
\sul{j=1}{N} A_{kj}^{(21)} \mc{N}^{(11)}\big( u_j^C, u_{\ell}^B \big) \; = \; 
 \i c \f{ h\big(u_{\ell}^{B},\ov{u}^{B}\big) }{  h\big(u_{\ell}^{B},\ov{u}^{C}\big) }
 \cdot \Bigg\{ \f{ f\big(\ov{v}^{B},u_{\ell}^{B}\big) }{ f\big(\ov{v}^{C},u_{\ell}^{B}\big) } \; - \;  \ex{ \be } \Bigg\}
\enq
and similarly 
\beq
\sul{j=1}{N} A_{kj}^{(21)} \mc{N}^{(12)}\big( u_j^C, v_{\ell}^C \big) \; = \; 
\ex{\be}  \i c \f{ h\big(\ov{v}^{C}, v_{\ell}^{C} \big) }{  h\big(\ov{v}^{B},v_{\ell}^{C}\big) }
 \cdot \Bigg\{ \f{ f\big(v_{\ell}^{C},\ov{u}^{C}\big) }{ f\big(v_{\ell}^{C},\ov{u}^{B}\big) } \; - \;  1 \Bigg\} \;. 
\enq
Finally, one gets that 
\beq
\sul{j=1}{M} A_{kj}^{(22)} \mc{N}^{(21)}\big( v_j^B, u_{\ell}^B \big) \; = \; 
\i c \f{ h(u_{\ell}^{B}, \ov{u}^{B})  }{ h(u_{\ell}^{B}, \ov{u}^{C}) }
 \Bigg\{  \Big( \f{ \i c }{ u_{\ell}^{B}-v_k^{C} - \i c   }  \Big)^{1-\de_{kM}}  \; - \; 
\Big( \f{ \i c }{ u_{\ell}^{B}-v_k^{C}   }  \Big)^{1-\de_{kM}}
\cdot \f{ f\big(\ov{v}^{B},u_{\ell}^{B}\big) }{ f\big(\ov{v}^{C}, u_{\ell}^{B} \big) }  
\Bigg\}
\enq
and similarly, for $k=1,\dots, M-1$
\bem
\sul{j=1}{M} A_{kj}^{(22)} \mc{N}^{(22)}\big( v_j^B, v_{\ell}^C \big) \; = \; 
- \Big[ 1-\ex{\be} V_{22}\big( v_{\ell}^C \big) \Big]  \de_{k \ell} \cdot \i c \cdot 
\f{ g\big(v_{k}^{C},\ov{v}^{B}\big) }{  \sla{g}\big(v_{k}^{C},\ov{v}^{C}\big) }
\\
\; - \; \i c \cdot \f{ h\big(\ov{v}^{C},v_{\ell}^{C}\big) }{ h\big(\ov{v}^{B},v_{\ell}^{C}\big) }  
 \cdot \Bigg\{  \f{ \i c }{v_k^C - v_{\ell}^C + \i c } \; + \;   \f{ \ex{\be} \i c }{v_{\ell}^C - v_{k}^C + \i c }  
 \cdot \f{ f\big(v_{\ell}^{C},\ov{u}^{C}\big) }{ f\big(v_{\ell}^{C}, \ov{u}^{B}\big) }  \Bigg\}
\end{multline}
whereas, for $k=M$, one has
\beq
\sul{j=1}{M} A_{Mj}^{(22)} \mc{N}^{(22)}\big( v_j^B, v_{\ell}^C \big) \; = \; 
 \i c \cdot \f{ h\big(\ov{v}^{C},v_{\ell}^{C}\big) }{ h\big(\ov{v}^{B},v_{\ell}^{C}\big) }  
 \cdot \Bigg\{ 1\; - \;   \ex{\be}  \cdot \f{ f\big(v_{\ell}^{C},\ov{u}^{C}\big) }{ f\big(v_{\ell}^{B}, \ov{u}^{B}\big) }  \Bigg\} \;. 
\enq

\vspace{3mm}
\noindent Upon recasting the entries of the matrix  $A\mc{N}$ with the help of these identities, we simplify its determinant by factoring 
\beq
 \i c \cdot \f{ h\big(u_{\ell}^{B},\ov{u}^{B}\big) }{ h\big(u_{\ell}^{B},\ov{u}^{C}\big) }  
\enq
out of all $u$-type columns and the terms
\beq
 \i c \cdot \f{ h\big(\ov{v}^{C},v_{\ell}^{C}\big) }{ h\big(\ov{v}^{B},v_{\ell}^{C}\big) }  
\enq
out of all $v$-type columns. This leads to 
\beq
\underset{N+M}{\det}\big[ A \mc{N} \big] \; = \; \big( \i c \big)^{N+M} \cdot 
\f{ h\big(\ov{u}^{B},\ov{u}^{B},\big) h\big(\ov{v}^{C},\ov{v}^{C}\big) }
{ h\big(\ov{u}^{B},\ov{u}^{C}\big)  h\big(\ov{v}^{B},\ov{v}^{C}\big) }  
\cdot \underset{N+M}{\det}\big[ \mc{B} \big] 
\enq
in which the matrix $\mc{B}$ given in block form
\beq
\mc{B} \; = \; \left( \ba{cc} \mc{B}^{(11)}\big( u_k^{B}, u_{\ell}^B \big)  &  
			\mc{B}^{(12)}\big( u_k^{B},v_{\ell}^C \big) \\
\mc{B}^{(21)}\big( v_k^{C}, u_{\ell}^B \big) &  \mc{B}^{(22)}\big( v_k^{C}, v_{\ell}^C \big)   \ea \right) 
\enq
has its block entries given by 
\beq
\mc{B}^{(11)}\big( u_k^{B}, u_{\ell}^B \big) \; = \; (\i c)^{-1} \cdot \Big[  \ex{\be} - V_{11}\big( u_{\ell}^B \big) \Big]  \de_{k \ell} \cdot 
\f{ f\big(u_{\ell}^{B},\ov{u}^{C}\big) }{  \sla{f}\big(u_{\ell}^{B},\ov{u}^{B}\big) }
-\f{ \ex{\be} }{u_{\ell}^B - u_{k}^B + \i c } \; - \;   \f{ 1 }{u_{k}^B - u_{\ell}^B + \i c }  
 \cdot \f{ f\big(\ov{v}^{B},u_{\ell}^{B}\big) }{ f\big(\ov{v}^{C}, u_{\ell}^{B}\big) } 
\enq
in what concerns the $(u,u)$ block, whereas the $(u,v)$ block reads 
\beq
\mc{B}^{(12)}\big( u_k^{B}, v_{\ell}^C \big) \; = \; \ex{\be} \cdot \Bigg\{ 
\f{ 1 }{v_{\ell}^C- u_{k}^B  } \cdot \f{ f\big(v_{\ell}^{C},\ov{u}^{C}\big) }{ f\big(v_{\ell}^{B},\ov{u}^{B}\big) } 
\; - \; \f{1}{v_{\ell}^C - u_{k}^B + \i c}  \Bigg\} \, .
\enq
Above, we made use of the same convention as in \eqref{definition produits slashes} relatively to $\sla{f}\big(u_{\ell}^{B},\ov{u}^{B}\big) $. 
Finally, for $k=1,\dots, M-1$, the $(v,u)$ and $(v,v)$ blocks are given by 
\beq
\mc{B}^{(21)}\big( v_k^{C}, u_{\ell}^B \big) \; = \;  \f{ \i c }{u_{\ell}^B - v_{k}^C + \i c} - \ex{\be}
 \; + \; f\big(v_k^C,u_{\ell}^B \big)  \cdot \f{ f\big(\ov{v}^{B},u_{\ell}^{B}\big) }{ f\big(\ov{v}^{C},u_{\ell}^{B}\big) } 
\enq
\beq
\mc{B}^{(22)}\big( v_k^{C}, v_{\ell}^C \big) \; = \;
\Big[ 1- \ex{\be}  V_{22}\big( v_{\ell}^C \big) \Big]  \de_{k \ell} \cdot 
\f{ f\big(\ov{v}^{B},v_{\ell}^{C}\big) }{  \sla{f}\big(\ov{v}^{C},v_{\ell}^{C}\big) }
\; + \; \f{ \ex{\be} }{ f\big(v_{\ell}^C, v_k^{C} \big) } \cdot \f{ f\big(v_{\ell}^{C},\ov{u}^{C}\big) }{ f\big(v_{\ell}^{C},\ov{u}^{B}\big) }  
\; -\; \Bigg\{ \f{ \i c }{ v_k^{C} - v_{\ell}^C + \i c } \; + \;    \ex{\be} \Bigg\}
\enq
whereas, at $k=M$ we have 
\beq
\mc{B}^{(u)}\big( v_M^{C}, u_{\ell}^B \big) \; =\; \mc{B}^{(v)}_1\big( v_M^{C}, v_{\ell}^C \big) \; = \; 1-\ex{\be} \;. 
\enq

We then subtract the last column of $\mc{B}$ from all the others. Upon factorising the diagonal elements, we do get that 
\beq
\underset{N+M}{\det} \big[ \mc{B} \big] \; = \; \f{1-\ex{\be}}{ (\i c)^{N}} \cdot 
\f{ f\big( \ov{v}^{B},\ov{v}^{C}\setminus\{v_M^C\} \big) f\big(\ov{u}^{B},\ov{u}^{C}\big) }
{  \sla{f}\big(\ov{v}^{C},\ov{v}^{C}\setminus\{v_M^C\}\big) \cdot \sla{f}\big(\ov{u}^{B},\ov{u}^{B}\big) }
\cdot \Big(\ex{\be} - V_{11}\big(\ov{u}^B\big) \Big) \cdot \Big(1-\ex{\be} V_{22}\big(\ov{v}^C\setminus \{v_M^C\} \big) \Big)
\underset{N+M}{\det} \big[ I_{N+M} \; + \; \wh{\mc{U}}_{v_M^C} \big] 
\label{equation pour factorisation determinant de B}
\enq
where the block decomposition of the matrix $\wh{\mc{U}}_{\th}$ takes the form 
\beq
\wh{\mc{U}}_{\th}  \; = \; 
\left( \ba{cc}  2 \i \pi \e{Res}_{ \om^{\prime} = u_{\ell}^B }\Big( \wh{\mc{U}}^{(11)}_{\th}
					\big( u_k^{B}, \om^{\prime} \big) \Big)  &  
		\qquad 	2 \i \pi \e{Res}_{\om^{\prime} = v_{\ell}^C}\Big( \wh{\mc{U}}^{(12)}_{\th}
							\big( u_k^{B},\om^{\prime} \big)   \Big)  \vspace{2mm} \\
2 \i\pi \e{Res}_{ \om^{\prime} = u_{\ell}^B}\Big( \wh{\mc{U}}^{(21)}_{\th}\big( v_k^{C}, \om^{\prime} \big)  \Big) 
		 & \qquad  2 \i \pi \e{Res}_{\om^{\prime} = v_{\ell}^C}\Big( \wh{\mc{U}}^{(22)}_{\th}
		\big( v_k^{C},\om^{\prime} \big)   \Big) \ea \right) 
\enq
and $\sla{f}$ means the product where \textit{all} coinciding elements between the first and second argument
of $f$ are omitted. 

The block entries of the matrix $ \wh{\mc{U}}_{\th} $ are given in terms of four auxiliary kernels. 
The ones of the first block column read 
\bem
\wh{\mc{U}}^{(11)}_{\th}\big( \om , \om^{\prime} \big)  \; = \; 
\f{ (2\i\pi)^{-1 } }{ 1-\ex{-\be} V_{11}\big( \om^{\prime} \big) }
\f{ f\big(\om^{\prime},\ov{u}^{B}\big) }{  f\big( \om^{\prime} ,\ov{u}^{C}\big) }
\Bigg\{  \bigg[ \f{ 1 }{ \th  - \om + \i c } \; - \;   \f{ 1 }{\om^{\prime} - \om + \i c }   \bigg]
\\
\; - \; \bigg[ \f{ \ex{-\be} }{ \om  - \om^{\prime} + \i c } \cdot
 \f{ f\big(\ov{v}^{B}, \om^{\prime}  \big) }{ f\big(\ov{v}^{C}, \om^{\prime} \big) }  
\; + \;   \f{ 1 }{\th - \om  }  \cdot \f{ f\big( \th, \ov{u}^{C}  \big) }{ f\big(\th ,\ov{u}^{B} \big) }   \bigg] \Bigg\}
\end{multline}
\beq
\wh{\mc{U}}^{(21)}_{\th}\big( \om , \om^{\prime} \big)  \; = \; 
\f{ (2 \i\pi)^{-1 }  }{ \ex{\be} -  V_{11}\big( \om^{\prime} \big) }
\f{ f\big(\om^{\prime},\ov{u}^{B}\big) }{  f\big( \om^{\prime} ,\ov{u}^{C}\big) }
\Bigg\{  f(\om, \om^{\prime})  \cdot \f{ f\big(\ov{v}^{B}, \om^{\prime}  \big) }{ f\big(\ov{v}^{C}, \om^{\prime} \big) }  
\; - \;\f{ \ex{\be} }{ f(\th, \om)   } \cdot \f{ f\big( \th, \ov{u}^{C}  \big) }{ f\big(\th ,\ov{u}^{B} \big) } 
\; + \; \f{ \i c }{\om^{\prime} - \om + \i c}  \; + \; \f{ \i c }{\om - \th + \i c} \Bigg\} 
\enq
whereas those of the second block column are given by 
\beq
\wh{\mc{U}}^{(12)}_{\th}\big( \om , \om^{\prime} \big)  \; = \; 
\f{  \ex{\be}  \cdot (-2\pi c)^{-1 } }{ 1 - \ex{\be} V_{22}\big( \om^{\prime} \big) }
\f{ f\big(\ov{v}^{C},\om^{\prime}\big) }{  f\big( \ov{v}^{B},\om^{\prime}\big) } \cdot
\Bigg\{  \f{1}{\om^{\prime} - \om } \cdot \f{ f\big( \om^{\prime}, \ov{u}^{C}  \big) }{ f\big(\om^{\prime} ,\ov{u}^{B} \big) } 
\; - \;\f{1}{\th - \om } \cdot \f{ f\big( \th, \ov{u}^{C}  \big) }{ f\big(\th ,\ov{u}^{B} \big) } 
\; + \; \f{1}{ \th - \om + \i c}  \; -\; \f{1}{\om^{\prime} - \om + \i c} \Bigg\} 
\enq
\beq
\wh{\mc{U}}^{(22)}_{\th}\big( \om , \om^{\prime} \big)  \; = \; 
\f{  (- 2\pi c)^{-1 }  }{ 1 - \ex{\be} V_{22}\big( \om^{\prime} \big) }
\f{ f\big(\ov{v}^{C},\om^{\prime}\big) }{  f\big( \ov{v}^{B},\om^{\prime}\big) } \cdot
\Bigg\{  \f{\ex{\be}}{ f\big(\om^{\prime}, \om \big) } 
			\cdot \f{ f\big( \om^{\prime}, \ov{u}^{C}  \big) }{ f\big(\om^{\prime} ,\ov{u}^{B} \big) } 
\; - \;\f{\ex{\be}}{f\big(\th,\om) } \cdot \f{ f\big( \th, \ov{u}^{C}  \big) }{ f\big(\th ,\ov{u}^{B} \big) } 
\; + \; \f{ \i c}{ \om - \th + \i  c}  \; -\; \f{ \i c}{\om - \om^{\prime} + \i c} \Bigg\}  \;. 
\enq

The determinant occurring in the \textit{rhs} of \eqref{equation pour factorisation determinant de B} 
is readily recast into the Fredholm determinant of the operator $\e{id}+\wh{\op{U}}_{\th}$, with $\wh{\op{U}}_{\th}$ being of finite rank and acting on the contour 
\beq
\msc{C}_{u,v} \; = \; \Ga\Big( \{u_a^B\}_1^N \Big) \cup \Ga\Big( \{v_a^C\}_1^M \Big) \;. 
\label{ecriture contour integration splitting}
\enq
More precisely, one has that 
\beq
\underset{N+M}{\det} \Big[ I_{N+M} \; + \; \mc{U}_{v_M^C} \Big] \; = \; 
\underset{\msc{C}_{u,v}}{\det} \; \Big[ \e{id}+\wh{\op{U}}_{v_M^{C}} \Big]
\enq
where the block decomposition of the kernel $\wh{U}_{\th}(\om,\om^{\prime})$ in respect to the splitting 
of the integration contour given in \eqref{ecriture contour integration splitting}
\beq
\wh{U}_{\th}  \; = \; 
\left( \ba{cc}  \wh{\mc{U}}^{(11)}_{\th}\big( \om, \om^{\prime} \big)   &  \wh{\mc{U}}^{(12)}_{\th}\big( \om, \om^{\prime} \big) 	\vspace{2mm} \\
\wh{\mc{U}}^{(21)}_{\th}\big( \om, \om^{\prime} \big) & \wh{\mc{U}}^{(22)}_{\th}\big( \om, \om^{\prime} \big)	  \ea \right)  \;. 
\enq

Hence, by putting together all the terms, one is led to 
\bem
\mc{S}_{\be}\big( \{ u^C \} , \{ v^C \} \mid  
		 \{ u^B \} , \{ v^B \}	 \big) 		\; = \;  
(1-\ex{\be})			 \cdot 			\f{ h\big(\ov{v}^{C},v^{C}_M\big) }{  h\big(\ov{v}^{B},v^{C}_M\big)   }
\cdot f\big(\ov{v}^{C},\ov{u}^B\cup \ov{u}^{C}\big)  \\
\times  f\big(\ov{v}^{B},\ov{u}^B\cup \ov{v}^{C}\big) \cdot f\big(\ov{u}^{B},\ov{u}^{C}\big) 
\cdot \Big(\ex{\be} - V_{11}\big(\ov{u}^B\big) \Big) \cdot \Big(1-\ex{\be} V_{22}\big(\ov{v}^C\setminus\{v_M^{C}\}\big) \Big)		 
	\cdot 	 \underset{\msc{C}_{u,v}}{\det} \; \Big[\e{id}+\wh{\op{U}}_{v_M^{C}} \Big] \;. 
\end{multline}
Thus, upon implementing the afore-discussed correspondence between the variables $(u,v)$
 and those corresponding to real valued solutions to the Bethe equations, one is led to the claim. 
Finally, observe that instead of choosing the $M^{\e{th}}$ line in the second block, one could have chosen any other. 
This entails that one can, in fact, do the substitution $v_M^{C} \hookrightarrow v_a^{C}$ in the above formula. \qed

\section{Main definitions  for $SU(3)$ invariant XXX model \label{sec:notations}}

 For the $SU(3)$ invariant XXX model, we use the following functions 
\beqa
&& g(x,y) \; = \; \f{ \i c}{ x - y}  \quad , \quad f(x,y) \; = \; \f{x - y + \i c}{ x - y}  \quad , \quad k(x,y) \; = \; \f{ x - y + \i \tf{c}{2} }{ x - y - \i \tf{c}{2} } \,, 
\label{fct-1}
\\
&& t(x,y) \; = \; \f{- c^2}{(x - y + \i c)(x - y)}  \quad , \quad  h(x,y) \; =  \; \f{x-y + \i c}{ \i c} \; .
\eeqa
 We also introduce the short hand notation for $\ov{u}=\{u_j,\,j=1,...,N\}$, $\ov{v}=\{v_j,\,j=1,...,M\}$ and any function $\ff$ of two variables
\beq
\ff(x,\ov{u}) \; = \; \pl{j=1}{N} \ff(x, u_j) \quad , \quad  \ff(\ov{u},\ov{v}) \; = \; \pl{j=1}{N} \pl{k=1}{M} \ff(u_j,v_k)
\quad , \quad  \sla{\!\!{\ff}}(\ov{u},\ov{u}) \; = \; \pl{j\neq k}{N} \ff(u_j,u_k)
\enq
as well as for $g$ given in \eqref{fct-1}
\beq
\De_N(\ov{u}) \; = \; \pl{j<k}{N} g(u_j, u_k)
\qquad \e{and} \qquad   \De_N^{\prime} (\ov{u}) \; = \; \pl{j>k}{N} g(u_j, u_k) \;. 
\enq
We also define the XXX bare  phase and momentum 
\beq
\vth_n(\om) \; = \; \f{1}{2 \i \pi } \ln \bigg( \f{ \tf{\i c}{n} +\om}{ \tf{\i c}{n} -\om  } \bigg) \qquad \e{and}\qquad 
p_{0;XXX}^{(1)}(\om) \; = \; \f{ \i }{ 2 \pi } \ln \bigg( \f{ \tf{\i c}{2} +\om}{ \tf{\i c}{2} -\om  } \bigg) \; . 
\enq

\end{document}